\documentclass[aps,prd,twocolumn,showpacs,nofootinbib,eqsecnum,
superscriptaddress]{revtex4-1}

\usepackage[centertags]{amsmath}
\usepackage{amssymb}
\usepackage{latexsym}
\usepackage{enumerate}
\usepackage{graphicx}
\usepackage{mathrsfs}
\usepackage[hidelinks]{hyperref}
\usepackage{stmaryrd}
\usepackage{import}
\usepackage{tensor}
\usepackage{color}
\usepackage{bm}
\usepackage{multirow}
\usepackage{breakurl}
\usepackage{float}
\usepackage{lipsum}
\usepackage{siunitx}
\usepackage{comment}
\usepackage{mathtools}

\mathtoolsset{centercolon}

\usepackage{stmaryrd}
\usepackage{color}
\usepackage{amsmath}
\usepackage{amssymb}
\usepackage{latexsym}
\usepackage{enumerate}
\usepackage{graphicx}
\usepackage{mathrsfs}
\usepackage{physics}
\usepackage{tabularx}
\usepackage{mathtools}
\makeindex

\newcommand{\be}{\begin{equation}}
\newcommand{\ee}{\end{equation}}
\newcommand{\bea}{\begin{eqnarray}}
\newcommand{\eea}{\end{eqnarray}}
\newcommand{\ba}{\begin{array}}
\newcommand{\ea}{\end{array}}
\newcommand{\bi}{\begin{itemize}}
\newcommand{\ei}{\end{itemize}}

\renewcommand{\l}{\left(}
\renewcommand{\r}{\right)}
\renewcommand{\a}{\alpha}
\renewcommand{\b}{\beta}
\newcommand{\g}{\gamma}
\newcommand{\G}{\Gamma}
\renewcommand{\d}{\delta}
\newcommand{\D}{\Delta}
\newcommand{\eps}{\epsilon}
\newcommand{\La}{\Lambda}
\newcommand{\la}{\lambda}
\newcommand{\vp}{\varphi}
\renewcommand{\O}{\Omega}
\renewcommand{\o}{\omega}
\renewcommand{\th}{\theta}

\newcommand{\s}{\sigma}
\newcommand{\Sig}{\Sigma}
\newcommand{\Up}{\Upsilon}

\newcommand{\ti}{\tilde}

\DeclareMathOperator{\sign}{sgn}

\newcommand{\lhat}{\hat{l}}
\newcommand{\etal}{et al.~}

\begin{document}

\title{Resonant self-force effects in extreme-mass-ratio binaries: A scalar model}
\author{Zachary Nasipak}
\affiliation{NASA Goddard Space Flight Center, 8800 Greenbelt Road, Greenbelt, Maryland 20771, USA}
\affiliation{Institute for Computational and Experimental Research in
Mathematics, Brown University, Providence,
Rhode Island 02903, USA}
\affiliation{Department of Physics and Astronomy, University of North
Carolina, Chapel Hill, North Carolina 27599, USA}
\author{Charles R. Evans}
\affiliation{Department of Physics and Astronomy, University of North
Carolina, Chapel Hill, North Carolina 27599, USA}

\begin{abstract}
Extreme-mass ratio inspirals (EMRIs), compact binaries with small mass ratios $\epsilon\ll 1$, will be important
sources for low-frequency gravitational wave detectors.  Almost all EMRIs will evolve through important transient
orbital $r\theta$ resonances, which will enhance or diminish their gravitational wave flux, thereby affecting the
phase evolution of the waveforms at $O(\epsilon^{1/2})$ relative to leading order.  While modeling the local
gravitational self-force (GSF) during resonances is essential for generating accurate EMRI waveforms, so far the full
GSF has not been calculated for an $r\theta$-resonant orbit owing to computational demands of the problem.  As a
first step we employ a simpler model, calculating the scalar self-force (SSF) along $r\theta$-resonant geodesics in
Kerr spacetime.  We demonstrate two ways of calculating the $r\theta$-resonant SSF (and likely GSF), with one method
leaving the radial and polar motions initially independent as if the geodesic is nonresonant.  We illustrate results
by calculating the SSF along geodesics defined by three $r\theta$-resonant ratios (1:3, 1:2, 2:3).  We show how the
SSF and averaged evolution of the orbital constants vary with the initial phase at which an EMRI enters resonance.
We then use our SSF data to test a previously proposed integrability conjecture, which argues that conservative
effects vanish at adiabatic order during resonances.  We find prominent contributions from the conservative SSF to
the secular evolution of the Carter constant, $\langle \dot{\mathcal{Q}}\rangle$, but these nonvanishing
contributions are on the order of, or less than, the estimated uncertainties of our self-force results.  The
uncertainties come from residual, incomplete removal of the singular field in the regularization process.  Higher
order regularization parameters, once available, will allow definitive tests of the integrability conjecture.
\end{abstract}


\maketitle

\section{Introduction}

The future LISA space mission \cite{ESA12,NASA11} will build upon the success
of current ground-based detectors by observing new gravitational wave sources in the milli-Hertz band
\cite{AmarETC13,AmarETC17,BerrETC19}.  Among these new sources are extreme-mass-ratio inspirals (EMRIs): binaries
composed of a stellar-mass compact object ($\mu \sim 10 M_\odot$) in bound orbit about a massive black hole
($M\sim 10^6 M_\odot$).  With their small mass ratios ($\eps\equiv\mu/M \ll 1$), EMRIs evolve adiabatically due
to orbit-averaged gravitational wave fluxes, with the smaller secondary body completing $\sim 10^5$ orbits around the
more massive primary as the binary emits gravitational waves visible to low-frequency detectors.  As a result of their
long durations, EMRI signals are expected to have cumulative signal-to-noise ratios (SNRs) of several tens to several
hundreds, allowing high-precision measurements that exceed the capabilities of current ground-based gravitational
wave observatories \cite{BerrETC19,BakeETC19}.  Such high SNR measurements will still require accurate waveform
models to assist in detecting events and in measuring physical source parameters \cite{BabaETC17}.

EMRIs are naturally modeled within the framework of black hole perturbation theory (BHPT), in which the small compact
object is treated as a perturbing body in the stationary background spacetime associated with the primary black hole
\cite{Mino03,PoisPounVega11}. On the orbital timescale $T_\mathrm{orb}\sim M$, the dynamics of the small body closely
approximates a geodesic with a trio of fundamental frequencies (the radial frequency $\O_r$, polar frequency
$\O_\th$, and azimuthal frequency $\O_\vp$) \cite{Schm02,DrasFlanHugh05,FujiHiki09}.  However, as the system evolves
the small body is gradually pushed away from this geodesic as it interacts with its own gravitational perturbation.
This behavior is typically described in terms of a local gravitational self-force (GSF) \cite{MinoSasaTana97,QuinWald97}
that acts on the secondary and, at leading perturbative-order, makes an $O(\eps)$ correction to its motion.  The
dissipative piece of the GSF drives the adiabatic inspiral of the secondary, while the conservative piece provides
nonsecular perturbations to the motion.  The secular effect of the averaged dissipative GSF dominates the phase
evolution of the orbit, which accumulates like $\sim \eps^{-1}$ at leading adiabatic order.  This defines the inspiral
or radiation timescale $T_\mathrm{rr}\sim M\eps^{-1}$, over which the orbital frequencies undergo order-unity
fractional changes. Furthermore, at a more subtle level, perturbations due to the oscillatory part of the dissipative
and conservative GSF produce small shifts in the orbital frequencies that affect the orbit and cumulative
phase at a level $O(\eps)$ relative to the leading order adiabatic inspiral \cite{HindFlan08}.  This correction is
the post-1 adiabatic order effect.

In almost all astrophysically-relevant EMRIs, the evolution of the orbital frequencies due to the GSF will cause
these systems to pass through one or more consequential orbital resonances at some point in their observed inspirals
\cite{RuanHugh14,BerrETC16}.  An orbital resonance occurs when at least two frequencies of orbital motion, $\O_1$
and $\O_2$, form a rational ratio, i.e., $\O_2/\O_1 =\b_2/\b_1$ with coprime $\b_1,\b_2\in \mathbb{Z}$.  The smaller
that the integers $\beta_1$ and $\beta_2$ are, the stronger the resonance.  In the solar system, orbital resonances
commonly occur among satellites sharing the same primary, such as the 2:3 resonance of Neptune and Pluto, and the
1:2 resonance of the Galilean satellites Io and Europa.

For EMRIs, resonances can form between any two of the three frequencies of the smaller body's orbital motion, with
different resonant combinations leading to different physical effects.\footnote{See
Refs.~\cite{GairYuneBend12,Vand14a,BaraPoun19} for comprehensive lists of research focused on orbital resonances in
Kerr spacetime, and \cite{GairYuneBend12,LukeWitz21} for more general discussions of EMRI resonances.}  For instance,
$r\vp$ \cite{Vand14b} and
$\th\vp$ resonances \cite{Hira11} can lead to anisotropic radiation of gravitational waves, resulting in resonant
`kicks' to the velocity of an EMRI's center-of-mass.  Such effects are expected to contribute to an EMRI's phase
evolution and waveform at $O(\eps^{3/2})$ relative to adiabatic order (i.e., post-$\tfrac{3}{2}$ adiabatic order)
\cite{Vand14b}.  Because LISA waveforms require a phase accuracy of $\sim 0.1$ radians, $r\vp$- and
$\th\vp$ resonances are presumably safe to neglect at present in EMRI models.  On the other hand, $r\th$ resonances,
which only arise in EMRIs with Kerr primaries, will enhance or diminish the time-averaged gravitational wave flux,
thereby influencing the evolution of the frequencies $\O_r$, $\O_\th$, and $\O_\vp$ (and similarly the orbital
energy, angular momentum, and Carter constant) \cite{FlanHind12,FlanHughRuan14,BerrETC16}.  The strongest
resonances will have frequency ratios such as 1:3, 1:2, or 2:3 \cite{RuanHugh14,MihaGair17} and will ordinarily
persist for a resonant period $T_\text{res}\sim M\eps^{-1/2}$ \cite{FlanHind12}.  Because these orbital resonances
are expected to be transient,\footnote{It may be possible for certain EMRIs to be caught in a sustained
$r\th$ resonance, but a system must meet stringent (if even possible) conditions for this to occur \cite{Vand14a}. }
their effect is at post-$\tfrac{1}{2}$ adiabatic order \cite{FlanHind12}.\footnote{While not the focus of this
work, tidal resonances, which may occur for EMRIs that are perturbed by one or more external bodies, can experience
similar post-$\tfrac{1}{2}$ adiabatic corrections to the phase evolution \cite{YangCasa17,BongYangHugh19, GuptETC21}.}

Incorporating this new resonant timescale into evolutionary models poses difficulties with, for example, near-identity
transformations \cite{VandWarb18} and multiscale expansions \cite{BaraPoun19,LukeWitz21}.  In effect, on a timescale
just larger than $T_\text{res}$, an EMRI's orbital parameters appear to experience $O(\eps^{1/2})$ shifts or jumps in
their values, as shown in \cite{FlanHind12,BerrETC16}.  These jumps, which are sensitive to the orbital phase of the
EMRI as it enters resonance \cite{FlanHughRuan14}, lead to an overall $O(\eps^{-1/2})$ shift in the cumulative phase
of the system.  Failing to account for these resonant phase shifts may
only slightly degrade EMRI detection rates \cite{BerrETC16}, but it will introduce significant systematic biases to
EMRI parameter estimation that will dominate over standard statistical errors \cite{SperGair21}.
Because nearly all EMRIs will encounter either a 1:3, 1:2, or 2:3 $r\th$ resonance as they emit gravitational wave
signals in the LISA passband \cite{RuanHugh14}, properly modeling transient $r\th$ resonances is essential to
producing subradian phase-accurate waveforms for the detection and characterization of EMRIs by LISA.

To date, numerical investigations of transient $r\th$ resonances have not incorporated local strong-field
conservative perturbations \cite{FlanHind12,FlanHughRuan14,BerrETC16,SperGair21}.
While conservative perturbations vanish at adiabatic order for nonresonant motion, it is not yet known if they may
contribute to the adiabatic evolution of EMRIs during $r\th$ resonances.
Flanagan and Hinderer \cite{FlanHind12} argue that conservative
effects will not contribute at leading order during resonances based on their \textit{integrability
conjecture}.\footnote{We follow \cite{FlanHughRuan14} in referring to Flanagan and Hinderer's argument as the
integrability conjecture.}  The validity of this conjecture remains unclear.  Isoyama
\etal \cite{IsoyETC13,IsoyETC19} have found the presence of potential conservative contributions to the evolution
of the orbit-averaged Carter constant through their Hamiltonian formulation of EMRI dynamics.  For the integrability
conjecture to hold, these terms would have to vanish when integrated over an entire orbit, which has not been
demonstrated analytically.

The integrability conjecture can potentially be tested through numerical calculations of conservative quantities,
such as a computation of the local GSF in Kerr spacetime during an $r\th$ resonance.  First-order GSF
calculations for generic bound orbits about a Schwarzschild black hole are well advanced, to the point of allowing
long-term evolutionary computations \cite{WarbETC12,OsbuWarbEvan16,WarbOsbuEvan17,VandWarb18}.  Indeed, even
second-order GSF calculations for restricted orbits in Schwarzschild are now emerging \cite{PounETC20}.  Recent work
by van de Meent provided the first GSF results for generic orbits in a Kerr EMRI \cite{Vand18}.  However, these
calculations are much more computationally demanding than their Schwarzschild counterparts, and long-term, self-forced
evolutions of Kerr EMRIs have not yet been produced.  Not even snapshots of the GSF in Kerr EMRIs during
$r\th$ resonances have been attempted.

Thus, as a first step in exploring local radiation-reaction effects driven by resonances, we consider a
scalar field analogue and calculate the scalar self-force (SSF) that arises due to a bound particle with scalar
charge $q$ in an $r\th$ resonance about a Kerr black hole.  This work builds off of a
previous paper \cite{NasiOsbuEvan19} in which we presented the first calculation of the SSF for nonresonant
inclined, eccentric orbits in Kerr spacetime.  By treating $q^2/(\mu M)$ as a small parameter, this SSF model mimics
the GSF problem.  As is common for adiabatic and osculating geodesic evolutions, we assume the motion of the particle
to be exactly geodesic and calculate the resulting \textit{geodesic scalar self-force}.  This assumption produces
an error that is on the order of the time-averaged dissipative part of the second-order self-force.  In this work
we are not yet concerned with applying the self-force and calculating a portion of the evolution based on the
backreaction.

\begin{figure*}[th]
	\includegraphics[width=0.35\textwidth]{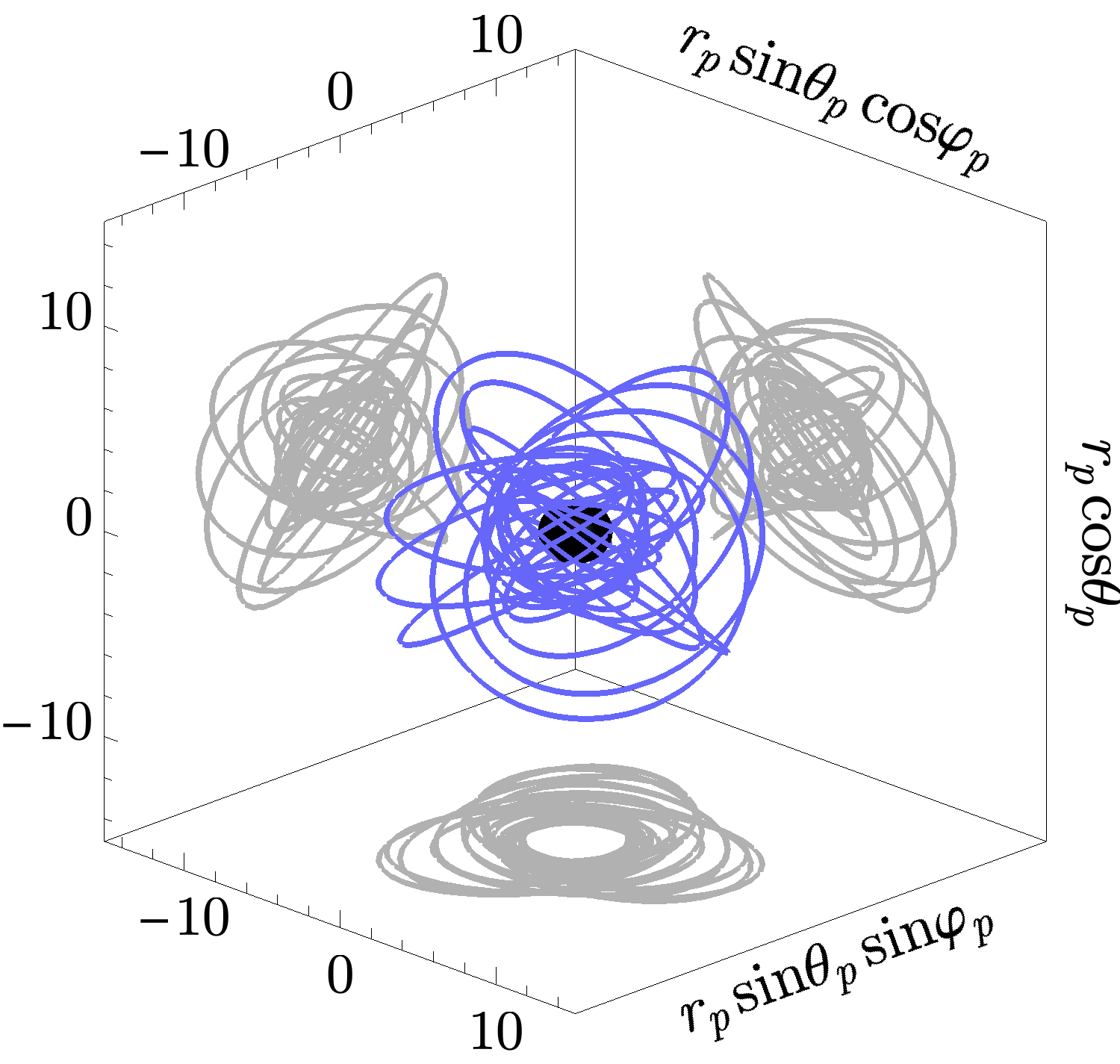}
	\hspace{1cm}
	\includegraphics[width=0.35\textwidth]{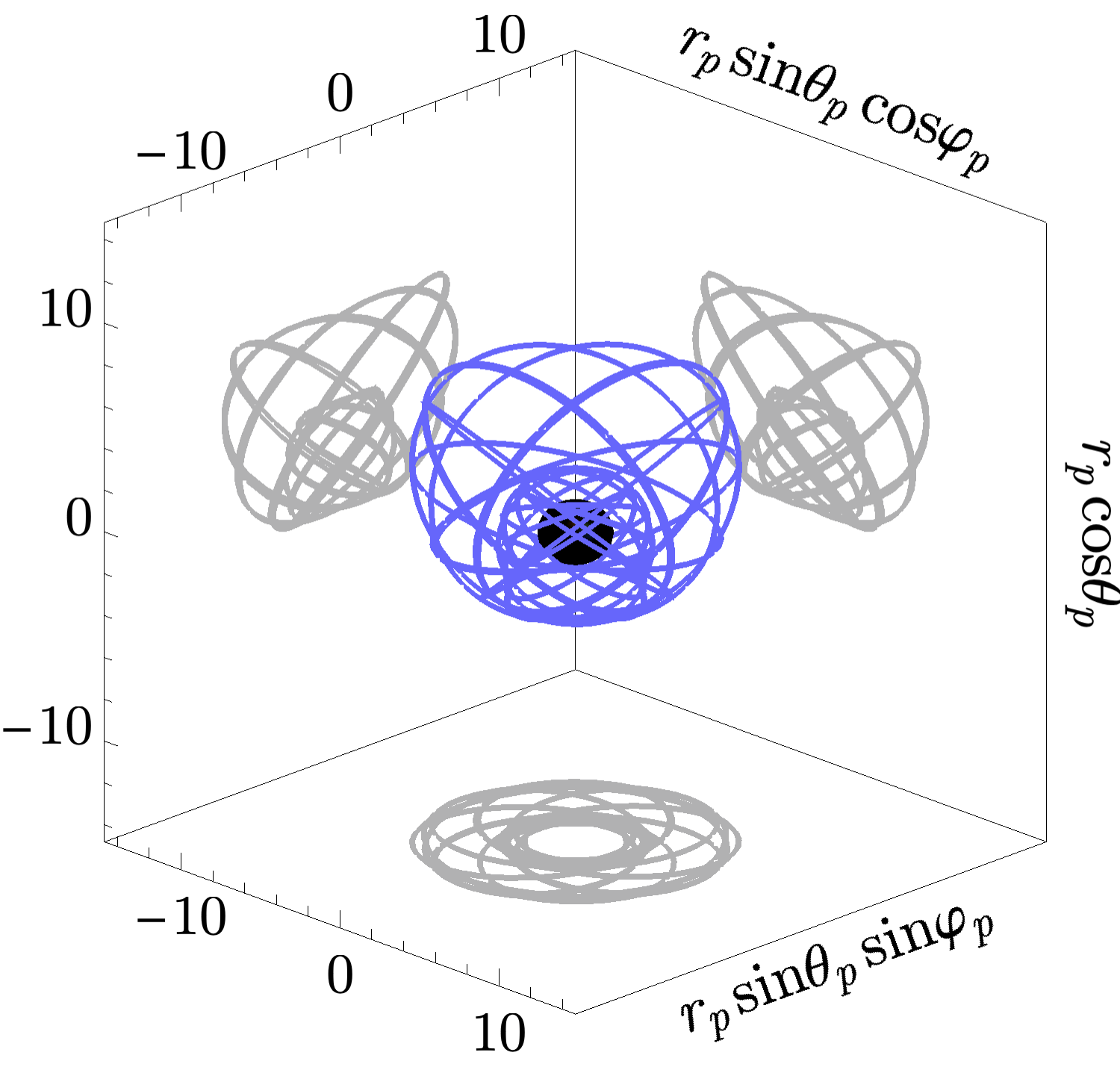}
	\caption{Inclined eccentric geodesics around a Kerr black hole with spin
	$a/M=0.9$ and mass $M=1$. The plot on the left is a nonresonant geodesic with
	orbital parameters $(p,e,x)=(4.700,0.5,\cos\pi/4)$. The plot on
	the right, in contrast, is a 1:2 $r\th$-resonant geodesic with orbital
	parameters $(p,e,x)\simeq (4.607,0.5,\cos\pi/4)$. The
	semilatus rectum $p$ is truncated at four significant digits and
	specifically chosen so that the particle undergoes one radial
	libration for every two polar librations. The solid (blue) curves trace out the
	three-dimensional motion of the two geodesics in Boyer-Lindquist coordinates
	$(r,\th,\vp)$.  The solid (grey) curves show various two-dimensional projections
	of the three-dimensional orbit.}
\label{fig:resGeoExample}
\end{figure*}

To set notation, in Sec.~\ref{sec:geo} we review nonresonant and resonant geodesics in Kerr spacetime, along with
various parametrizations of geodesic functions. In Sec.~\ref{sec:ssf} we review the scalar self-force model
presented in \cite{NasiOsbuEvan19}.  In Sec.~\ref{sec:constructSSF}, we extend our previous methods for calculating
the SSF along nonresonant orbits to the case of $r\th$ resonances by making use of a simple shifting relation
based on symmetries (Killing vectors) of the Kerr spacetime.  In Sec.~\ref{sec:results}, we numerically calculate
the SSF experienced by a scalar-charged particle for six different $r\th$-resonant geodesics.  We use these results
in Sec.~\ref{sec:flux} to compute the secular evolution (orbit-averaged time rate-of-change) of the orbital constants
(e.g., $\langle \dot{\mathcal{E}}\rangle$, $\langle \dot{\mathcal{L}}_z\rangle$, $\langle \dot{\mathcal{Q}}\rangle$).
We demonstrate that the conservative components of the SSF do not contribute to the evolution of the orbital energy and
angular momentum, $\langle \dot{\mathcal{E}}\rangle$ and $\langle \dot{\mathcal{L}}_z\rangle$, respectively,
as expected from flux-balance arguments.  We do find nonzero contributions from the conservative SSF
to the secular evolution of the Carter constant $\langle \dot{\mathcal{Q}}\rangle$, though these
contributions are consistent with the level of systematic numerical errors produced by our regularization scheme.
We conclude with a discussion of these results in Sec.~\ref{sec:conclusion}.  Because we only focus on
$r\th$ resonances in this work, henceforth we will occasionally refer to these as simply resonances.
For this paper we use units such that $c=G=1$, use metric signature $(-+++)$ and sign conventions, where
applicable, of Misner, Thorne, and Wheeler \cite{MisnThorWhee73}.

\section{Geodesics and resonant motion in Kerr spacetime}
\label{sec:geo}

The instantaneous motion of a small mass orbiting a much more massive rotating black hole is approximated by that
of a bound, timelike geodesic in Kerr spacetime.  This is the zeroth-order motion in BHPT.  Bound Kerr geodesics
librate at different frequencies in the radial and polar directions, as shown in the left panel of
Fig.~\ref{fig:resGeoExample}.  Generally, the instantaneous periods of a Kerr EMRI's radial and polar motions are
incommensurate.  In these circumstances, the geodesic is ergodic: given an infinite time, the motion passes through
every point in a finite, bounded region in $r$ and $\theta$.  For certain orbital parameters, however, these radial
and polar periods will be in a rational-number ratio, giving rise to a resonance.  In these cases the motion is
not ergodic. Instead it loops back on itself, and it does not fill the corresponding bounded region in $r$ and $\theta$
\cite{GrosLeviPere12} (see the right panel of Fig.~\ref{fig:resGeoExample}).

To better understand these different behaviors, and to establish notation, we review the analytic framework for
studying bound geodesics in the Kerr spacetime, primarily following the work of
\cite{Cart68,MisnThorWhee73,Schm02,DrasFlanHugh05,DrasHugh06,FlanHughRuan14}, though we have consolidated and
adapted notation for consistency.  We also discuss various prescriptions for parametrizing geodesics and how those
parametrizations must be handled when describing resonances.

\subsection{Separation of the geodesic equations}

Consider a point particle with mass $\mu$ on a bound geodesic $x_p^\mu(\tau)$ with four-velocity $u^\alpha \equiv
dx_p^\mu/d\tau$, where $\tau$
is the particle's proper time.  The background spacetime, described by the metric $g_{\a\b}$, is parametrized by
the black hole spin $a$ and mass $M$.  Adopting Boyer-Lindquist coordinates $(t,r,\th,\vp)$, the Kerr line element
reads
\begin{multline}
	ds^2 = -\left(1-\frac{2Mr}{\Sig}\right)dt^2 + \frac{\Sig}{\D}dr^2
	-\frac{4Mar\sin^2\th}{\Sig}dtd\vp
	\\
	+ \Sig d\th^2 + \frac{\sin^2\th}{\Sig}(\varpi^4-a^2\D\sin^2\th)d\vp^2,
\end{multline}
where $\Sig \equiv r^2 + a^2 \cos^2\th$, $\D \equiv r^2 - 2 M r + a^2$, and $\varpi^2 \equiv r^2+a^2$.

Geodesic motion in Kerr spacetime is completely integrable, leading to three constants of motion---the specific
energy $\mathcal{E}$, the $z$ component of the specific angular momentum $\mathcal{L}_z$, and the (scaled) Carter
constant $\mathcal{Q}$ \cite{Cart68}---all of which can be related to the Killing symmetries of Kerr.  The specific
energy and angular momentum correspond to the Killing vectors $\xi^\mu_{(t)}$ and $\xi^\mu_{(\vp)}$,
\begin{align}
\label{eqn:EAndLzConstants}
	\mathcal{E} &\equiv -\xi^{\mu}_{(t)}u_\mu = -u_t,
	\\
	\mathcal{L}_z &\equiv + \xi^{\mu}_{(\vp)}u_\mu = + u_\vp,
\end{align}
while the Carter constant is associated with the Killing tensor $K^{\mu\nu}$ \cite{WalkPenr70},
\begin{gather}
\label{eqn:Q}
	\mathcal{Q} \equiv K^{\mu\nu}u_\mu u_\nu -(\mathcal{L}_z-a\mathcal{E})^2,
\end{gather}
which is discussed further in Appendix \ref{app:Qdot}.

Introducing the (Carter-)Mino time parameter $\la$ \cite{Cart68,MisnThorWhee73,Mino03,DrasFlanHugh05}, defined by
$d\la \equiv \Sig_p^{-1}d\tau$, the geodesic equations separate
\begin{align}
\label{eqn:geoT}
	\frac{dt_p}{d\la} &= V_{tr}(r_p) + V_{t\th}(\th_p),
	\\  \label{eqn:geoR}
	\frac{dr_p}{d\la} &= \pm \sqrt{V_r(r_p)},
	\\  \label{eqn:geoTh}
	\frac{d\th_p}{d\la} &= \pm \sqrt{V_\th(\th_p)},
	\\ \label{eqn:geoPhi}
	\frac{d\vp_p}{d\la} &= V_{\vp r}(r_p) + V_{\vp\th}(\th_p),
\end{align}
where the various $V$ functions are defined in \cite{DrasFlanHugh05,NasiOsbuEvan19} and the subscript $p$ denotes
that a function is evaluated on the particle's worldline, $x_p^\mu$.

Rather than specifying a geodesic by directly choosing values for $\mathcal{E}$, $\mathcal{L}_z$, and $\mathcal{Q}$,
we use relativistic definitions of semilatus rectum $p$ and orbital eccentricity $e$ that are analogous to those of
Keplerian orbits,
\begin{gather}
	p \equiv \frac{2r_\text{min}r_\text{max}}{M(r_\text{max}+r_\text{min})},
	\qquad
	e \equiv \frac{r_\text{max}-r_\text{min}}{r_\text{max}+r_\text{min}} .
\end{gather}
We add to that the projection of the orbital inclination
\begin{equation}
	x \equiv \cos(\frac{\pi}{2}-\th_\text{min}) ,
\end{equation}
to round out the parametrization of the orbits.  Here, $r_\text{min}$ and $r_\text{max}$ are the minimum and maximum
radii reached by the point mass and $\th_\text{min} = \pi - \th_\text{max}$ is its minimum polar angle.  These are
the turning points of the geodesic where
$V_r(r_\text{min})=V_r(r_\text{max})=V_\th(\th_\text{min})=V_\th(\pi-\th_\text{min})=0$.
Once $p$, $e$, and $x$ are specified for an orbit, it is straightforward to determine the corresponding $\mathcal{E}$,
$\mathcal{L}_z$, and $\mathcal{Q}$ \cite{Schm02,DrasHugh04}.  One can then solve \eqref{eqn:geoT}-\eqref{eqn:geoPhi}
using spectral integration methods \cite{HoppETC15,NasiOsbuEvan19} or analytic special functions
\cite{DrasHugh04,FujiHiki09}.  In this work we use a hybrid scheme: we sample the analytic geodesics solutions
presented in \cite{FujiHiki09}, then construct their discrete Fourier representations, which provide an
exponentially-convergent numerical approximation of the geodesics.

\subsection{Frequencies of generic bound motion}
\label{sec:freq}

For inclined, eccentric (bound) geodesics the point mass librates in $r$ and $\th$ with radial and polar Mino time
periods
\begin{align}
	\La_r &\equiv 2 \int_{r_\text{min}}^{r_\text{max}} \frac{dr}{\sqrt{V_r(r)}},
	\\
	\La_\th &\equiv 2 \int_{\th_\text{min}}^{\pi-\th_\text{min}}
	\frac{d\th}{\sqrt{V_\th(\th)}},
\end{align}
and corresponding Mino time frequencies
\begin{equation}
	\Up_{r} \equiv \frac{2\pi}{\La_r} , \qquad \Up_{\th} \equiv
	\frac{2\pi}{\La_\th}.
\end{equation}
In Kerr spacetime, $\Up_\th$ is always greater than $\Up_r$ for bound motion.  The time and azimuthal
locations, which depend on the radial and polar motions [see \eqref{eqn:geoT} and \eqref{eqn:geoPhi}], accumulate
at average rates in $\la$ denoted by $\G$ and $\Up_\vp$ and are given by
\begin{align}
	\G &=\frac{1}{\La_r} \int_0^{\La_r} V_{tr}(r_p)d\la
	+ \frac{1}{\La_\th} \int_0^{\La_\th} V_{t\th} (\th_p)d\la ,
	\\
	\Up_\vp &=\frac{1}{\La_r} \int_0^{\La_r} V_{\vp r} (r_p)d\la
	+ \frac{1}{\La_\th} \int_0^{\La_\th} V_{\vp\th}(\th_p)d\la,
\end{align}
where $r_p$ and $\th_p$ are understood to be functions of $\la$.  Together, these form the complete set of Mino
time frequencies $\Up_\a=(\G,\Up_r,\Up_\th,\Up_\vp)$.  They are related to the fundamental coordinate-time
frequencies by
\begin{equation}
	\O_r = \frac{\Up_r}{\G},
	\qquad \O_\th = \frac{\Up_\th}{\G},
	\qquad \O_\th = \frac{\Up_\vp}{\G},
\end{equation}
which then define the discrete frequencies
\begin{equation}
\label{eqn:genFreq}
	\o_{mkn} \equiv m\O_\vp +k\O_\th + n\O_r ,
\end{equation}
in the multiperiodic Fourier spectrum of any variable made time dependent by the geodesic motion.  Note that these
coordinate-time frequencies $\O_\a$ do not uniquely define a geodesic due to the existence of isofrequency
pairings \cite{WarbBaraSago13}, though it is well argued that (up to initial conditions) geodesics are uniquely
defined by their Mino time frequencies $\Up_\a$ \cite{Vand14a}.

\subsection{Analytic structure of geodesic solutions and their dependence on initial conditions}

We consider next families of geodesics and their dependence on initial conditions.  Let an inclined eccentric
geodesic be parametrized by $\la$, $\hat{x}^\mu = \hat{x}^\mu_p(\la)$, and have the following initial conditions
\begin{gather}
\label{eqn:fiducialConditions}
	\hat{x}_p^\mu(0)=(0,r_\text{min},\th_\text{min},0),
	\\
	\hat{u}^r(0)=\hat{u}^\th(0)=0 .
\end{gather}
Following the nomenclature of \cite{DrasFlanHugh05}, we refer to a trajectory with these initial conditions as a
fiducial geodesic and distinguish it with a hat.  Integrating \eqref{eqn:geoT}-\eqref{eqn:geoPhi} and enforcing these
fiducial initial conditions, the periodicity in the motion gives rise to solutions that have the form
\begin{align}
\label{eqn:geoTFid}
	\hat{t}_p(\la) &= \G\la
	+ \D\hat{t}^{(r)}(\Up_r\la) + \D\hat{t}^{(\th)}({\Up_\th\la})
	\\ \label{eqn:geoRFid}
	\hat{r}_p(\la)& = r_\text{min} + \D\hat{r}^{(r)}(\Up_r\la),
	\\ \label{eqn:geoThFid}
	\hat{\th}_p(\la)& = \th_\text{min} + \D\hat{\th}^{(\th)}(\Up_\th\la),
	\\ \label{eqn:geoPhiFid}
	\hat{\vp}_p(\la)&= \Up_\vp\la
	+ \D\hat{\vp}^{(r)}(\Up_r\la) + \D\hat{\vp}^{(\th)}({\Up_\th\la}),
\end{align}
where the various $\Delta \hat{x}$ terms are oscillatory, periodic functions.  Here we use $\D\hat{x}$ to represent
$\D \hat{t}$, $\D \hat{r}$, $\D \hat{\th}$, and $\D \hat{\vp}$, which have the
following properties
\begin{align} \label{eqn:periodicRTh}
	\D\hat{x}^{(r)}(2\pi+\Up_{r}\la) &= \D\hat{x}^{(r)}(\Up_{r}\la),
	& &\D\hat{x}^{(r)}(0) = 0,
	\\
	\D\hat{x}^{(\th)}(2\pi+\Up_{\th}\la) &= \D\hat{x}^{(\th)}(\Up_{\th}\la),
	& &\D\hat{x}^{(\th)}(0) = 0 .
\end{align}
Furthermore, from the way a fiducial geodesic is defined, all of these periodic functions are either even or odd with
respect to $\la=0$, with $\Delta \hat{t}$ and $\Delta \hat{\vp}$ being odd (antisymmetric) and $\Delta \hat{r}^{(r)}$ and
$\Delta \hat{\th}^{(\th)}$ being even (symmetric).  Exact definitions of these geodesic functions are provided in
\cite{DrasFlanHugh05,FujiHiki09,NasiOsbuEvan19}.

Next, we consider an inclined, eccentric geodesic $x^\mu_p(\la)$ with arbitrary initial conditions
\begin{gather}
	x_p^\mu(0)=(t_0,r_0,\th_0,\vp_0),
	\\
	u^r(0)=u^r_0, \qquad \qquad u^\th(0)=u^\th_0 .
\end{gather}
An arbitrary geodesic can be expressed in terms of the fiducial solutions by shifting the arguments of the periodic functions,
\begin{align}
\label{eqn:geoTGen}
	t_p&(\la;t_0,\la^{(r)}_0,\la^{(\th)}_0) = t_0
	- \D\hat{t}(\Up_r\la^{(r)}_0,\Up_\th\la^{(\th)}_0) \notag
	\\
	& \quad
	+ \G\la +
	\D\hat{t}\left(\Up_r(\la+\la^{(r)}_0),\Up_\th(\la+\la^{(\th)}_0)\right),
	\\
	r_p&(\la;\la^{(r)}_0)= r_\text{min}
	+ \D\hat{r}^{(r)}(\Up_r\la+\Up_r\la^{(r)}_0),
	\\
	\th_p&(\la; \la^{(\th)}_0) = \th_\text{min}
	+ \D\hat{\th}^{(\th)}(\Up_\th\la+\Up_\th\la^{(\th)}_0),
	\\ \label{eqn:geoPhiGen}
	\vp_p&(\la; \vp_0,\la^{(r)}_0,\la^{(\th)}_0)= \vp_0
	- \D\hat{\vp}(\Up_r\la^{(r)}_0,\Up_\th\la^{(\th)}_0) \notag
	\\ & \;\;\;
	+ \Up_\vp\la
	+ \D\hat{\vp}\left(\Up_r(\la+\la^{(r)}_0),\Up_\th(\la+\la^{(\th)}_0)\right).
\end{align}
In the above expressions, we introduced the compact notation,
\begin{align}
	\D\hat{t}(\Up_r\la,\Up_\th\ti{\la})&\equiv \D\hat{t}^{(r)}(\Up_r\la) + \D\hat{t}^{(\th)}(\Up_\th\ti{\la}),
	\\
	\D\hat{\vp}(\Up_r\la,\Up_\th\ti{\la})&\equiv \D\hat{\vp}^{(r)}(\Up_r\la) +
	\D\hat{\vp}^{(\th)}(\Up_\th\ti{\la}),
\end{align}
for the sums of the radial and polar dependencies of the time and azimuthal components.  The initial orbital offsets,
$\la^{(r)}_0$ and $\la^{(\th)}_0$, are defined in terms of the initial radial and polar positions and velocities by
\begin{align*}
 \D\hat{r}^{(r)}(\Up_r\la^{(r)}_0) &= r_0 - r_\text{min},
 \;\; \sign\left(\sin\Up_r \la^{(r)}_0\right) = \sign u^r_0,
 \\
 \D\hat{\th}^{(\th)}(\Up_\th\la^{(\th)}_0) &= \th_0 - \th_\text{min},
 \;\; \sign\left(\sin\Up_\th \la^{(\th)}_0\right) = \sign u^\th_0.
\end{align*}
Here, $\sign$ represents the sign function. The fiducial case is recovered by setting
$t_0=\vp_0=\la^{(r)}_0=\la^{(\th)}_0=0$.

\begin{figure*}[!th]
\centering
\includegraphics[width=\textwidth]{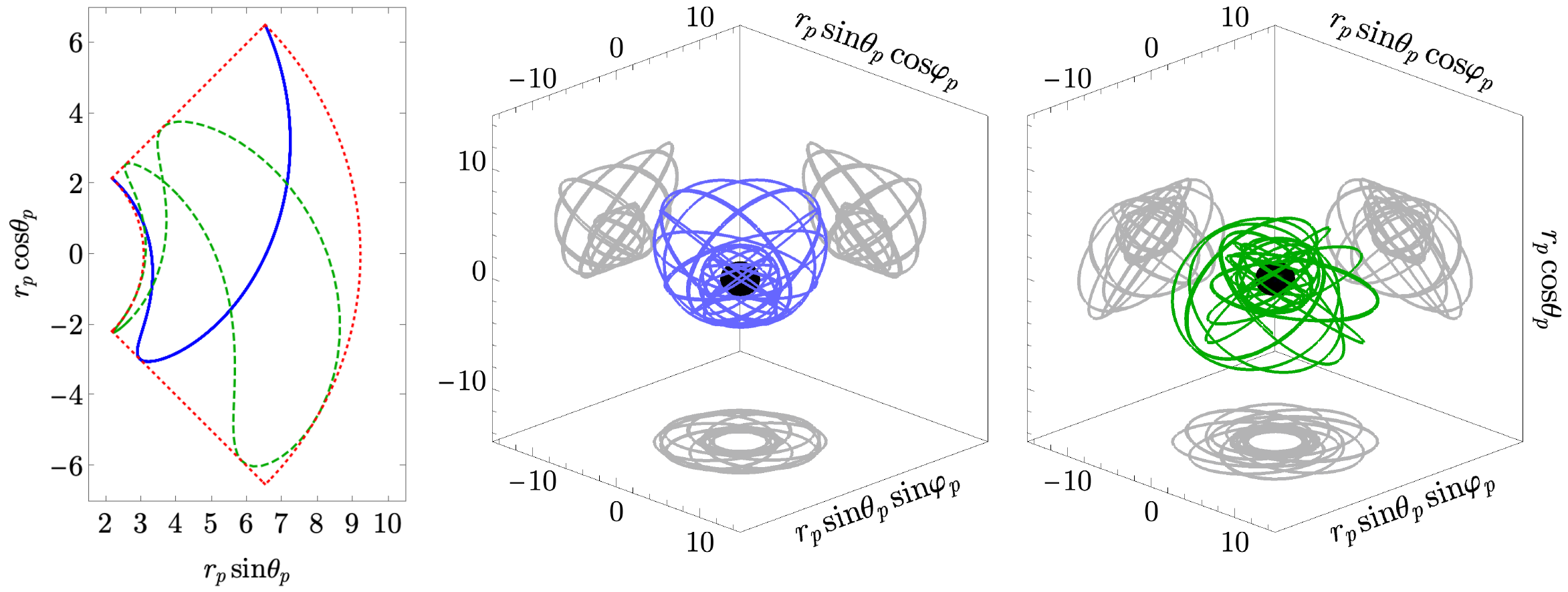}
\caption{Projected motion for two 1:2 $r\th$-resonant geodesics around a Kerr black hole with spin $a/M=0.9$
and mass $M=1$.  The left plot suppresses the azimuthal motion and depicts the radial and polar motion of both
geodesics in the poloidal plane.  Three-dimensional projections of these two geodesics are mapped separately in
the center and right plots. Both geodesics share the orbital parameters $(p,e,x)\approx(4.607,0.5,\cos\pi/4)$ and
the same $r\th$-resonant frequencies.  While they both start at $r_p(0)=r_\text{min}$, they differ in their initial
polar positions and polar velocities.  The solid (blue) line in the left plot and the center plot trace out a
geodesic with the initial offset $\la^{(\th)}_0=0$ ($q_{\th 0}=\bar{q}_0=0$), while the dashed (green) line in the left
plot and the far right plot represent a geodesic with the initial offset $\la^{(\th)}_0=-3\La_\th/8$
($q_{\th 0}=\b_\theta \bar{q}_0=-3\pi/4$). The motion for each plot is shown from $\la=0$ to $\la=47$.  }
\label{fig:resInitOrbits}
\end{figure*}

All bound geodesics can be described by \eqref{eqn:geoTGen}-\eqref{eqn:geoPhiGen}, though any geodesic that passes
through a simultaneous minimum in the radial and polar motion (i.e., $r_p=r_\text{min}$ and $\th_p=\th_\text{min}$)
can be mapped to a fiducial geodesic via trivial offsets in $t_0$ and $\vp_0$.  As long as there exist integers
$k'$ and $n'$ such that
\begin{equation}
\label{eqn:fiducialCondition}
	\la^{(\th)}_0 - \la^{(r)}_0 = n'\La_r-k'\La_\th,
\end{equation}
is satisfied, then a simultaneous turning point at these minimum positions will occur.\footnote{A similar symmetry
also exists for other simultaneous turning points [e.g., $(r_\mathrm{min},\th_\mathrm{max})$,
$(r_\mathrm{max},\th_\mathrm{max})$], which leads to the alternate constraint of Eq.~(4.9) in \cite{DrasFlanHugh05}.
For simplicity we only focus on the case of a simultaneous minimum turning point $(r_\mathrm{min},\th_\mathrm{min})$,
which is sufficient for our discussion.}  Because nonresonant eccentric inclined geodesics are ergodic, this
simultaneous turning point always exists (i.e., on a long enough timescale), and, therefore, nonresonant geodesics
can be described by the fiducial expressions in \eqref{eqn:geoTFid}-\eqref{eqn:geoPhiFid}, without loss of
generality up to a trivial shifting of Mino time $\la$.

\subsection{Special case of $r\th$-resonant geodesics}
\label{sec:resGeo}

We classify $r\th$ resonances in terms of the (relatively prime) integers $\b_r$ and $\b_\th$ that define the ratio
between the radial and polar frequencies, i.e.,
\begin{equation}
	\frac{\O_r}{\O_\th} = \frac{\b_r}{\b_\th} .
\end{equation}
Low-integer ratios (i.e., ones with small integers like $\b_r$:$\b_\th=$ 1:2, 2:3) are referred to as strong
resonances, while high-integer ratios (e.g., 10:11, 1:20) are weak resonances.  Because the radial and polar
frequencies are commensurate during an $r\th$ resonance, the discrete frequency spectrum of resonant geodesics
reduces to
\begin{equation}
\label{eqn:resFreq}
	\o_{mN}\equiv m\O_\vp + N\O,
\end{equation}
where $\O\equiv \O_r/\b_r=\O_\th/\b_\th$ and $N\equiv k\b_\th+n\b_r \in \mathbb{Z}$.  In other words, for an
$r\th$ resonance the normally separate sets of harmonics of the radial and polar fundamental frequencies merge into
one set of harmonics, $N$, of a new, lower net frequency, $\Omega$.

Resonant geodesics can also be described by
\eqref{eqn:geoTGen}-\eqref{eqn:geoPhiGen}.  However, unlike nonresonant geodesics which are ergodic,
$r\theta$ resonances follow restricted paths through the poloidal plane---as shown in
Fig.~\ref{fig:resInitOrbits}---and these paths are sensitive to the initial conditions $\la^{(r)}_0$ and
$\la^{(\th)}_0$.  For a resonance, the radial and polar motions oscillate with the shared (net) Mino time frequency
and period
\begin{equation}
	\Up\equiv \frac{\Up_r}{\b_r}=\frac{\Up_\th}{\b_\th}, \qquad \qquad \La\equiv \b_r\La_r=\b_\th\La_\th.
\end{equation}
For simultaneous minimum turning points to occur during a resonance, \eqref{eqn:fiducialCondition} reduces to the
more stringent restriction that $\la^{(\th)}_0-\la^{(r)}_0=N'\La$, for some integer $N'$, which will not hold true
for most choices of $\la^{(r)}_0$ and $\la^{(\th)}_0$.  Therefore, most resonant orbits cannot be mapped to the
fiducial geodesic with the same frequencies.

Nevertheless, the symmetries of Kerr spacetime allow the same functions describing the fiducial orbit to be applied
more broadly to the general resonant case.  This process starts with defining the initial resonant offset
$\la_0\equiv \la^{(\th)}_0-\la^{(r)}_0$.  Any two $r\th$ resonances that share the same value of $\la_0$ (modulo
$\La$) can be mapped onto one another.  Taking advantage of this mapping, one can simplify the description of geodesic
orbits by setting $\la^{(r)}_0$ or $\la^{(\th)}_0$ to 0.  In this work, we choose $\la^{(r)}_0=0$ so that
$\la_0=\la^{(\th)}_0$, without loss of generality.  The parametrization of the resonant motion in terms of the
offset $\la_0$ is then
\begin{align}
\label{eqn:geoTRes}
	t_p&(\la;\la_0) =
	\G\la - \D\hat{t}(0,\b_\th\Up\la_0)
	\\
	& \qquad \qquad \quad \;\,
	+ \D\hat{t}\left(\b_r\Up\la,\b_\th\Up(\la+\la_0)\right) \notag
	\\
	r_p&(\la;\la_0)= r_\text{min}
	+ \D\hat{r}^{(r)}(\b_r\Up\la),
	\\
	\th_p&(\la; \la_0) = \th_\text{min}
	+ \D\hat{\th}^{(\th)}(\b_\th\Up(\la+\la_0)),
	\\ \label{eqn:geoPhiRes}
	\vp_p&(\la; \la_0)=
	\Up_\vp\la - \D\hat{\vp}(0,\b_\th\Up\la_0)
	\\
	& \qquad \qquad \qquad \,
	+ \D\hat{\vp}\left(\b_r\Up\la,\b_\th\Up(\la+\la_0)\right). \notag
\end{align}
By varying the value of the offset in the range $0\leq\la_0 < \La$, we can generate all resonant paths through
the poloidal plane that are characterized by the same orbital parameters and frequencies but have different
initial positions (see Fig.~\ref{fig:resInitOrbits}).

\begin{figure*}[!th]
	\centering
	\includegraphics[width=0.37\textwidth]{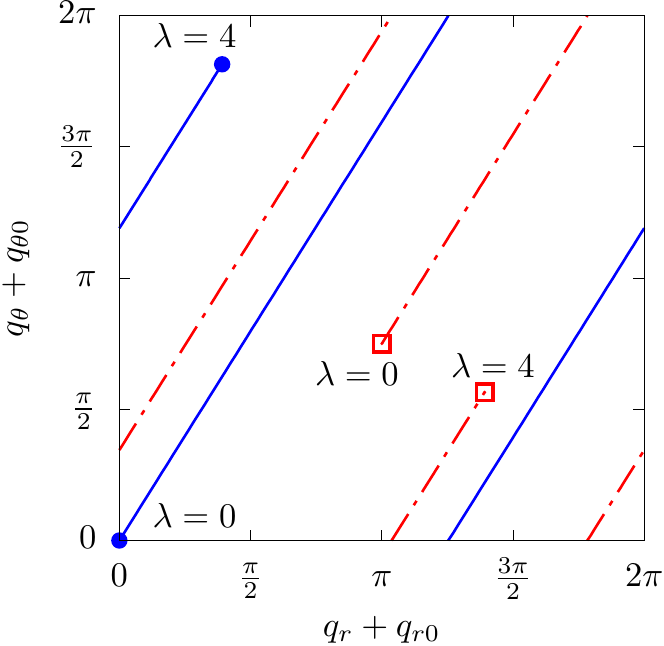}
	\hspace{6mm}
	\includegraphics[width=0.37\textwidth]{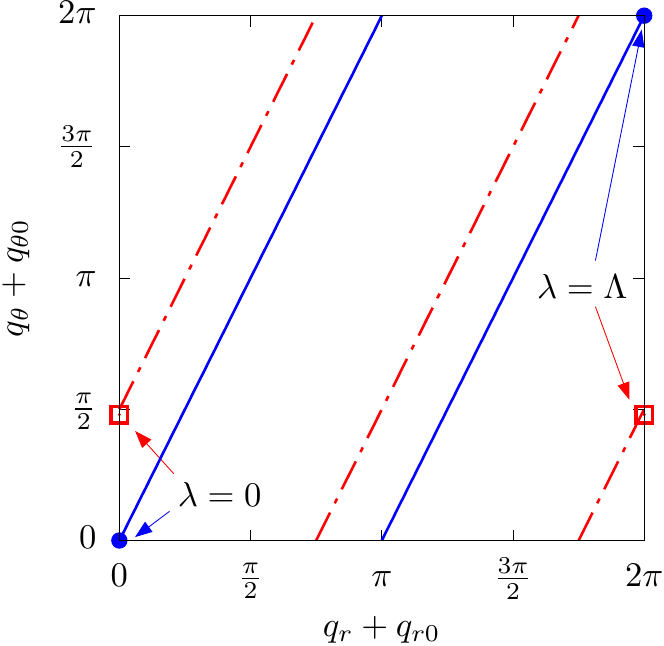}
	\caption{Poloidal motion depicted on the torus for orbits with parameters
	$(a/M,p,e,x)=(0.9,6,0.5,\cos\pi/4)$ (left panel) and
	$(a/M,p,e,x)=(0.9,4.607,0.5,\cos\pi/4)$ (right panel).  In the right plot parameters
	are chosen to generate a 1:2 $r\th$ resonance.  The orbit in the left panel is nonresonant.
	There the solid (blue) line follows the path of a geodesic with fiducial initial
	conditions $q_{(a)0}=(t_0,q_{r0},q_{\th 0},\vp_0)=(0,0,0,0)$, while the dashed (red) line has
	initial conditions $q_{(a)0}=(0,\pi,\pi/2,0)$. (Both paths are plotted over the Mino time interval $\la\in [0,4]$,
	with $M=1$.)  Given sufficient time, both paths will fill the entire torus and completely overlap.
	In the right panel the solid (blue) line follows
	the path of a geodesic with the initial resonant phase $\bar{q}_0=0$ on the Mino time
	interval $\la\in [0,\La]$, where $\La\simeq 4.494$ is the (net) resonant Mino time period.  The
	dashed (red) line follows a geodesic with initial phase
	$\bar{q}_0=\pi/2$ ($\th_p(\la=0)=\pi-\th_\text{min}$) on the same time interval. Unlike the nonresonant paths,
	the resonant geodesic flows in the right panel will never overlap with one another.}
	\label{fig:twoTorus}
\end{figure*}

\subsection{Parametrizing geodesics with angle variables}
\label{sec:angleVar}

The integrability of geodesic motion in Kerr spacetime also leads to a natural representation of the motion in terms
of action-angle variables.  This formalism forms the basis of the two-timescale description of EMRI dynamics
\cite{HindFlan08} and provides a characteristic parametrization for functions that depend on the librating radial
and polar motions.  The angle variable parametrization is also used for presenting gravitational self-force results
\cite{Vand18}, which we adopt in similar fashion.

The angle variables are related to $\la$ by the Mino frequencies,
\begin{equation}
	q_r =  \Up_r \la, \qquad \qquad q_\th = \Up_\th \la .
\end{equation}
We similarly define the initial orbital phases
\begin{align} \label{eqn:initialPhasesTuple}
	q_{(\a) 0} = \{ q_{t0},q_{r0},q_{\th 0},q_{\vp 0} \} =
	\{ t_0, \Up_r\la^{(r)}_0, \Up_\th\la^{(\th)}_0, \vp_0 \} .
\end{align}
(Technically, for $q_{t0}$ to represent a dimensionless phase it should be rescaled, i.e., $\Omega t_0$, but the
distinction is unimportant for our present purposes.)  Functions that are periodic with respect to the Mino time
periods, $\La_{r}$ and $\La_{\th}$, can then be parametrized in terms of the corresponding angle variables,
$q_{r}$ and $q_{\th}$, e.g.,
\begin{gather*}
	\D\hat{r}^{(r)}\left(\Up_r(\la+\la^{(r)}_0)\right)
	\rightarrow \D\hat{r}^{(r)}(q_r+q_{r0}),
	\\
	\D\hat{\th}^{(\th)}\left(\Up_\th(\la+\la^{(\th)}_0)\right)
	\rightarrow \D\hat{\th}^{(\th)}(q_\th+q_{\th 0}).
\end{gather*}

Upon parametrizing functions in terms of the angle variables, it is straightforward to project function values on
the two-torus spanned by the two angle variables, such as the tori depicted in Fig.~\ref{fig:twoTorus}.  Each
(invariant) two-torus forms a section of configuration space for the radial and polar motion of the small body.  The
geodesic flow on the torus then describes the evolution of the motion through this configuration
space, as discussed in \cite{BrinGeyeHind15b}.  For a nonresonant orbit, starting these paths at different points on
the torus is equivalent to choosing different initial conditions.  The effect of initial conditions on possible paths
is shown in the top panel of Fig.~\ref{fig:twoTorus}.  Given an infinite amount of Mino time, a system following a
nonresonant geodesic will sample every point on the torus.

In the case of a resonance, the system executes a closed, repeating motion through the domain, as seen in the
bottom panel of Fig.~\ref{fig:twoTorus}.  Choosing different initial positions on the torus can generate unique
paths.  The only way to sample all of the points on the torus is to consider an infinite number of resonant orbits
that share the same frequencies $\Up_\a$ but different initial offsets $\la_0$.

To distinguish resonant geodesics, we define a
single angle variable $\bar{q}$ for resonant motion and then a single angle parameter $\bar{q}_0$ for the initial resonant
phase\footnote{Our resonant angle $\bar{q}$ differs from the resonant variable $q_\perp = \beta_\theta q_r - \beta_r q_\theta$ used by other authors (e.g., \cite{Vand14a}).},
\begin{equation}
\label{eqn:resAngleVar}
	\bar{q} \equiv  \Up \la = \frac{q_r}{\b_r} = \frac{q_\th}{\b_\th},
	\qquad
	\bar{q}_0 \equiv  \Up\la_0 = \frac{q_{\th 0}}{\b_\th} - \frac{q_{r 0}}{\b_r} .
\end{equation}
In this mapping, $\bar{q}_0$ is a constant as the system evolves.

In terms of these resonant angles, we denote reparametrized functions with an overbar, such that
\begin{align}
	\D\bar{r}^{(r)}(\bar{q}) &\equiv \D\hat{r}^{(r)}(\b_r \bar{q}) ,
	\\
	\D\bar{\th}^{(\th)}(\bar{q};\bar{q}_{0}) &\equiv \D\hat{\th}^{(\th)}(\b_\th \bar{q}+\b_\th \bar{q}_0),
	\\
	\D\bar{t}(\bar{q};\bar{q}_{0}) &\equiv \D\hat{t}(\b_r \bar{q}, \b_\th \bar{q}+\b_\th \bar{q}_0),
	\\
	\D\bar{\vp}(\bar{q};\bar{q}_{0}) &\equiv \D\hat{\vp}(\b_r \bar{q}, \b_\th \bar{q}+\b_\th \bar{q}_0).
\end{align}
This description is particularly useful for calculating and visualizing the SSF in later sections.

\section{Scalar self-force problem}
\label{sec:ssf}

\subsection{Overview}

We use the same scalar model and resulting SSF formalism as outlined in our previous paper \cite{NasiOsbuEvan19}.
We give a brief summary here to establish notation.  The small body is treated as a point particle with mass $\mu$
following an (arbitrary) geodesic about a Kerr black hole with mass $M$ and spin $a$, but with the particle endowed
with a scalar charge $q \sim \mu \ll M$.  We neglect gravitational perturbations and the GSF due to the mass $\mu$.
The motion of the charge sources a radiative scalar field $\Phi$, which satisfies the curved-space Klein-Gordon
equation \cite{Teuk73}
\begin{equation}
\label{eqn:KG}
	g^{\a\b}\nabla_\a \nabla_\b \Phi = -4\pi \rho,
\end{equation}
where $\rho$ is the scalar charge density and the covariant derivative $\nabla_\a$ is taken with respect to the
stationary Kerr background $g_{\a\b}$.  The charge density takes the form
\begin{equation}
\label{eqn:pointSource}
\rho =
q\frac{\d(r-r_p(t))\d(\cos\th-\cos\th_p(t))\d(\vp-\vp_p(t))}
{V_{tr}(r)+V_{t\th}(\th)} .
\end{equation}

The scalar field produces a backreaction on the scalar charge in the form of a SSF $F_\a$ (per unit charge) that
drives its motion off of a background geodesic \cite{Quin00},
\begin{equation}
	u^\b \, \nabla_\b \l \mu u^\a \r = q^2 F^\a.
\end{equation}
Unlike the gravitational case, the SSF is not orthogonal to the four-velocity and contributes to a variation in the
rest mass
\begin{equation}
	\frac{d\mu}{d\tau} = -q^2 u^\a F_\a ,
\end{equation}
requiring all four components of the SSF to be computed.

The contribution of $\Phi$ to the SSF can be completely encoded in the Detweiler-Whiting regular field
$\Phi^\text{R}$ \cite{Quin00,DetwWhit03}
\begin{equation}
	q^2 F^\a = \lim_{x^\mu\rightarrow x_p^\mu}
		q g^{\a\b}\nabla_\b \Phi^\text{R} .
\end{equation}
The regular field $\Phi^\text{R}=\Phi^\text{ret}-\Phi^\text{S}$ is defined as the difference between the retarded
field $\Phi^\text{ret}$, which satisfies Eq.~\eqref{eqn:KG} with causal boundary conditions, and the singular field
$\Phi^\text{S}$, which also satisfies Eq.~\eqref{eqn:KG} (but with different boundary conditions) and which captures
the local, singular behavior.

This makes both $\Phi^\text{ret}$ and $\Phi^\text{S}$ divergent along the point-particle worldline, and their
subtraction from one another requires a careful regularization procedure.  We make use of mode-sum regularization
\cite{BaraOri00,BaraOri03a}
\begin{align}
\label{eqn:modeSumSSF}
	F_\a &= \lim_{x^\mu \rightarrow \pm x^\mu_p} \sum_{l=0}^\infty
	(F_{\a\pm}^{\text{ret},l}-F_{\a\pm}^{\text{S},l}) ,
\end{align}
where $F_{\a\pm}^{\text{ret},l}$ and $F_{\a\pm}^{\text{S},l}$ are the finite, spherical harmonic moments of the
full divergent quantities $q^2F_\a^\text{ret}\equiv q\nabla_\a\Phi^\text{ret}$ and
$q^2F_\a^\text{S}\equiv q\nabla_\a\Phi^\text{S}$.  The $\pm$ notation accounts for the fact that the moments for some
vector components are discontinuous at the point source, with their value depending upon the direction in $r$ in
which the limit is taken.

In the following subsections we outline how $F_\a^{\text{ret},l}$ and $F_\a^{\text{S},l}$ are constructed for a
point source following an arbitrary geodesic.  We calculate the geodesic SSF, finding the force along a background
geodesic, and not the self-consistent SSF that would result from applying the backreaction continuously.

\subsection{The retarded field $\Phi^\text{ret}$}
\label{sec:phi}

In the Kerr background Eq.~\eqref{eqn:KG} is amenable to separation of variables if we both decompose
$\Phi^\text{ret}$ in azimuthal $m$ modes and transform to the frequency domain \cite{Teuk73,BrilETC72}.  The
discrete spectrum of the bound source motion reduces the frequency-domain representation of the field to a set
of discrete sums
\begin{equation} \label{eqn:fieldDecomposition}
	\Phi^\text{ret} = q
	\sum_{\lhat mkn} R_{\lhat mkn}(r) S_{\lhat mkn}(\th)e^{im\vp}e^{-i\o_{mkn}t},
\end{equation}
where the discrete frequency spectrum $\o_{mkn}$ is defined in Eq.~\eqref{eqn:genFreq} and the sum
\begin{equation}
	\sum_{\lhat mkn} \equiv \sum_{\lhat=0}^{+\infty} \sum_{m = -\lhat}^{\lhat}
	\sum_{k=-\infty}^{+\infty} \sum_{n=-\infty}^{+\infty},
\end{equation}
is a compact notation for the sums over $\lhat$ and $m$ and the harmonics of the polar and radial motions.

In the decomposition, $S_{\lhat mkn}(\th)$ is the scalar spheroidal harmonic (spin weight 0) with spheroidicity
$\s^2=-a^2\o_{mkn}^2$, and $R_{\lhat mkn}(r)$ is the solution to the (scalar) radial inhomogeneous Teukolsky equation
\cite{Teuk73}.  We distinguish between spheroidal and spherical harmonic indices using $\lhat$ and $l$, respectively.
In our calculations, we construct $S_{\lhat mkn}$ as a sum over spherical harmonics
\begin{equation}
	S_{\lhat mkn}(\th)e^{im\vp} = \sum_{l=|m|}^\infty b^l_{\lhat mkn}
	Y_{lm}(\th,\vp),
\end{equation}
where the coefficients $b^{l}_{\lhat mkn}$ satisfy a three-term recurrence relation described in
\cite{NasiOsbuEvan19}.

\subsection{Radial mode functions and extended homogeneous solutions}

The radial mode functions $R_{\lhat mkn}$ are solved in a way that allows us to apply the method of extended
homogeneous solutions \cite{BaraOriSago08,HoppEvan10,WarbBara10,WarbBara11,Warb15}.  This technique circumvents the
appearance of Gibbs ringing in the time-domain retarded field \eqref{eqn:fieldDecomposition} when the source terms
are pointlike distributions.  The method starts with calculating the radial mode functions in the frequency
domain.  We first transform to the tortoise coordinate $r_*$ by integrating
\begin{equation}
	\frac{dr_*}{dr}=\frac{\varpi^2}{\D},
\end{equation}
and then introduce scaled radial functions
\begin{equation} \label{eqn:gsnTransform}
	X_{\lhat mkn}= \varpi R_{\lhat mkn}.
\end{equation}
The new radial functions satisfy a radial wave equation
\begin{equation} \label{eqn:gsnODE}
	\left[\frac{d^2}{dr_*^2}-U_{\lhat mkn}(r)\right]X_{\lhat mkn}(r)
	= Z_{\lhat mkn}(r),
\end{equation}
where the radial potential $U_{\lhat mkn}$ and source term $Z_{\lhat mkn}$ are defined in Sec.~II C of
\cite{NasiOsbuEvan19} and Appendix \ref{app:normC} of this paper.

We then construct (unit-normalized) homogeneous solutions $\ti{X}^\pm_{\lhat mkn}$ that satisfy downgoing ($-$) and
outgoing ($+$) wave boundary conditions at the horizon and infinity, respectively (also referred to as the in-wave
and up-wave \cite{Galt82}).  In practice, rather than numerically constructing solutions by solving \eqref{eqn:gsnODE},
we directly compute the homogeneous radial Teukolsky solutions $\tilde{R}^\pm_{\lhat mkn}$ using the Mano-Suzuki-Takasugi function
expansions \cite{ManoSuzuTaka96b,SasaTago03}, and then obtain $\tilde{X}^\pm_{\lhat mkn}$ from \eqref{eqn:gsnTransform}.

Through variation of parameters, we calculate the normalization
coefficients (or Teukolsky amplitudes) ${C}^\pm_{\lhat mkn}$ that relate the homogeneous and inhomogeneous solutions
in the source-free regions
\begin{align}
	X_{\lhat mkn}(r<r_\text{min}) &= {C}^-_{\lhat mkn}\ti{X}^-_{\lhat mkn}(r),
	\\
	X_{\lhat mkn}(r>r_\text{max}) &= {C}^+_{\lhat mkn}\ti{X}^+_{\lhat mkn}(r).
\end{align}
The calculation of ${C}^\pm_{\lhat mkn}$ is described in Appendix \ref{app:normC} and \cite{NasiOsbuEvan19}.  As
noted previously \cite{DrasFlanHugh05,FlanHughRuan14}, varying the initial conditions of the source changes
${C}^\pm_{\lhat mkn}$ by a phase factor \cite{DrasFlanHugh05}
\begin{equation}
\label{eqn:fidToGenClmkn}
	C^\pm_{\lhat mkn}(q_{(\a)0}) =
	e^{i\xi_{mkn}(q_{(\a)0})}\hat{C}^\pm_{\lhat mkn},
\end{equation}
where the set of initial conditions $q_{(\a)0}$ is defined in \eqref{eqn:initialPhasesTuple}. The phase offset takes the form
\begin{multline}
	\xi_{mkn}(q_{(\a)0})\equiv m(\D \hat{\vp}(q_{r0},q_{\th0})-\vp_0)
	\\
	-\o_{mkn}(\D \hat{t}(q_{r0},q_{\th0})-t_0)
		\\
		- k q_{\th0} - n q_{r0} ,
\end{multline}
and hatted normalization constants $\hat{C}^\pm_{\lhat mkn}$ are calculated assuming the fiducial orbit.
We provide a derivation of this relationship in Appendix \ref{app:normC}.  Equation \eqref{eqn:fidToGenClmkn}
also holds true for the Teukolsky amplitudes calculated for gravitational perturbations \cite{FlanHughRuan14}.

With these normalized homogeneous solutions, we define the following extended homogeneous functions
\begin{align}
\label{eqn:EHS}
	&{\phi}^\pm_{lm}(t,r)\equiv \sum_{\lhat=|m|}^{+\infty}
	\sum_{k=-\infty}^{+\infty}\sum_{n=-\infty}^{+\infty}
	{\phi}^\pm_{l\lhat mkn}(r)e^{-i\o_{mkn}t},
	\\ \label{eqn:EHSmodes}
	&{\phi}^\pm_{l\lhat mkn}(r)\equiv \frac{1}{\varpi}
	b^l_{\lhat mkn}{C}^\pm_{\lhat mkn}
	\ti{X}^\pm_{\lhat mkn}(r) ,
\end{align}
for each spherical harmonic $Y_{lm}(\th,\vp)$.  We refer to ${\phi}^\pm_{lm}(t,r)$ as extended homogeneous
functions, not extended homogeneous solutions, since by construction there are no time domain wave equations that
they satisfy (the Teukolsky equation does not separate with ordinary spherical harmonics).  These functions do
however have the advantage that the sums in \eqref{eqn:EHS}
converge exponentially, unlike those in \eqref{eqn:fieldDecomposition}.  Furthermore, even though the individual
$l\lhat mkn$ modes of \eqref{eqn:EHSmodes} are not valid solutions of the inhomogeneous Teukolsky
equation in the source region $r_\text{min}\leq r \leq r_\text{max}$, once the full field is reconstructed in the
time domain by summing over all modes, we are left with extended homogeneous solutions that provide an accurate and
convergent representation of the retarded field up to the location of the point charge
\begin{align}
\label{eqn:phiRet}
	{\Phi}^\text{ret}(t,r,\th,\vp)&=
	{\Phi}^{-}(t,r,\th,\vp)\Theta\left(r_p(t)-r\right)
 \\
 &\qquad + \hat{\Phi}^{+}(t,r,\th,\vp)\Theta\left(r-r_p(t)\right), \notag
 \\  \label{eqn:phiPM}
 {\Phi}^{\pm}(t,r,\th,\vp) &=
 q \sum_{lm} {\phi}^\pm_{lm}(t,r) Y_{lm}(\th,\vp).
\end{align}

\subsection{Retarded and singular contributions to the SSF}
\label{sec:ssfReg}

Using \eqref{eqn:phiRet} and \eqref{eqn:phiPM}, we construct the $l$ modes of the force ${F}^{\text{ret},l}_{\a\pm}$,
along the particle worldline
\begin{align}
\label{eqn:ssfRetMino}
	{F}^{\text{ret},l}_{\a\pm}(\la) &=
	\sum_{m=-l}^l
	(\mathcal{D}^{lm}_\a {\phi}_{lm}^{\pm})({t}_p,{r}_p)
	Y_{lm}({\th}_p,{\vp}_p) ,
\end{align}
where the coordinate positions of the particle are understood to be functions of Mino time [e.g., ${t}_p={t}_p(\la)$],
and the operator $\mathcal{D}^{lm}_\a$ performs the following operations on the extended homogeneous functions:
\begin{align}
	\mathcal{D}_t^{lm}\phi^\pm_{lm}&\equiv\partial_t\phi^\pm_{lm},
	\\
	\mathcal{D}_r^{lm}\phi^\pm_{lm}&\equiv\partial_r\phi^\pm_{lm},
	\\ \label{eqn:delTheta}
	\mathcal{D}_\th^{lm}\phi^\pm_{lm}&\equiv\b^{(-3)}_{l+3,m}\phi^\pm_{l+3,m}
	+\b^{(-1)}_{l+1,m}\phi^\pm_{l+1,m}
	\\
	&\qquad
	+\b^{(+1)}_{l-1,m}\phi^\pm_{l-1,m}
	+\b^{(+3)}_{l-3,m}\phi^\pm_{l-3,m}, \notag
	\\
	\mathcal{D}_\vp^{lm}\phi^\pm_{lm}&\equiv im\phi^\pm_{lm} .
\end{align}
The coefficients $\b_{lm}^{(\pm i)}$ are defined in Appendix A of \cite{NasiOsbuEvan19}\footnote{A minor error exists in
Eq. (A4) of [39]. The coefficients $\beta^{(\pm 3)}_{lm}$ are missing a minus sign in front of the
parentheses on the righthand side of Eq. (A4).} and are
obtained by first applying the window function proposed by Warburton \cite{Warb15} and then reprojecting the
derivatives $\partial_\th Y_{lm}$ onto the $Y_{lm}$ basis.  Details of this operation can be found in
\cite{Warb15}, with a correction added in \cite{NasiOsbuEvan19}.

The singular contribution is obtained through a local analytic expansion in the neighborhood of the source worldline
\cite{BaraOri00,BaraOri03a,DetwMessWhit03}
\begin{equation}
\label{eqn:regParameters}
	F^{\text{S},l}_{\a\pm} = A^\pm_{\a}L + B_\a
	+\sum_{n=1}^{+\infty} \frac{D_{\a,2n}}{\prod_{k=1}^n (2L-2k)(2L+2k)},
\end{equation}
where $L \equiv l+1/2$, and the regularization parameters $A_{\a\pm}$, $B_\a$, and
$D_{\a,2n}$ are independent of $l$ but functions of $r_p$, $\th_p$, $u^r$, $u^\th$, $\mathcal{E}$, $\mathcal{L}_z$,
and $\mathcal{Q}$ (as well as $a$ and $M$).  Only $A_{\a\pm}$ and $B_\a$ are known analytically for generic bound
orbits in Kerr spacetime \cite{BaraOri03a}, while $D_{\a,2}$ is known analytically for equatorial orbits in Kerr
\cite{HeffOtteWard14}.

The terms with higher-order parameters $D_{\a,2n}$ have the useful property that their $l$-dependent weights vanish
upon summing over all $l$,
\begin{equation}
	\sum_{l=0}^{+\infty} \left[\prod_{k=1}^n (2L-2k)(2L+2k)\right]^{-1} = 0 .
\end{equation}
Only $A_{\a\pm}$ and $B_\a$ are needed for convergent results, but if we neglect the $D_{\a,2n}$ terms upon combining
Eqs.~\eqref{eqn:modeSumSSF} and \eqref{eqn:regParameters}, $F_\a$ converges at a rate $\sim l^{-2}$.  Each
$D_{\a,2n}$ term reintroduced to the regularization procedure improves the convergence rate by another factor of
$l^{-2}$.  Since we truncate the sum over $l$ modes around $l_\text{max}\sim 20$, we must numerically fit for the
higher-order regularization parameters to improve the convergence of the mode-sum regularization.  Our fitting
procedure is described in \cite{NasiOsbuEvan19}.  The uncertainties associated with this fitting procedure often
dominate other numerical errors in the calculation.  We use this to obtain uncertainty estimates for our SSF
results.

\section{Constructing the SSF for resonant and nonresonant sources}
\label{sec:constructSSF}

\subsection{Nonresonant sources}
\label{sec:genSSF}

With a generic nonresonant orbit, the SSF is multiply periodic and never repeats over the entire interval
$-\infty<\la<\infty$.  Rather than sampling the SSF over this infinite domain in $\la$, we map the SSF
to the angle variables introduced in Sec.~\ref{sec:angleVar},
\begin{equation} \label{eqn:angleVarSSF}
	\hat{F}^{\text{ret},l}_{\a\pm}(q_r,q_\th) =
	\sum_{m=-l}^l (\mathcal{D}^{lm}_\a \hat{\phi}_{lm}^{\pm})(q_r,q_\th)
	\hat{Y}_{lm}(q_r,q_\th) .
\end{equation}
Note that we have placed hats on all functions that are evaluated using the fiducial geodesic solutions of
\eqref{eqn:geoTFid}-\eqref{eqn:geoPhiFid}.  The evolution of $\vp$ is dependent on the motion of both $r$ and
$\theta$, so in a slight abuse of notation we reparametrize functions to have the following meaning
\begin{align}
\label{eqn:angleVarYlm}
	\hat{Y}_{lm}(q_r,q_\th)\equiv Y_{lm}(\hat{\th}_p(q_\th),
	\D\hat{\vp}(q_r,q_\th)) ,
\end{align}
and
\begin{align}
	&\hat{\phi}^\pm_{lm}(q_r,q_\th)\equiv \sum_{\lhat kn}
	{\hat{\phi}^\pm_{l\lhat mkn}(q_r)}
	 e^{-i(\o_{mkn}\D\hat{t}(q_r,q_\th)+kq_\th+nq_r)}, \notag
	\\   \label{eqn:angleVarEHS}
	&\hat{\phi}^\pm_{l\lhat mkn}(q_r)\equiv \hat{\varpi}^{-1}_p(q_r)b^l_{\lhat mkn}\hat{C}^\pm_{\lhat mkn}
	\ti{X}^\pm_{\lhat mkn}(\hat{r}_p(q_r)) .
\end{align}
Here $\hat{\varpi}_p(q_r)=(\hat{r}_p^2(q_r)+a^2)^{1/2}$ and the operator $\mathcal{D}^{lm}_{\a}$ performs the
same function as before.  Because the regularization parameters only vary with respect to $r_p$, $\th_p$, $u^r$,
and $u^\th$ (assuming the orbital constants are fixed), the singular field can also be translated into this angle
variable parametrization, ultimately providing a description of the SSF in terms of $q_r$ and $q_\th$
\begin{equation}
\label{eqn:ssfGenAngleVar}
	\hat{F}_\a(q_r,q_\th) = \sum_{l=0}^\infty \left(
	\hat{F}_{\a\pm}^{\text{ret},l}(q_r,q_\th)
	-\hat{F}^{\text{S},l}_{\a\pm}(q_r,q_\th)
	\right) .
\end{equation}

The angle variable parametrization maps the entire self-force history onto the finite domain of the invariant
two-torus visualized in Fig.~\ref{fig:twoTorus}.  The SSF, projected onto this torus, can then also be represented
by the (double) Fourier series
\begin{align}
\label{eqn:ssfFourier}
	F_{\a}&(q_r,q_\th)= \sum_{k=-\infty}^{+\infty}\sum_{n=-\infty}^{+\infty}
	g^{kn}_\a e^{-i(kq_\th+nq_r)},
	\\
	g^{kn}_\a &= \frac{1}{4\pi^2}\int_0^{2\pi} dq_r\int_0^{2\pi} dq_\th\;
	F_\a\left(q_{r}, q_{\th}\right)e^{i(kq_{\th}+nq_{r})}. \notag
\end{align}
By densely sampling values of $q_r$ and $q_\th$ over the torus at evenly-spaced points $q_{r,i}=2\pi i/N_r$ and
$q_{\th,j}=2\pi j/N_\th$ (where $N_r, N_\th \in \mathbb{Z}$), we can construct a discrete Fourier representation of
the SSF
\begin{align}
\label{eqn:ssfDiscreteFourier}
	F_{\a}&(q_r,q_\th)\simeq \sum_{k=0}^{N_\th-1}\sum_{n=0}^{N_r-1}
	f^{kn}_\a e^{-i(kq_\th+nq_r)},
	\\
	f^{kn}_\a &= \frac{1}{N_r N_\th}\sum_{i=0}^{N_\th-1}\sum_{j=0}^{N_r-1}
	F_\a\left(q_{r,i}, q_{\th,j}\right)e^{i(kq_{\th,j}+nq_{r,i})} . \notag
\end{align}
Given $N_r$ and $N_\th$ that are large enough such that $\text{max}|f^{kn}_\a-g^{kn}_\a|<\eps_\text{FS}$, where
$\eps_\text{FS}$ is some predefined accuracy goal, the discrete representation will provide an accurate
approximation of Eq.~\eqref{eqn:ssfFourier} \cite{HoppETC15,NasiOsbuEvan19}.  We found that sample numbers of
$N_r=N_\th=2^8$ were typically sufficient for constructing a discrete representation that was accurate to about
$\eps_\text{FS}\sim 10^{-8}-10^{-10}$.  The discrete Fourier series provides an efficient method for storing and
interpolating SSF data.

We can easily generalize our results to geodesics with arbitrary initial conditions by applying the following
shifting relation,
\begin{equation}
\label{eqn:ssfFid2Gen}
	F_\a(q_r,q_\th;q_{a0})
	= \hat{F}_\a(q_r+q_{r 0},q_\th+q_{\th 0}).
\end{equation}
A proof of Eq.~\eqref{eqn:ssfFid2Gen}, which applies for both the SSF and GSF, is provided in Appendix
\ref{app:ssfInvariant}.  While this result seems almost trivial for the nonresonant case, it surprisingly plays a
role in improving the efficiency of SSF calculations for resonant orbits as well, as discussed in
Sec.~\ref{sec:resSSF}.

\subsection{Resonant sources}
\label{sec:resSSF}

The SSF experienced by a charge following an $r\th$-resonant geodesic requires a different treatment.  The worldline
of the charge is described by \eqref{eqn:geoTRes}-\eqref{eqn:geoPhiRes}.  In contrast to the SSF for a nonresonant
orbit [e.g., $\hat{F}_\a(q_r,q_\th)$], we construct the resonant-SSF $\bar{F}_\a$ to be a function of the single
resonant angle variable $\bar{q}$ and the initial resonant phase $\bar{q}_0$ [defined in Eq.~\eqref{eqn:resAngleVar}].
We describe here two methods of calculating $\bar{F}^\mathrm{res}_\a(\bar{q},\bar{q}_0)$: the first uses the reduced
mode spectrum $\o_{mN}$ defined in \eqref{eqn:resFreq} to construct the SSF on an $l\lhat mN$ basis, while the second
uses the generic mode spectrum $\o_{mkn}$ to construct the SSF on the $l\lhat mkn$ basis, just as we outlined in the
previous section for nonresonant orbits.  These two approaches are similar to the two approaches for calculating
gravitational wave fluxes discussed in \cite{FlanHughRuan14}.

\subsubsection{Constructing the resonant SSF on an $l\lhat m N$ basis}

The retarded SSF sourced by an $r\th$ resonant geodesic, when parametrized in terms of the resonant angle variable
and resonant phase, takes the form
\begin{equation}
\label{eqn:fRetRes}
	\bar{F}^{\text{ret},l}_{\a\pm}(\bar{q};\bar{q}_0) =
	\sum_{m=-l}^l
	(\mathcal{D}^{lm}_\a \bar{\phi}_{p,lm}^{\pm})(\bar{q};\bar{q}_0)\bar{Y}_{lm}(\bar{q};\bar{q}_0),
\end{equation}
where, in contrast to Eqs.~\eqref{eqn:angleVarYlm} and \eqref{eqn:angleVarEHS},
\begin{multline}
	\bar{Y}_{lm}(\bar{q};\bar{q}_0)\equiv
	Y_{lm}(\bar{\th}_p(\bar{q};\bar{q}_0),\D\bar{\vp}(\bar{q};\bar{q}_0)-\D\bar{\vp}(0;\bar{q}_0)),
\end{multline}
and
\begin{align}
\label{eqn:resEHS}
	&\bar{\phi}^\pm_{lm}(\bar{q};\bar{q}_0)\equiv \sum_{\lhat=|m|}^{+\infty} \sum_{N=-\infty}^{+\infty}
	{\bar{\phi}^\pm_{l\lhat mN}(\bar{q};\bar{q}_0)}
	\\
	&\qquad \qquad \qquad
	 \times e^{-i\o_{mN}(\D\bar{t}(\bar{q};\bar{q}_0)-\D\bar{t}(0;\bar{q}_0)+N\bar{q})}, \notag
	\\
	&\bar{\phi}^\pm_{l\lhat mN}(\bar{q};\bar{q}_0)\equiv
	\bar{\varpi}^{-1}_p(\bar{q})b^l_{\lhat mN}\bar{C}^\pm_{\lhat mN}(\bar{q}_0)
	\ti{X}^\pm_{\lhat mN}(\bar{r}_p(\bar{q})). \notag
\end{align}
All functions and coefficients with an overbar are evaluated using the resonant geodesic solutions described by
\eqref{eqn:geoTRes}-\eqref{eqn:geoPhiRes}.  The $\bar{C}^\pm_{\lhat mN}$ are defined in Appendix \ref{app:normC} and
vary with the resonant phase parameter $\bar{q}_0$.  Unlike $C^\pm_{\lhat mkn}(\bar{q}_0)$,
$\bar{C}^\pm_{\lhat mN}(\bar{q}_0)$ is not related to the fiducial case $\bar{C}^\pm_{\lhat mN}(0)$ by a simple phase
factor.  Each time we calculate the SSF for a new value of $\bar{q}_0$, the source term must be integrated over a
new resonant orbit.  Since source integration is a computationally-intensive aspect of the SSF calculation, needing
to repeat this operation is not ideal.  Thus, the advantage of reduced dimensionality in the mode spectrum must be
weighed against the disadvantage of repeated source integration.

\subsubsection{Constructing the resonant SSF on an $l\lhat m k n$ basis}

Alternatively, we first construct the fiducial SSF $\hat{F}_\a(q_r,q_\th)$ using the methods outlined in
Sec.~\ref{sec:genSSF}.  Combining \eqref{eqn:ssfGenAngleVar} and \eqref{eqn:ssfFid2Gen}, we can then relate the
resonant SSF $\bar{F}^\mathrm{res}_\a(\bar{q};\bar{q}_0)$ to the fiducial result by fixing the relationship between
$q_r$ and $q_\theta$
\begin{equation} \label{eqn:ssfqq0}
	\bar{F}^\mathrm{res}_{\a}
	(\bar{q};\bar{q}_0) =
	\hat{F}_{\a}
	(\b_r \bar{q}, \b_\th \bar{q} + \b_\th \bar{q}_0) .
\end{equation}
In this way, we simply construct the fiducial SSF $\hat{F}_{\a}(q_r,q_\th)$ on an $l\lhat m k n$ basis by relating
the $\lhat mN$-mode functions and constants to their $\lhat mkn$-mode counterparts
\begin{gather}
\label{eqn:freqMKNfreqMN}
\o_{mN}=\o_{m(kn)_N} ,
\\
\ti{X}_{\lhat mN}=\ti{X}_{\lhat m(kn)_N}, \qquad b^l_{\lhat mN}=b^l_{\lhat m(kn)_N} ,
\end{gather}
where one must be careful to understand that $(k,n)_N$ represents the set of all $k$ and $n$ values that produce
the same value $N$ that satisfies $N=k\b_\th+n\b_r$.  Significant computational time is saved by recycling values
of the homogeneous radial functions for different values of $k$ and $n$, provided
they share the same frequency and spheroidal mode numbers $(\lhat, m)$.

The normalization coefficients are related by a coherent sum over all $k$ and $n$ modes that share the same frequency
(given by $N$)
\begin{align}
\label{eqn:ClmN2Clmkn}
	\bar{C}^\pm_{\lhat mN}(\bar{q}_0)&=  \sum_{(k,n)_N}
	e^{i\ti{\xi}_{mkn}(\bar{q}_0)}\hat{C}^\pm_{\lhat mkn} ,
\end{align}
as demonstrated in Appendix A of \cite{FlanHughRuan14} and Sec.~III D of \cite{GrosLeviPere13}.  In this way,
each $\bar{C}^\pm_{\lhat mN}(\bar{q}_0)$ is a superposition of many amplitudes $\hat{C}^\pm_{\lhat mkn}$ that
would have been regarded as independent in the nonresonant case.  In a complex square, this superposition leads to
constructive or destructive interference terms in the fluxes.  Note that
$\ti{\xi}_{mkn}(\bar{q}_0) \equiv{\xi}_{mkn}(0,0,\b_\th q_{\th 0},0)$.  Substituting
Eqs.~\eqref{eqn:freqMKNfreqMN}-\eqref{eqn:ClmN2Clmkn} into Eqs.~\eqref{eqn:fRetRes}-\eqref{eqn:resEHS}, brings them
into the same form as Eqs.~\eqref{eqn:angleVarSSF}-\eqref{eqn:angleVarEHS}.  Unlike $\ti{X}_{\lhat mkn}$, each
$C_{\lhat mkn}$ must be calculated independently, even if they share the same frequencies and spheroidal harmonic
mode numbers.  Essentially, by introducing the more generic mode spectrum $\o_{mkn}$, we circumvent the need to
repeatedly evaluate each $lmN$ mode at different initial phases, but at the expense of summing over an additional
mode number.  The advantage of this approach is that, once a code has already been built to calculate the fiducial
SSF for nonresonant orbits, it can be easily modified to produce the SSF for resonant sources and avoids the need
to construct an entirely separate code.

\subsubsection{Discrete Fourier representation of the resonant SSF}

The resonant SSF is periodic with respect to $\bar{q}$ and $\bar{q}_0$, and therefore can be expressed as a multiple
Fourier series.  By sampling the resonant SSF on an evenly spaced two-dimensional grid in $\bar{q}$ and $\bar{q}_0$,
the discrete Fourier representation of $\bar{F}^\mathrm{res}_\a$ is
\begin{align}
\label{eqn:FresFourier}
	&\bar{F}^\text{res}_\a(\bar{q};\bar{q}_0) \simeq \sum_{K=0}^{N_0-1}\sum_{N=0}^{N_\text{res}-1}
	\bar{g}_\a^{KN} e^{-iN\bar{q}}e^{-iK \bar{q}_0},
	\\
	&\bar{g}_\a^{KN} = \frac{1}{N_0 N_\text{res}}
	\sum_{\jmath=0}^{N_0-1}\sum_{\imath=0}^{N_\text{res}-1}
	\bar{F}^\text{res}_\a(\bar{q}_\imath;\bar{q}_{0\jmath}) e^{iN\bar{q}_\imath}e^{iK \bar{q}_{0\jmath}} ,
\end{align}
where $\bar{q}_{i}=2\pi i/N_\mathrm{res}$ and $\bar{q}_{0,j}=2\pi j/N_0$, with $N_\mathrm{res}, N_0 \in \mathbb{Z}$.
By comparing \eqref{eqn:FresFourier} with \eqref{eqn:ssfDiscreteFourier} and \eqref{eqn:ssfqq0}, we can relate
$\hat{f}_\a^{kn}$ and $\bar{g}_\a^{KN}$ by
\begin{equation}
	\bar{g}_\a^{KN} = \hat{f}_\a^{K/\b_\th,(N-K)/\b_r} .
\end{equation}
From this relation, we see that $\bar{g}_\a^{KN}=0$ unless $K$ is a multiple of $\b_\th$ and $N-K$ is a multiple of
$\b_r$.  Thus, while the resonant angle variable and the initial resonant phase more naturally capture both the coupled nature of
the radial and polar motion and the sensitivity of the source to initial conditions, this parametrization is less
efficient at capturing the behavior of the self-force.  For example, if one wants to calculate $\hat{f}^{kn}_\a$
for $0\leq k < N_\th$, $0\leq n < N_r$, then one would need to sample $N_r \times N_\th$ points in the
$q_r$-$q_\th$ domain, but $\b_rN_r \times (\b_\th N_\th+\b_r N_r)$ points in the $\bar{q}$-$\bar{q}_0$ domain.  This
oversampling occurs because the resonant parametrization does not take full advantage of the symmetries of the
orbit, which are better captured by the separation of the radial and polar motion in the $q_r$-$q_\th$ angle
parametrization.

\subsection{Dissipative and conservative SSF}

Irrespective of the type of orbit, the self-force can be decomposed into conservative and dissipative parts,
$F_\a^\text{cons}$ and $F_\a^\text{diss}$.  These parts impact the evolution of EMRIs in different ways
\cite{Bara09,DiazETC04,Mino03,HindFlan08} and computationally converge at different rates in the mode-sum
regularization procedure.  The dissipative part $F_\a^\text{diss}$ does not require regularization and converges
exponentially.  The conservative part $F_\a^\text{cons}$ requires regularization and converges as a power law in
the number of $l$ modes.

Summarizing our previous discussion \cite{NasiOsbuEvan19} of this decomposition, the split depends on both the
retarded force and the advanced force $F^\text{adv}_\a$, which depends on the advanced scalar field solution.
The decomposition is made in terms of spherical harmonic elements, e.g., $F^{\text{adv},l}_\a$ and is given by
\begin{align}
	F^\text{diss}_{\a} &=\sum_{l=0}^{+\infty}\frac{1}{2}
	\left( F^{\text{ret},l}_{\a\pm}-
	F^{\text{adv},l}_{\a\pm}\right),
	\\ \label{eqn:consSSF}
	F^\text{cons}_{\a} &=\sum_{l=0}^{+\infty}\left\{\frac{1}{2}
	\left( F^{\text{ret},l}_{\a\pm}+
	F^{\text{adv},l}_{\a\pm}\right)-F^{\text{S},l}_{\a\pm} \right\}.
\end{align}
The inconvenience of calculating the advanced scalar field solution is avoided by using symmetries of Kerr
geodesics \cite{Mino03,HindFlan08,Bara09} (summarized also in \cite{NasiOsbuEvan19}), which lead to convenient
relationships between spacetime components of $F^{\text{ret},l}_\a$ and $F^{\text{adv},l}_\a$,
\begin{equation}
\label{eqn:ssfRet2Adv}
	F^{\text{adv},l}_\a(q_r,q_\th) = \eps_{(\a)}
	F^{\text{ret},l}_\a(2\pi-q_r,2\pi-q_\th) ,
\end{equation}
where $\eps_{(\a)}=(-1,1,1,-1)$.  Thus, $F_t^\text{diss}$, $F_r^\text{cons}$, $F_\th^\text{cons}$, and
$F_\vp^\text{diss}$ are symmetric (even) functions on the $q_r-q_\th$ two-torus, while $F_t^\text{cons}$,
$F_r^\text{diss}$, $F_\th^\text{diss}$, and $F_\vp^\text{cons}$ are antisymmetric (odd).  These relationships
between advanced and retarded solutions have been previously discussed \cite{WarbBara11,Warb15,ThorWard17} in the
context of restricted orbits but, in fact, Eq.~\eqref{eqn:ssfRet2Adv} holds for arbitrary geodesic motion.

{\renewcommand{\arraystretch}{1.25}
\begin{table}[t!]
	\caption{Summary of the resonant orbits considered in Sec.~\ref{sec:results}.  In all cases the primary
	spin is $a=0.9$ (with $M=1$). The real number values are truncated in the table to four significant figures for brevity.}
	\label{tab:orbits}
	\centering
	\begin{tabular*}{\columnwidth}{c @{\extracolsep{\fill}} c c c c c}
		\hline
		\hline
		\multicolumn{2}{c }{\quad Model} & $p$ & $e$ & $x_\text{inc}$ & $\b_r$:$\b_\th$
		\\
		\hline
		\multicolumn{2}{c}{\quad $e02.13$} & 3.622 & 0.2 & $\cos(\pi/4)$ & 1:3
		\\
		\multicolumn{2}{c}{\quad $e02.12$} & 4.508 & 0.2 & $\cos(\pi/4)$ & 1:2
		\\
		\multicolumn{2}{c}{\quad $e02.23$} & 6.643 & 0.2 & $\cos(\pi/4)$ & 2:3
		\\
		\multicolumn{2}{c}{\quad $e05.13$} & 3.804 & 0.5 & $\cos(\pi/4)$ & 1:3
		\\
		\multicolumn{2}{c}{\quad $e05.12$} & 4.607 & 0.5 & $\cos(\pi/4)$ & 1:2
		\\
		\multicolumn{2}{c}{\quad $e05.23$} & 6.707 & 0.5 & $\cos(\pi/4)$ & 2:3
		\\
		\hline
		\hline
	\end{tabular*}
\end{table}
}

\section{Resonant SSF results}
\label{sec:results}

Using the methods outlined in the prior sections, we generated new results for the SSF on six different
resonant orbits, the orbital parameters of which are listed in Table \ref{tab:orbits}.  These calculations were made
with a \textsc{Mathematica} code first described in \cite{NasiOsbuEvan19}.  These calculations also made use of
software from the Black Hole Perturbation Toolkit \cite{BHPTK18}, specifically the \textsc{KerrGeodesics} and
\textsc{SpinWeightedSpheroidalHarmonics} packages.

In generating numerical results we set $M=1$, which is assumed for the remainder of this work.  Each resonant orbit
had primary spin $a=0.9$.  We focused on 1:3, 1:2, and 2:3 $r\th$ resonances, the three resonances an EMRI is most
likely to encounter during its final years of evolution when its signal falls within the LISA passband
\cite{RuanHugh14,BerrETC16}.  To pick orbital parameters $(p,e,x)$ that produce $r\th$-resonant frequencies, we
follow the approach of Brink, Geyer, and Hinderer \cite{BrinGeyeHind15a,BrinGeyeHind15b}.  Specified values of $e$
and $x$ are chosen first, and then $p$ is numerically calculated using the root-finding method described in Sec.~V$\,$E
of \cite{BrinGeyeHind15b}.  In our work all of the orbits share the same inclination, $x=\cos(\pi/4)$, while two
different eccentricities, $e=0.2$ and $e=0.5$, are considered.  The resulting values of $p$ (to four places) for
each resonant orbit are listed in Table \ref{tab:orbits}.

\begin{figure*}[th!]
		\includegraphics[width=0.7\textwidth]{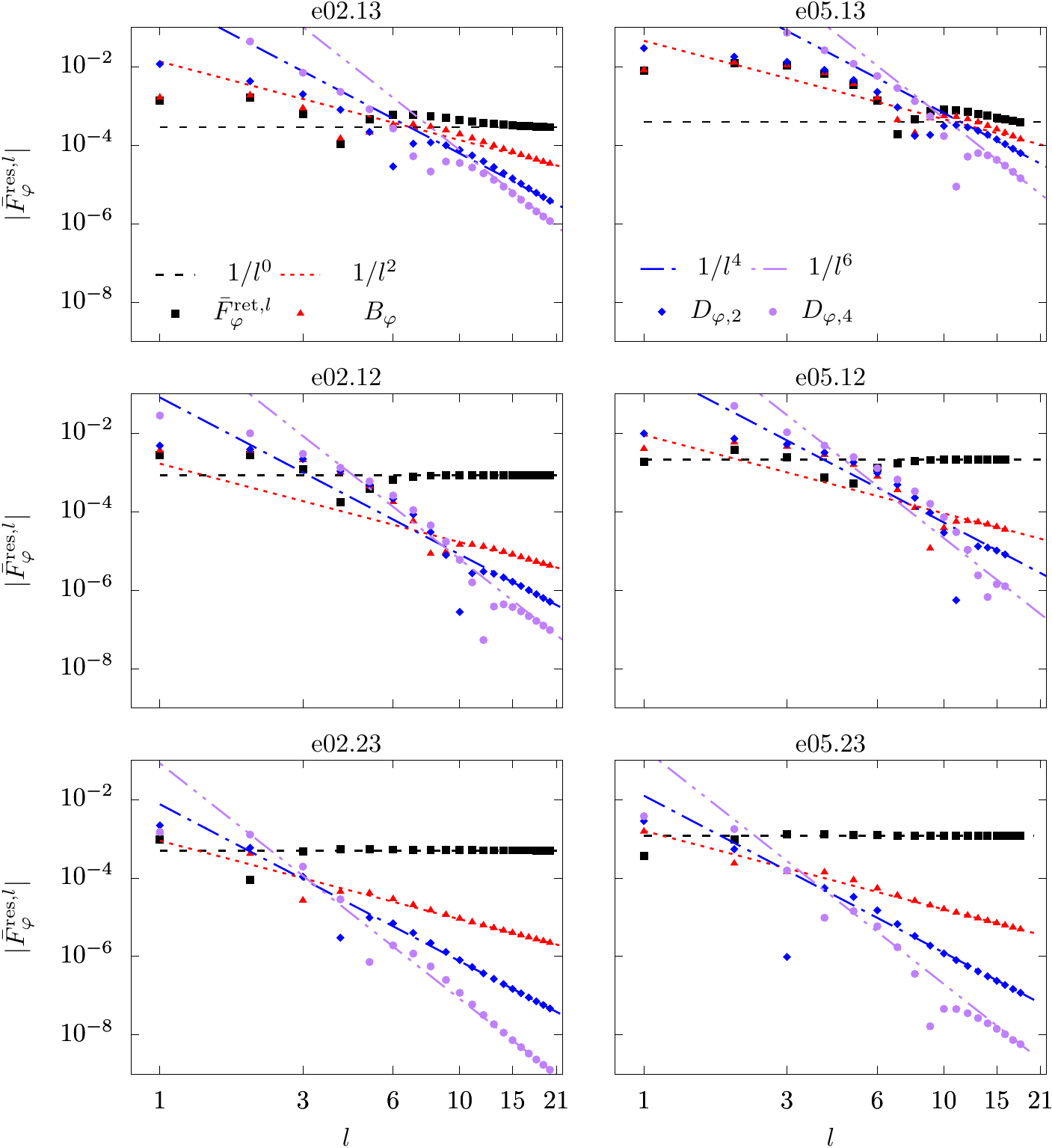}
		\caption{Convergence of the SSF $l$ modes for resonant models listed in Table \ref{tab:orbits}.
		The dashed and dotted lines depict comparative power-law rates of
		convergence for $\bar{F}^\text{res}_\vp(\bar{q}=5\pi/16;\bar{q}_0=5\pi/32/\b_\th)$ as more
		regularization terms are incorporated.  The (black) squares represent individual $l$ modes
		of the unregularized SSF, the sum of which clearly diverges.  The (red) triangles are the residuals
		from subtracting $A_\vp$ and $B_\vp$.  The (blue) diamonds represent the residuals after subtracting
		$D_{\vp,2}$, obtained through numerical fitting.  The (purple) circles represent the inclusion of
		$D_{\vp,4}$, also approximated via a numerical fit.}
		\label{fig:convergenceFph}
\end{figure*}

As discussed in Sec.~\ref{sec:constructSSF}, for resonant orbits we express the SSF as a function of the
resonant angle variable $\bar{q}$ and the resonant phase parameter $\bar{q}_0$, i.e.,
$\bar{F}^\text{res}_\a(\bar{q};\bar{q}_0)$, or (as convenient) as a function of the more general angle
variables $q_r$ and $q_\th$ and the initial phases $q_{r0}$ and $q_{\th 0}$, i.e.,
$\hat{F}_\a(\b_r \bar{q},\b_\th \bar{q}+\b_\th \bar{q}_{0})=\hat{F}_\a(q_r,q_\th+q_{\th 0})$.  Plotting the SSF as
a function of $\bar{q}$, as shown in Sec.~\ref{sec:FresQQ0}, highlights the periodicity of the SSF during resonances
and is qualitatively representative of the Mino or coordinate time dependence of the SSF.  On the other hand,
plotting the SSF as a function of $q_r$ and $q_\th$, as shown in Sec.~\ref{sec:FresQrQth}, separates the dependence
of the SSF on the radial and polar motion of the orbit.  This way of depicting the SSF mirrors the parametrizations
used for nonresonant orbits, as seen in \cite{Vand18,NasiOsbuEvan19}.  To better analyze the impact of different
orbital parameters and types of resonances, we present each spacetime component of the self-force separately.

\subsection{Regularization and convergence of results}
\label{sec:regAndConvergeRes}

The SSF is constructed by mode-sum regularization and the numerical fitting procedures discussed in
Sec.~\ref{sec:ssfReg}.  The convergence of the mode-sum regularization procedure is well understood: subtracting
the analytically known regularization parameters, $A_\a$ and $B_\a$, produces residuals that fall off as
$\sim l^{-2}$ for large $l$.  There is no fundamental difference when an orbit is on resonance.  In
Fig.~\ref{fig:convergenceFph} we plot the mode-sum convergence of $\bar{F}^\text{res}_\vp$ at the point
$(\bar{q}=5\pi/16,\b_\th \bar{q}_0=5\pi/32)$ for all six resonant configurations.  Points refer to the $l$-mode
residuals that result from subtracting the analytically known and numerically fitted regularization parameters, while
the lines depict expected power-law convergence rates for large $l$.  In each resonance that we consider, the
residuals approach their expected asymptotic rates of convergence.

While all of the models have the same asymptotic behavior at large $l$, Fig.~\ref{fig:convergenceFph} demonstrates
that for low $l$ modes the $e=0.2$ sources converge faster than those with $e=0.5$, the 2:3 resonances converge
faster than the 1:2 resonances, and the 1:2 resonances converge faster than the 1:3 ones.  Higher eccentricities
require a broader frequency spectrum to capture the radial motion.  Additionally, sources that orbit farther into
the strong field excite larger perturbations and require higher frequency modes to capture the behavior of the
self-force.  The 1:3 resonances have the smallest pericentric separations, the 1:2 resonances have the next smallest,
and the 2:3 resonances have the largest, which is reflected in varying rates of convergence at low $l$.

Given these factors, the $e05.13$ orbit presents the greatest challenge.  For this model it takes thousands of
additional modes to capture the behavior of the SSF compared to other resonant configurations.  Because of the slow
convergence at low multipoles, truncating mode summations at the same value of $l_\text{max}$ as the other orbits
will introduce larger numerical errors in the retarded SSF contributions.  While these numerical errors are still
relatively small, they are significant enough that they make it much more difficult to fit for higher-order
regularization parameters.  The accuracy of the conservative component of the SSF suffers because of this.  In
consequence, the conservative SSF is only known to $\gtrsim 2$ digits of accuracy for the $e05.13$ orbit, with the
numerical error greatest when a component of the SSF is in the vicinity of passing through zero.  Fortunately, the
dissipative component typically dominates over the conservative contribution in regions of the orbit where the
conservative contribution is known less accurately.

\begin{figure*}
	\centering
	\includegraphics[width=0.95\textwidth]{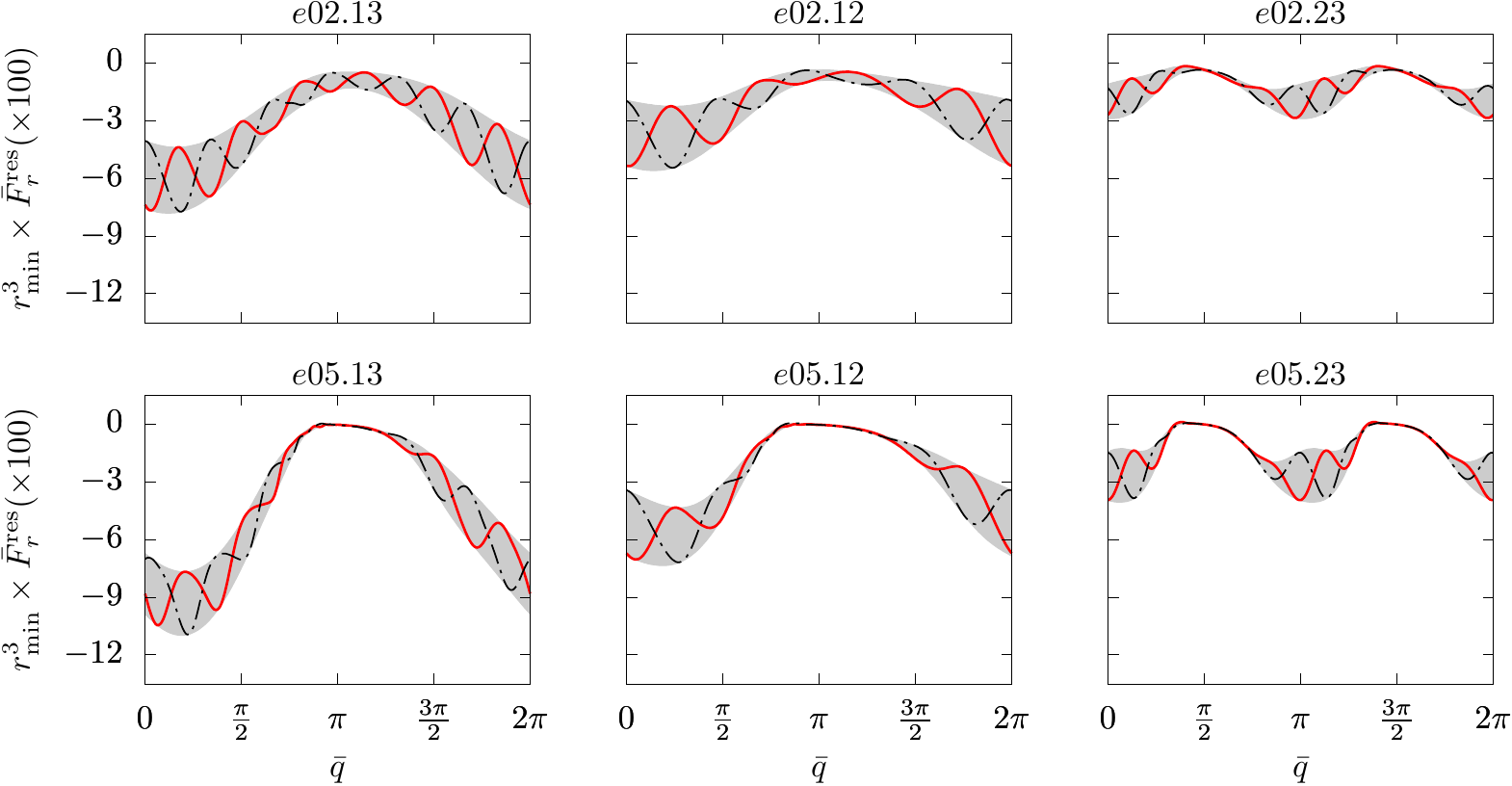}
	\caption{Radial component of the SSF as a function of the resonant angle variable $\bar{q}$,
	i.e., $\bar{F}^\text{res}_r(\bar{q};\bar{q}_0)$, for the six resonant geodesics listed in Table
	\ref{tab:orbits}.  The SSF is weighted by the cube of the pericentric radius, $r_\text{min}^3$, so that
	all six orbits are of comparable magnitude.  The dot-dashed (black) line represents the SSF for a
	resonant geodesic with an initial resonant phase of $\b_\th \bar{q}_0=q_{\th 0}=0$, while the solid (red)
	line represents the SSF for a resonance with the same orbital parameters but an initial resonant phase
	of $\b_\th \bar{q}_0=q_{\th 0}=-\pi/2$.  The shaded grey region represents all of the SSF values produced
	by varying the initial phase parameter $\bar{q}_0$ from $0$ to $2\pi$.}
\label{fig:ssfRVar}
\end{figure*}

\begin{figure*}
	\centering
	\includegraphics[width=0.95\textwidth]{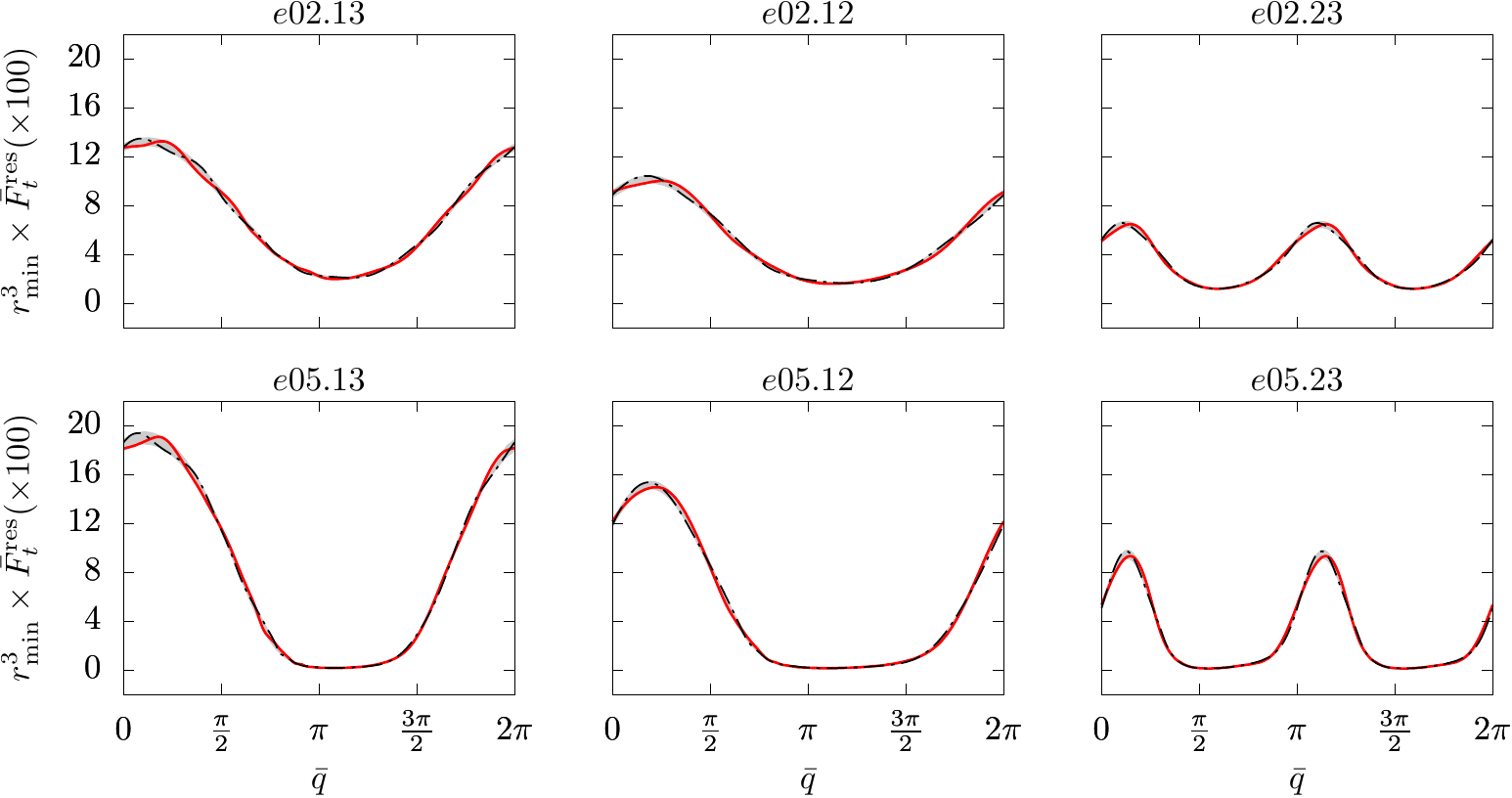}
	\caption{Time component of the SSF as a function of the resonant
	angle variable $\bar{q}$, i.e., $\bar{F}^\text{res}_t(\bar{q};\bar{q}_0)$,
	for the six resonant geodesics listed in Table \ref{tab:orbits}.  The dot-dashed
	(black) line represents the SSF for a resonant geodesic with an initial resonant
	phase of $\b_\th \bar{q}_0=q_{\th 0}=0$, the solid (red) line represents an initial
	resonant phase of $\b_\th \bar{q}_0=q_{\th 0}=-\pi/2$, and the shaded grey region
	represents all of the SSF values produced by varying $\bar{q}_0$ from $0$ to $2\pi$.}
	\label{fig:ssfTVar}
\end{figure*}

\begin{figure*}
	\centering
	\includegraphics[width=0.95\textwidth]{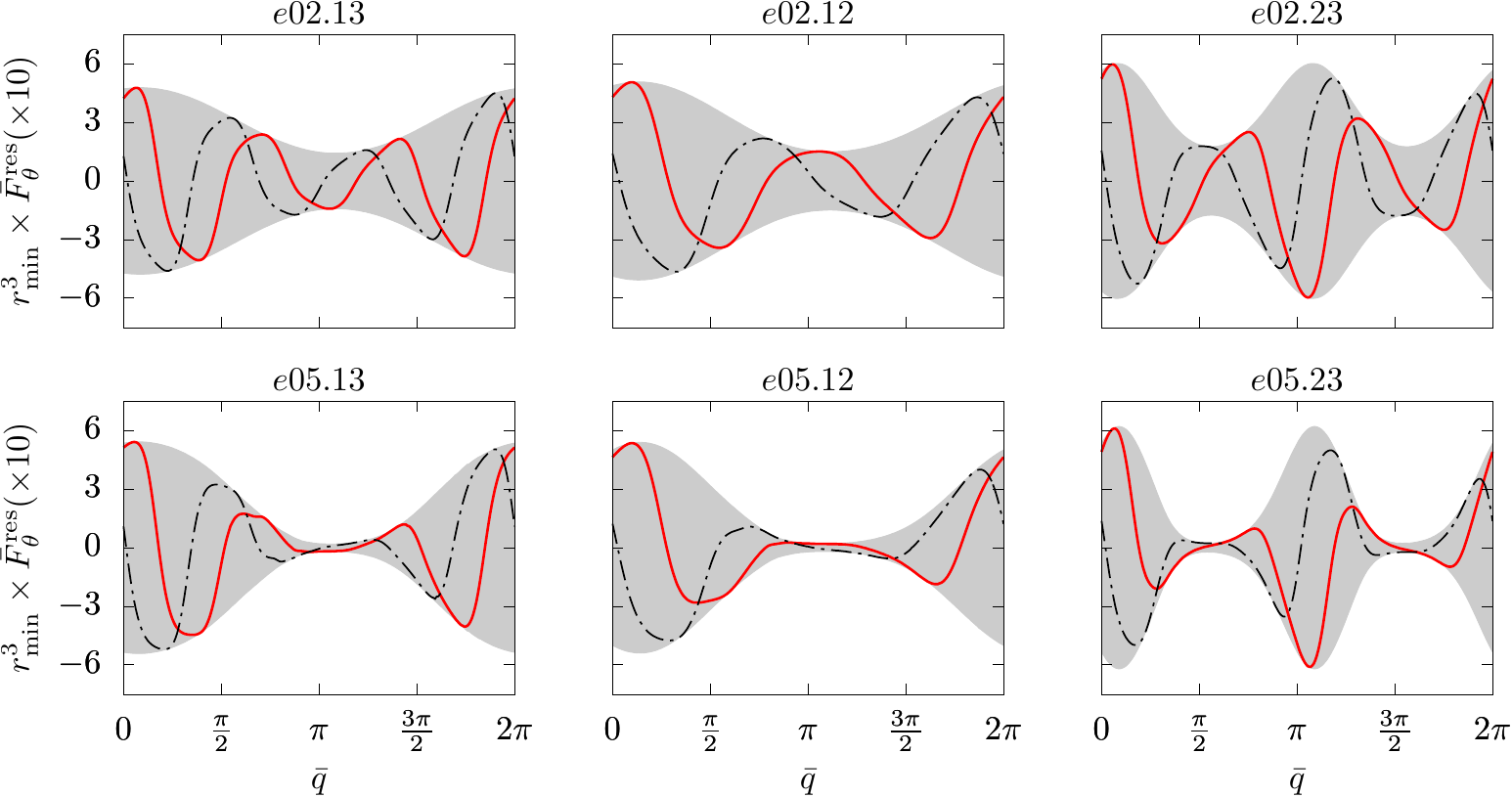}
	\caption{Polar component of the SSF as a function of the resonant angle
	variable $\bar{q}$, i.e., $\bar{F}^\text{res}_\th(\bar{q};\bar{q}_0)$, for
	the six resonant geodesics listed in Table \ref{tab:orbits}.  The dot-dashed
	(black) line represents the SSF for a resonant geodesic with an initial resonant
	phase of $\b_\th \bar{q}_0=q_{\th 0}=0$, the solid (red) line represents an initial
	resonant phase of $\b_\th \bar{q}_0=q_{\th 0}=-\pi/2$, and the shaded grey region
	represents all of the SSF values produced by varying $\bar{q}_0$ from $0$ to $2\pi$.}
	\label{fig:ssfThVar}
\end{figure*}

\begin{figure*}
	\centering
	\includegraphics[width=0.95\textwidth]{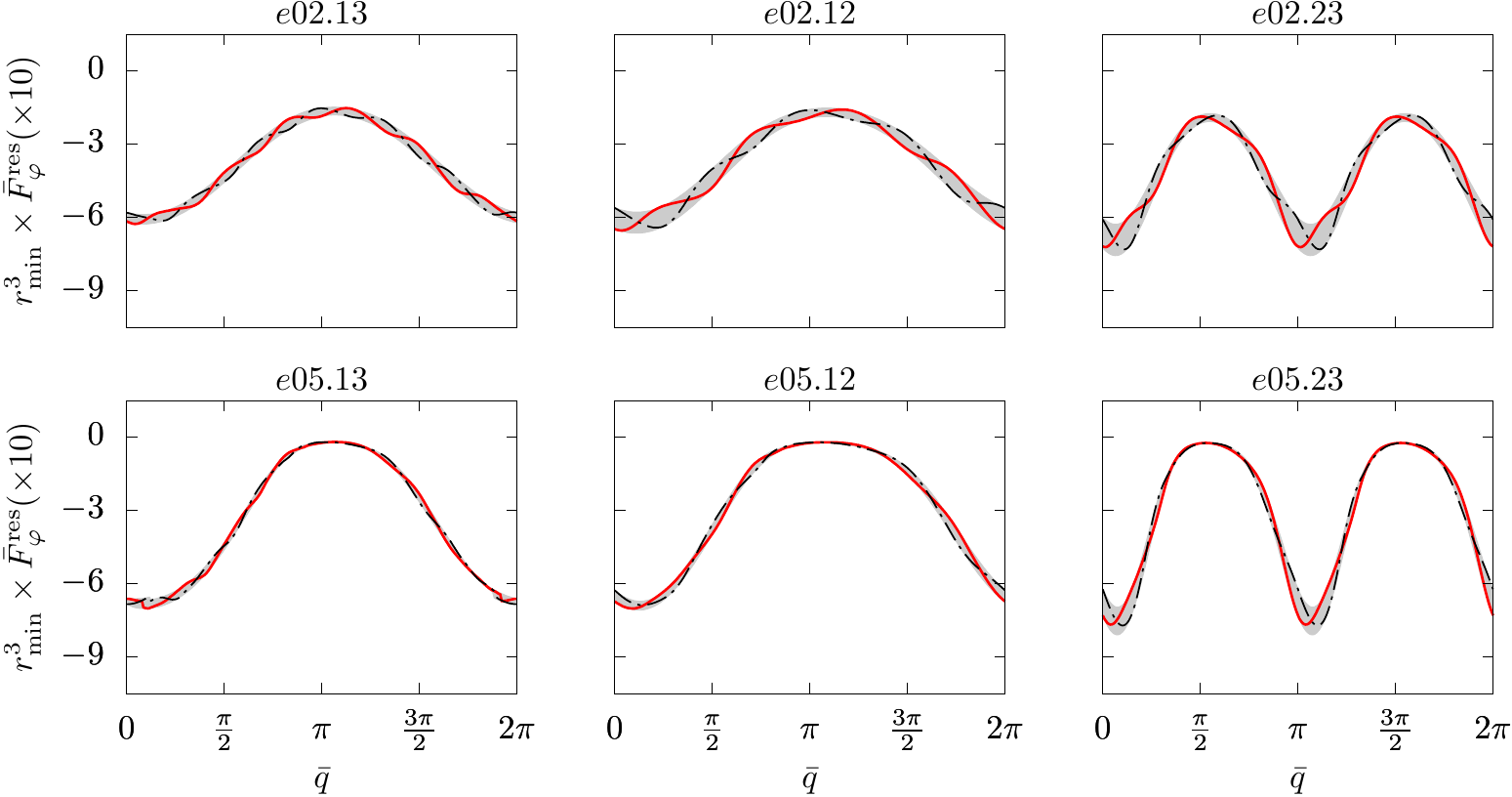}
	\caption{Azimuthal component of the SSF as a function of the resonant angle variable
	$\bar{q}$, i.e., $\bar{F}^\text{res}_\vp(\bar{q};\bar{q}_0)$, for the six resonant geodesics
	listed in Table \ref{tab:orbits}.  The dot-dashed (black) line represents the SSF for a
	resonant geodesic with an initial resonant phase of $\b_\th \bar{q}_0=q_{\th 0}=0$, the
	solid (red) line represents an initial resonant phase of $\b_\th \bar{q}_0=q_{\th 0}=-\pi/2$,
	and the shaded grey region represents all of the SSF values produced by varying $\bar{q}_0$
	from $0$ to $2\pi$.}
	\label{fig:ssfPhVar}
\end{figure*}

\subsection{Scalar self-force as a function of $\bar{q}$}
\label{sec:FresQQ0}

For a resonant orbit we can present the SSF in a simple line plot as a function of the net angle variable
$\bar{q}$, as depicted in Figs.~\ref{fig:ssfRVar}, \ref{fig:ssfTVar}, \ref{fig:ssfThVar}, and \ref{fig:ssfPhVar}.
In these plots the SSF has been weighted by the cube of the pericentric radius of the orbit (i.e.,
$r_\text{min}^3\,\bar{F}^\text{res}_\a$), which more tightly bounds the variations in the SSF and facilitates
comparisons across different models.  Each plot shows the SSF variation with $\bar{q}$ for two different initial
conditions (i.e., values of $\bar{q}_0$).  The dot-dashed (black) curves show the SSF when the initial polar phase
is $\b_\th \bar{q}_0=q_{\th 0}=0$ (i.e., initial conditions $x^\mu_p(\la=0)=(0,r_\text{min},\th_\text{min},0)$ and
$u^r(0)=u^\th(0)=0$), while the solid (red) curves show the SSF when $\b_\th \bar{q}_0=q_{\th 0}=-\pi/2$ (i.e., initial
conditions $x^\mu_p(\la=0)=(0,r_\text{min},\pi/2,0)$, $u^r(0)=0$, and $u^\th(0)<0$).\footnote{Note that $q_{\th 0}$
is held constant rather than $\bar{q}_0$, because the same value of $\bar{q}_0$ will generate different initial
conditions for resonances with different values of $\b_\th$.}  The shaded grey regions depict the range of SSF
values that result from varying the initial phases---either $q_{\th 0}$ or $\bar{q}_0$---through their entire range.

The SSF is, of course, periodic with respect to $\bar{q}$, but interestingly for the 2:3 resonances
$\bar{F}^\text{res}_t$, $\bar{F}^\text{res}_r$, and $\bar{F}^\text{res}_\vp$ are additionally periodic on the half
interval $[0,\pi]$.  This behavior arises in the Kerr background because the time, radial, and azimuthal components
of the SSF are invariant under parity transformations (i.e., reflections $\th_p \rightarrow \pi-\th_p$), while the
polar component flips sign \cite{Warb15} (equally true of the gravitational self-force \cite{Vand18}).  For a
2:3 resonance, the radial motion of the orbit is identical on the intervals $[0,\pi]$ and $[\pi,2\pi]$, while the
polar motion is related by the parity transformation.  From this fact follows the repetition in $\bar{F}^\text{res}_t$,
$\bar{F}^\text{res}_r$, and $\bar{F}^\text{res}_\vp$, while also giving the reflection behavior
$\bar{F}^\text{res}_\th(\bar{q};\bar{q}_0)=-\bar{F}^\text{res}_\th(\bar{q}+\pi;\bar{q}_0)$.

These symmetries in the geodesic motion also manifest themselves in the number of low-frequency oscillations
that appear in the SSF components, particularly in the low-eccentricity orbits.  Focusing on $\bar{F}^\text{res}_r$
in Fig.~\ref{fig:ssfRVar}, the SSF locally peaks 6 times for the $e02.13$ and $e02.23$ models and 4 times in
the $e02.12$ case.  The peaks closely align with the epochs at which each orbit passes through its polar extrema.
A similar behavior is also seen for $\bar{F}^\text{res}_t$, $\bar{F}^\text{res}_\vp$, and the higher eccentricity
models, though for $e=0.5$ it is more difficult to identify local peaks, particularly as the orbit approaches
apocenter.  For $\bar{F}^\text{res}_\th$ in Fig.~\ref{fig:ssfThVar}, the peaks align with the passage of the source
through $\th_\text{min}$, while the troughs align with its passages through $\pi-\th_\text{min}$.

The degree to which the SSF varies with respect to changes in initial phase depends primarily on which component
of the SSF vector we consider.  The time component, $\bar{F}^\text{res}_t$ (Fig.~\ref{fig:ssfTVar}), displays the
least effect of varying the initial conditions.  The azimuthal component, $\bar{F}^\text{res}_\vp$
(Fig.~\ref{fig:ssfPhVar}), shows slightly greater variations with respect to initial conditions.  The radial
component, $\bar{F}^\text{res}_r$ (Fig.~\ref{fig:ssfRVar}), is still more affected.  Finally, the polar angular
component, $\bar{F}^\text{res}_\th$ (Fig.~\ref{fig:ssfThVar}), displays the most significant variations.  To
understand these variations, recall that the radial and polar position of the resonant source, $\bar{r}_p$
and $\bar{\th}_p$, depend on the angle variables according to
\begin{align}
	\bar{r}_p&=\hat{r}_p(q_r)=\hat{r}(\b_r \bar{q}), \\
	\bar{\th}_p&=\hat{\th}_p(q_\th+q_{\th 0})=\hat{\th}(\b_\th \bar{q}+\b_\th \bar{q}_0).
\end{align}
Consequently, a broader grey band indicates a stronger dependence on the polar motion.  Thus, $\bar{F}^\text{res}_t$
primarily depends on the radial motion of the source, while $\bar{F}^\text{res}_r$ is sensitive to both polar and
radial motions.  In behavior opposite of $\bar{F}^\text{res}_t$, $\bar{F}^\text{res}_\th$ is primarily dependent on
the polar motion of the orbit.  Finally, $\bar{F}^\text{res}_\vp$ depends mostly on radial motion of the source,
though the polar position becomes important near pericenter.

\begin{figure*}
	\centering
	\includegraphics[width=0.89\textwidth]{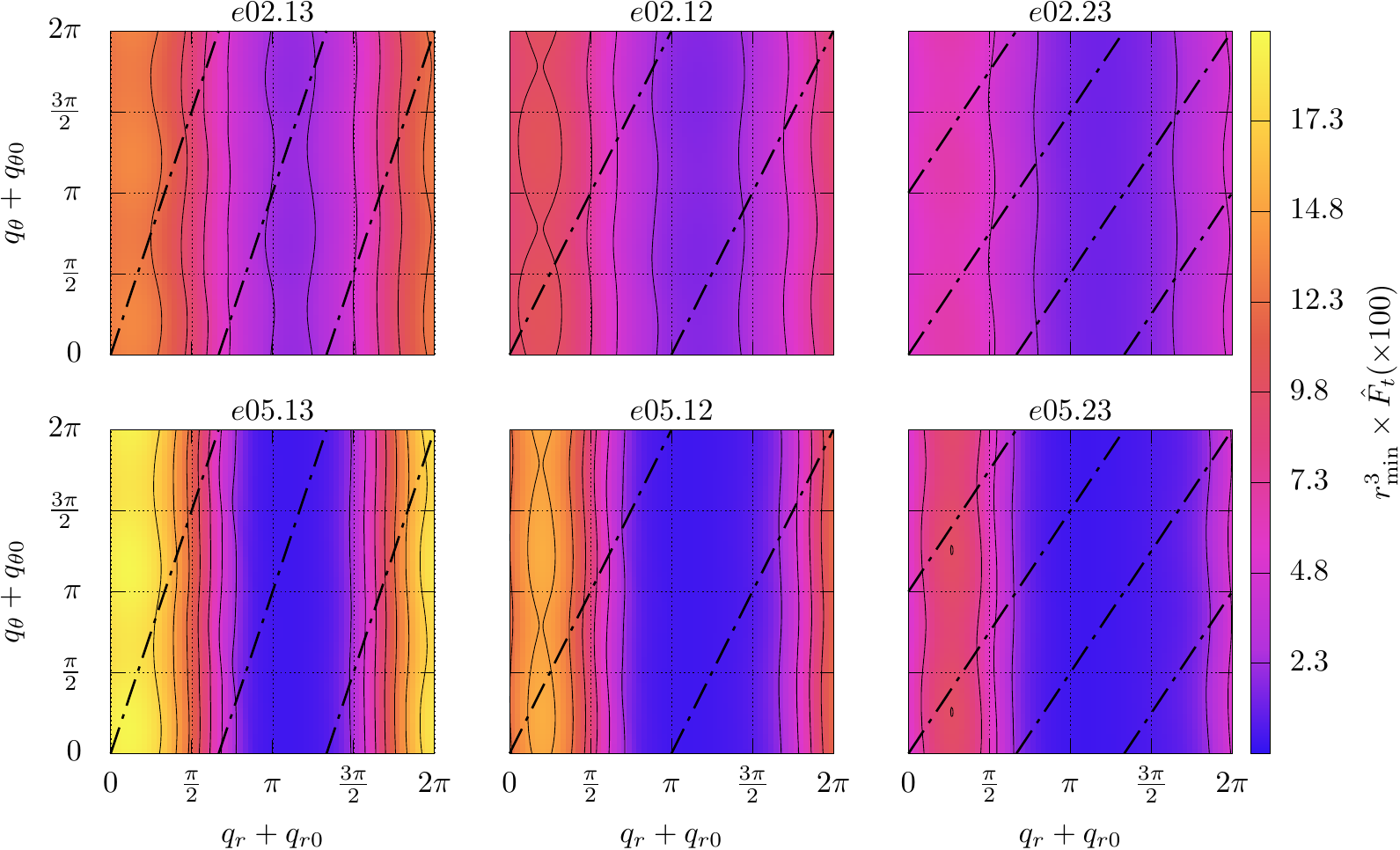}
	\caption{Time component of the SSF $\hat{F}_t$ projected on the poloidal motion two-torus for the six
sources listed in Table~\ref{tab:orbits}.  The SSF is normalized by the cube of each source's pericenter distance.
Colors correspond to values of the self-force (see colorbar).  The self-force is constant along each (solid) contour
line with tic labels in the colorbar giving the values on those contours.  The dot-dashed lines depict the resonant
motion for fiducial initial conditions ($\bar{q}_0=0$).}
\label{fig:ssfTTorus}
\end{figure*}

\begin{figure*}
	\centering
	\includegraphics[width=0.89\textwidth]{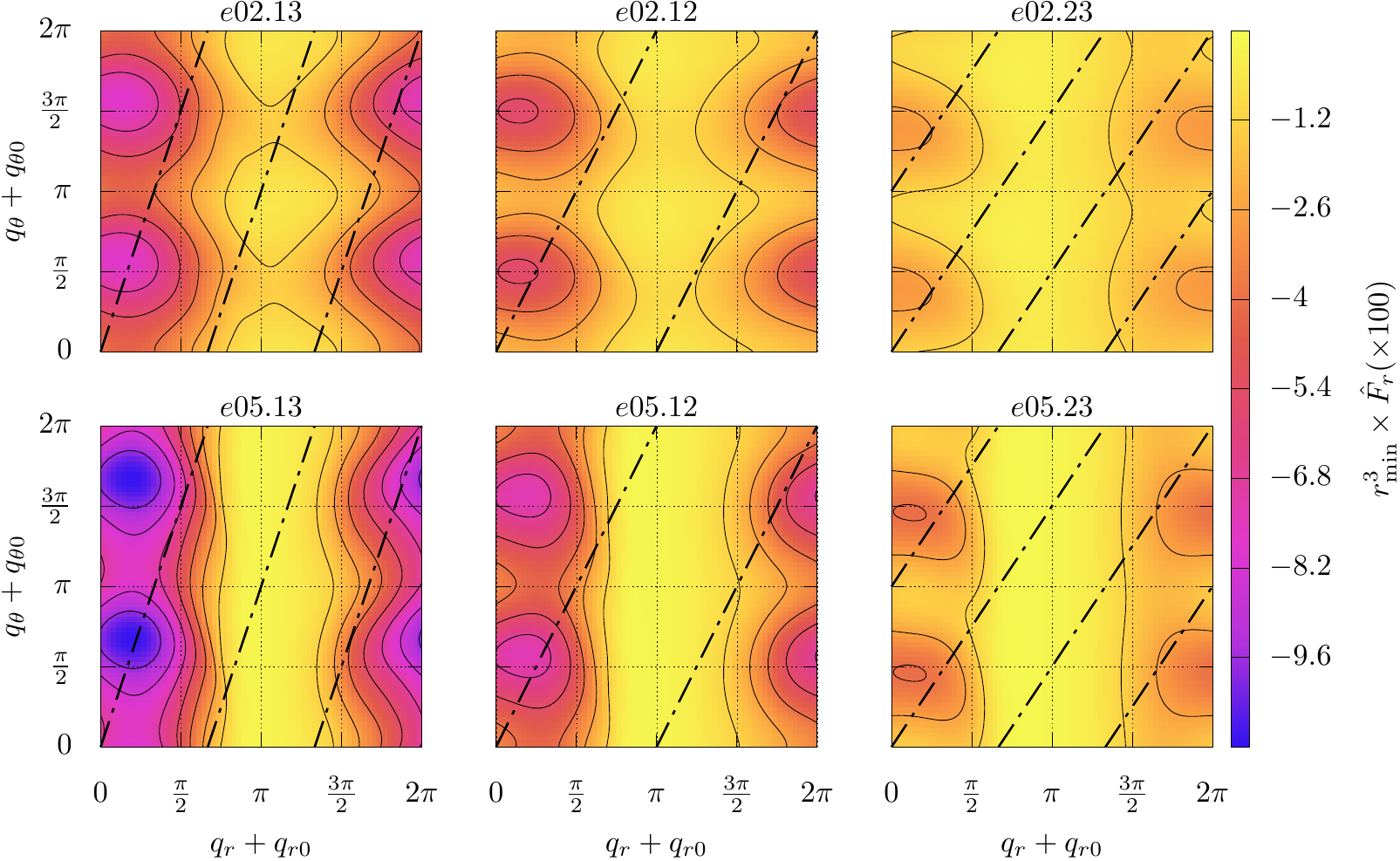}
	\caption{Radial component of the scalar self-force $\hat{F}_r$ for the six orbits listed in
Table~\ref{tab:orbits}.}
	\label{fig:ssfRTorus}
\end{figure*}

\begin{figure*}
	\centering
	\includegraphics[width=0.9\textwidth]{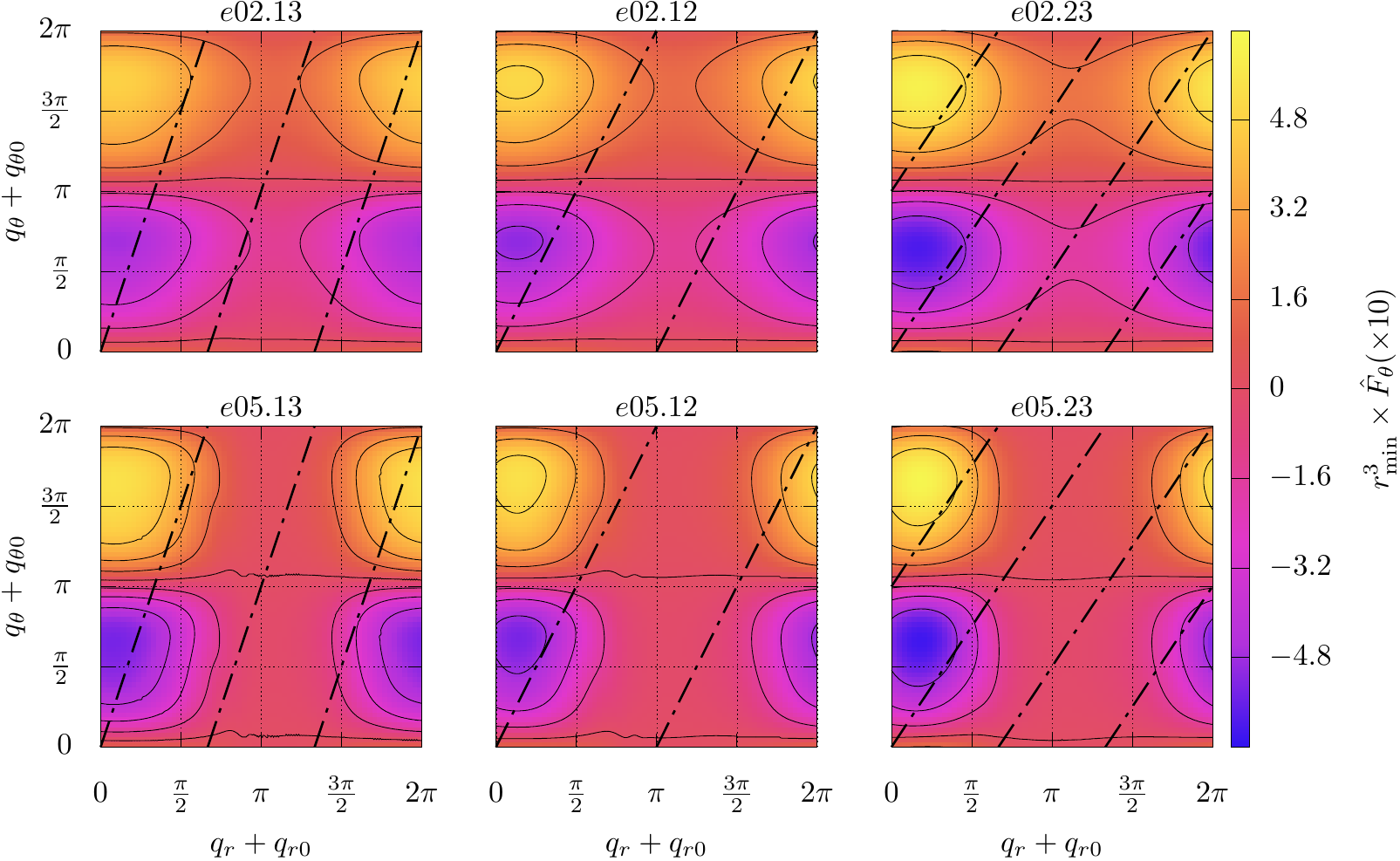}
	\caption{Polar component of the scalar self-force $\hat{F}_\th$ for the six orbits listed in
	Table~\ref{tab:orbits}.}
	\label{fig:ssfThTorus}
\end{figure*}

\begin{figure*}
	\centering
	\includegraphics[width=0.9\textwidth]{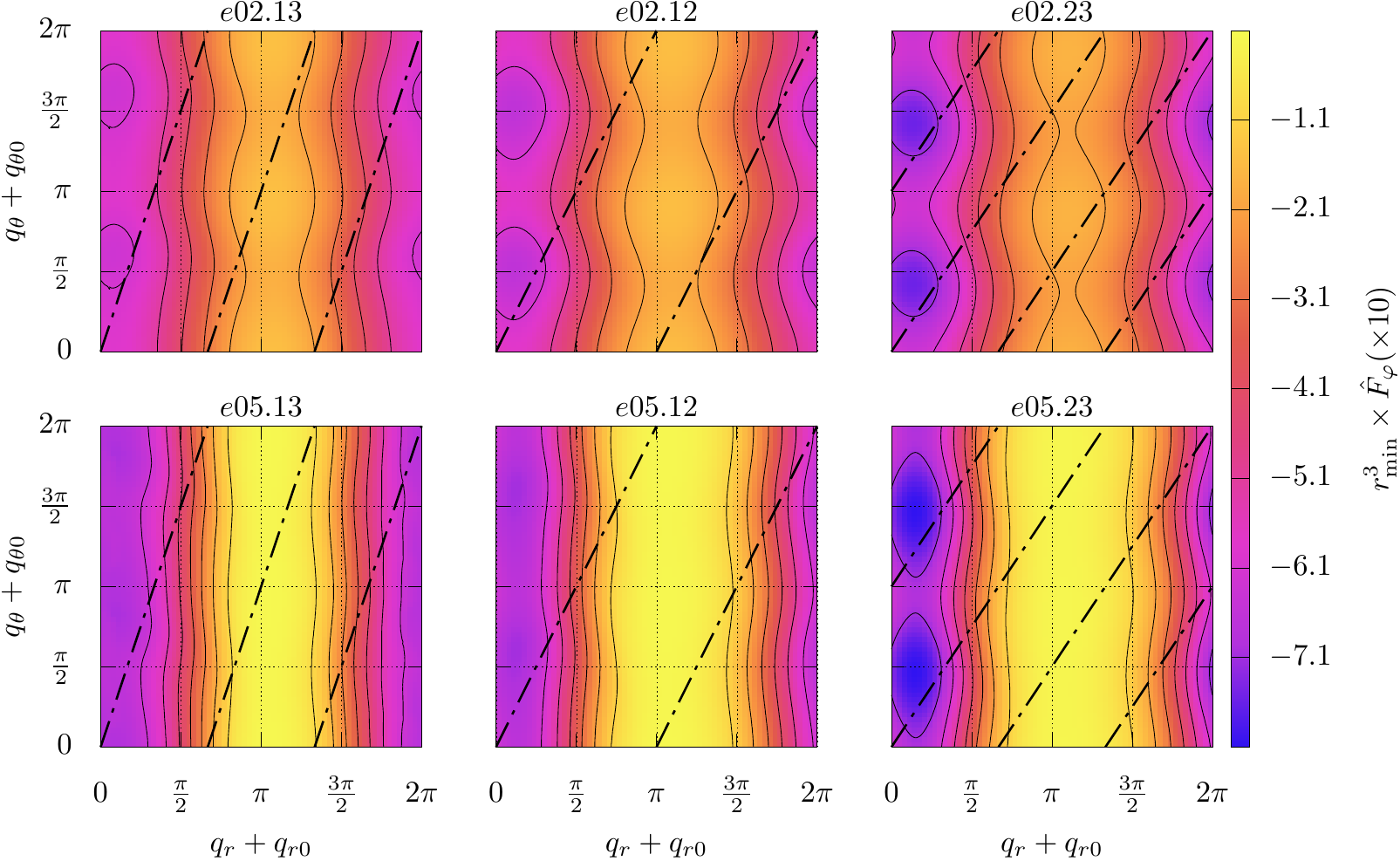}
	\caption{Azimuthal component of the scalar self-force $\hat{F}_\vp$ for the six orbits listed in
	Table~\ref{tab:orbits}.}
	\label{fig:ssfPhTorus}
\end{figure*}

\subsection{Scalar self-force as a function of $q_r$ and $q_\th$}
\label{sec:FresQrQth}

An alternative way to visualize the dependence of the SSF on the radial and polar motion of resonant orbits is
to project the SSF components onto the two-torus spanned by $q_r+q_{r0}$ and $q_\th+q_{\th 0}$.  This projection
is depicted in Figs.~\ref{fig:ssfTTorus}, \ref{fig:ssfRTorus}, \ref{fig:ssfThTorus}, and \ref{fig:ssfPhTorus}.
We again weight the SSF components by the cube of the pericentric radius of the orbit.  The dot-dashed (black) lines
trace the motion of an orbit with initial conditions $\b_\th \bar{q}_0=q_{\th 0}=0$.  Sampling the SSF as the source
moves along these tracks reproduces the black dot-dashed curves in Figs.~\ref{fig:ssfRVar}, \ref{fig:ssfTVar},
\ref{fig:ssfThVar}, and \ref{fig:ssfPhVar}.  Maintaining previous notation, we refer to the SSF parametrized by
$q_r+q_{r0}$ and $q_\th+q_{\th 0}$ as $\hat{F}_\a$.

As observed in Sec.~\ref{sec:FresQQ0}, $\hat{F}_t$ (shown in Fig.~\ref{fig:ssfTTorus}) primarily depends on the
radial motion, with little variation as the orbit advances along the $q_\th$ axis.  As the contours show in
Figs.~\ref{fig:ssfRTorus} and \ref{fig:ssfPhTorus}, the radial and azimuthal components, $\hat{F}_r$ and
$\hat{F}_\vp$ are sensitive to both the radial and polar motion, especially near pericenter.  Finally, the contours
of the SSF seen in Fig.~\ref{fig:ssfThTorus} clearly demonstrate the antisymmetry across the equatorial plane
of $\hat{F}_\th$, as discussed in the previous section.

In agreement with previous investigations \cite{WarbBara10,WarbBara11,Warb15,NasiOsbuEvan19} of the SSF, we see
that $\hat{F}_t$ is strictly positive.  This contrasts with the gravitational self-force case where the time component
can become negative in both radiation and Lorenz gauge \cite{Vand16,Vand18}.\footnote{We do not try to draw any
physical interpretation from this behavior since the SSF and gravitational self-force are coordinate and (in the
GSF case) gauge-dependent results.}  On the other hand, $\hat{F}_r$ is predominantly negative across the entire
torus, though it becomes slightly positive near apocenter.  This behavior is consistent with the observation
\cite{WarbBara10} that higher black hole spin leads to an attractive radial SSF.  Large inclinations, on the other
hand, lead predominantly to positive values of the SSF, as seen in SSF results for spherical orbits \cite{Warb15}.
However, those prior observations involved inclinations $x\gtrsim 0.5$, which we did not consider here.

Interestingly, while all of the SSF components peak in magnitude following pericenter passage, the magnitude of
these peaks grows for $\hat{F}_t$ and $\hat{F}_r$ as $r_\text{min}$ decreases, while the peaks grow for
$\hat{F}_\th$ and $\hat{F}_\vp$ as $r_\text{min}$ increases.  The latter behavior is actually due to the factor of
$r_\text{min}^3$.  If one removes this weighting, then closer pericenter passages excite larger peaks in the SSF
for all components.  This suggests that the leading order behavior of $\hat{F}_\th$ and $\hat{F}_\vp$ is closer
to $1/r^2$, which one might expect based on dimensional analysis ($[F_{\th,\vp}/F_{t,r}]_\text{dim}
\sim [M]_\text{dim}$).

\section{Evolution of the orbital constants}
\label{sec:flux}

\subsection{Overview}

In the presence of radiative losses and the self-force, the ordinarily constant quantities $\mathcal{E}$,
$\mathcal{L}_z$, and $\mathcal{Q}$ are perturbed and gradually evolve according to
\begin{gather}
\label{eqn:edot}
	\dot{\mathcal{E}} = - \frac{q^2}{u^t} a_t, \qquad \qquad
	\dot{\mathcal{L}}_z = \frac{q^2}{u^t} a_\vp,
	\\ \label{eqn:qdot}
	\dot{\mathcal{Q}}=
	 \frac{2}{u^t} \left[q^2 K_{\mu\nu}u^\mu a^\nu - (\mathcal{L}_z
	-a\mathcal{E})(\dot{\mathcal{L}}_z-a\dot{\mathcal{E}})
	\right],
\end{gather}
where an overdot represents a derivative with respect to Boyer-Lindquist time, and the self-acceleration $a^\mu$ is given
by $\mu a^\nu = (g^{\mu\nu}+u^\mu u^\nu)F_\mu = F^\nu - q^{-2}u^\nu d\mu/d\tau$.  Note that the lack of orthogonality between
$F_\nu$ and $u^\nu$ drives changes in the mass $\mu$ (see
\ref{app:RestMass}).

The changes $\dot{\mathcal{E}}$, $\dot{\mathcal{L}}_z$, and $\dot{\mathcal{Q}}$ consist of both secularly growing
and oscillating parts, with the secular piece found by orbit-averaging \eqref{eqn:edot} and \eqref{eqn:qdot} with
respect to $t$.  For a nonresonant orbit, the averaging is over a long timescale,
\begin{equation} \label{eqn:orbitAvg}
	\langle \dot{\mathcal{X}} \rangle \equiv \lim_{T\rightarrow \infty}
	 \frac{1}{T} \int_0^T \dot{\mathcal{X}} dt,
	\qquad \mathcal{X} =
	\mathcal{E},\, \mathcal{L}_z,\, \mathcal{Q} .
\end{equation}
These averages produce the leading order, adiabatic evolution of the system \cite{HindFlan08}.  The time integrals
can be reexpressed in terms of the angle variables that are used to parametrize the self-force.  Then the averaging
is done over the motion on the torus \cite{DrasHugh04,GrosLeviPere13}.  For nonresonant orbits, $\dot{\mathcal{E}}$,
$\dot{\mathcal{L}}_z$, and $\dot{\mathcal{Q}}$ are averaged over the entire two-torus by integrating with equal weight
over all $q_r$ and $q_\th$.  For resonances, these orbit-averages are carried out over a single, one-dimensional
closed track on the torus, reducing \eqref{eqn:orbitAvg} to a single integral over the resonant phase variable
$\bar{q}$,
\begin{align} \label{eqn:work}
	\mu \langle \dot{\mathcal{E}} \rangle & =
	- \frac{q^2}{\G} \int_0^{2\pi}  \frac{d\bar{q}}{2\pi}\,
	\bar{\Sig}_p\,
	\bar{F}^\text{res}_t = q^2\mathcal{W},
	\\ \label{eqn:torque}
	\mu \langle \dot{\mathcal{L}}_z \rangle &=
	\frac{q^2}{\G} \int_0^{2\pi}  \frac{d\bar{q}}{2\pi}\,
	\bar{\Sig}_p\,
	\bar{F}^\text{res}_\vp = q^2\mathcal{T},
	\\  \label{eqn:qdotAvg}
	\mu \langle \dot{\mathcal{Q}} \rangle &=
	2q^2\Big[- (\mathcal{L}_z
	-a\mathcal{E})(\dot{\mathcal{T}}-a\dot{\mathcal{W}})
	\\
	& \qquad \qquad \; +
	\frac{1}{\G} \int_0^{2\pi}  \frac{d\bar{q}}{2\pi}\,
	\bar{\Sig}_p\,\bar{K}_p^{\mu\nu}
	\bar{u}_\mu \bar{F}^\text{res}_\nu \Big] .
	\notag
\end{align}
In the expressions above, all quantities with an overbar are understood to be functions of $\bar{q}$ and
parametrized by $\bar{q}_0$ [e.g.,
$\bar{\Sig}_p=\bar{\Sig}(\bar{q}; \bar{q}_0)=\bar{r}_p^2(\bar{q})+a^2\cos^2\bar{\th}_p(\bar{q}+\bar{q}_0)$].
The changes $\langle \dot{\mathcal{E}} \rangle$ and $\langle \dot{\mathcal{L}}_z \rangle$ are directly related to
the average rate of work $\mathcal{W}$ and torque $\mathcal{T}$ done on the small body by the SSF (per charge squared)
and incorporate the fact that the average change in $\mu$ vanishes (see \ref{app:RestMass}).

For nonresonant orbits the conservative components of the self-force vanish when averaged over the entire torus.
This fact can be seen from the symmetries of \eqref{eqn:consSSF} and \eqref{eqn:ssfRet2Adv}, combined with the
expressions for $\dot{\mathcal{E}}$, $\dot{\mathcal{L}}_z$, and $\dot{\mathcal{Q}}$.  Only the dissipative
self-force contributes to the leading order adiabatic evolution of the system when it is off resonance.  When on
resonance, we cannot make use of these same symmetries to discard the conservative component of the self-force in
\eqref{eqn:work}, \eqref{eqn:torque}, and \eqref{eqn:qdotAvg}.  However, flux-balance conditions do confirm that
conservative contributions to $\langle \dot{\mathcal{E}}\rangle$ and $\langle \dot{\mathcal{L}}_z\rangle$ continue
to vanish, as we further discuss in Sec.~\ref{sec:EnAndLz}.  Additionally, the averages over an $r\th$ resonance
retain their dependence on $\bar{q}_0$, meaning that they vary according to the initial phase at which the system
enters a resonance, as demonstrated previously \cite{FlanHughRuan14,BerrETC16}.  Thus different initial conditions
can either diminish or enhance the averaged evolution of $\mathcal{E}$, $\mathcal{L}_z$, and $\mathcal{Q}$ during
a resonance.  The following subsections detail this behavior in the scalar case and provide numerical data on how
the conservative and dissipative components of the SSF contribute to $\langle \dot{\mathcal{E}} \rangle$,
$\langle \dot{\mathcal{L}}_z \rangle$, and $\langle \dot{\mathcal{Q}} \rangle$.

\subsection{Energy and angular momentum changes for a resonant orbit}
\label{sec:EnAndLz}

Flux-balance equates the average changes in the orbital energy and angular momentum,
$\langle \dot{\mathcal{E}} \rangle$ and $\langle \dot{\mathcal{L}}_z \rangle$, to the average radiative fluxes
\cite{Galt82,QuinWald99,Mino03}.  For energy, the average work $\mathcal{W}$ done by the SSF balances the total flux
$\langle \dot{E}\rangle^\mathrm{tot}$ radiated by the scalar field to infinity and down the horizon, with the
on-resonance fluxes having slightly modified expressions
\begin{align}
\label{eqn:fluxEBalance}
	-\mathcal{W} &= \langle \dot{E} \rangle^\mathrm{tot} \equiv
	\langle \dot{E} \rangle^\mathcal{H}
	+ \langle \dot{E} \rangle^\infty,
	\\
	\langle \dot{E} \rangle^\mathcal{H} & = \frac{1}{4\pi}
	\sum_{\lhat = 0}^\infty \sum_{m=-\lhat}^{\lhat} \sum_{N=-\infty}^\infty
	\o_{mN} \g_{mN} |\bar{C}^-_{\lhat mN}|^2,
	\\
	\langle \dot{E} \rangle^\infty & = \frac{1}{4\pi}
	\sum_{\lhat = 0}^\infty \sum_{m=-\lhat}^{\lhat} \sum_{N=-\infty}^\infty
	\o_{mN}^2 |\bar{C}^+_{\lhat mN}|^2 .
\end{align}
Here $\langle \dot{E} \rangle^\mathcal{H}$ is the energy flux (per charge squared) through the horizon, and
$\langle \dot{E} \rangle^\infty$ is the energy flux (per charge squared) at infinity, with
$\g_{mN}\equiv \o_{mN} - ma/(2Mr_+)$ being the spatial frequency of the radial modes at the horizon and
$r_+ \equiv M+\sqrt{M^2-a^2}$ denoting the radius of the outer horizon.  In the resonant case, the fluxes include
a single sum over the net harmonic number $N$ of the net amplitudes $\bar{C}^\pm_{\lhat mN}$.  The net amplitudes
are themselves sums \eqref{eqn:ClmN2Clmkn} over amplitudes $\hat{C}^\pm_{\lhat mkn}$ with individual radial and
polar harmonic numbers $n$ and $k$.  These underlying sums reflect the coherence between all harmonics of the
radial and polar librations that contribute to a given $N$.  In this way, interference terms appear in the flux
that would otherwise average to zero in the off-resonance case.

\begin{figure}[tb]
	\centering
	\includegraphics[width=0.42\textwidth]{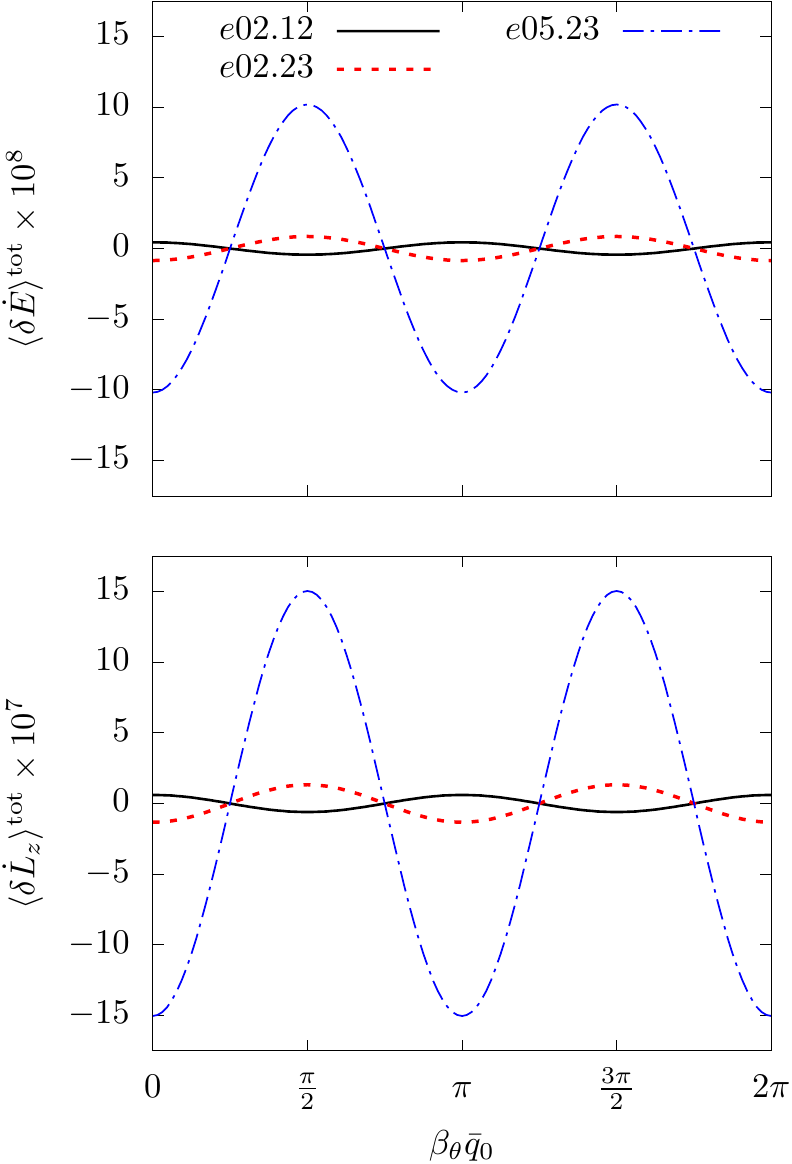}
	\caption{Residual variations in the energy (top panel) and angular
	momentum (bottom panel) fluxes as a function of the initial phase
	$q_{\th 0}=\b_\th \bar{q}_0$ for the $e02.23$ (solid black curves),
	$e05.23$ (dot-dashed blue curves), and
	$e05.12$ (dotted red curves) orbits in in Table \ref{tab:orbits}.
	}
	\label{fig:fluxRes}
\end{figure}

{\renewcommand{\arraystretch}{1.75}
\begin{table*}[!bthp]
\caption{Energy and angular momentum fluxes for the resonant-orbit models listed in Table \ref{tab:orbits}.
Fluxes through the horizon, $\langle\dot{E}\rangle^\mathcal{H}$ and $\langle\dot{L}_z\rangle^\mathcal{H}$,
and infinity, $\langle\dot{E}\rangle^\infty$ and $\langle\dot{L}_z\rangle^\infty$, are included.  Each model contains
a row of fluxes for an orbit with initial phase $q_{\th 0} = \b_\th\bar{q}_0=0$ and a row of fluxes for an orbit with
$q_{\th 0} = \b_\th\bar{q}_0= -\pi/2$.  A third row in each case shows the $\bar{q}_0$ averages of the fluxes as
defined by \eqref{eqn:fluxAvg}, which ignores constructive and destructive interference terms.  The reported
precision in each flux indicates the accuracy of each calculation (though we truncate more accurate results at nine
decimal places).  The total fluxes are also compared to the local work and torque due to the SSF, $\mathcal{W}$ and
$\mathcal{T}$, to illustrate the (orbit-averaged) fractional errors in the flux balance relations.  These errors
range from $\sim 10^{-11}$ to $\sim 10^{-5}$, reflecting the numerical accuracy of our SSF results.}
	\label{tab:fluxes}
	\centering
	\begin{tabular*}{\textwidth}{c @{\extracolsep{\fill}} c c c c c c c}
		\hline
		\hline
		\vspace{-10pt}
		\\
		Model
		& $\b_\th\bar{q}_0$
		& $\langle\dot{E}\rangle^\mathcal{H}\times {10^5}$
		& $\langle\dot{L}_z\rangle^\mathcal{H}\times {10^4}$
		& $\langle\dot{E}\rangle^\infty\times {10^3}$
		& $\langle\dot{L}_z\rangle^\infty\times {10^3}$
		& $\left|1+\frac{\langle\dot{E}\rangle^\text{tot}}{\mathcal{W}}\right|$
		& $\left|1+\frac{\langle\dot{L}_z\rangle^\text{tot}}
		{\mathcal{T}}\right|$
		\\
		\vspace{-10pt}
		\\
		\hline
		$e02.13$ & $0$ & $-4.411457095$ & $-7.017966266$
		& $1.301535\phantom{000}$ & $7.677846\phantom{000}$ & $9\times 10^{-7}$ &
		$8\times 10^{-7}$
		\\
		& $-\pi/2$ & $-4.411497781$ & $-7.017992874$
		& $1.301534\phantom{000}$ & $7.677831\phantom{000}$ & $7\times 10^{-7}$ &
		$6\times 10^{-7}$
		\\
		& avg & $-4.411477437$ & $-7.017979570$
		& $1.301535\phantom{000}$ & $7.677838\phantom{000}$ & - & -
		\\ \hline
		$e02.12$ & 0 & $-2.021123696$ & $-3.395925026$
		& $0.5737075\phantom{00}$ & $4.843929\phantom{000}$ & $2\times 10^{-8}$ &
		$2\times 10^{-8}$
		\\
		& $-\pi/2$ & $-2.021127357$ & $-3.396083610$
		& $0.5736988\phantom{00}$ & $4.843824\phantom{000}$ & $2\times 10^{-8}$ &
		$2\times 10^{-8}$
		\\
		& avg & $-2.021125529$ & $-3.396004318$
		& $0.5737031\phantom{00}$ & $4.483877\phantom{000}$ & - & -
		\\ \hline
		$e02.23$ & $0$ & $-0.324830139$ & $-0.877896699$
		& $0.134964247$ & $1.762343845$ & $6\times 10^{-9}$ &
		${}^{\phantom{1}}3\times 10^{-11}$
		\\
		& $-\pi/2$ & $-0.325170299$ & $-0.877675562$
		& $0.134984611$ & $1.762586419$ & $6\times 10^{-9}$ &
		${}^{\phantom{1}}3\times 10^{-11}$
		\\
		& avg & $-0.325000220$ & $-0.877786129$
		& $0.134974429$ & $1.762465132$ & - & -
		\\ \hline
		$e05.13$ & $0$ & $-0.482340196$ & $-6.445882073$
		& $1.45739\phantom{0000}$ & $7.28846\phantom{0000}$ & $9\times 10^{-5}$ &
		$8\times 10^{-5}$
		\\
		& $-\pi/2$ & $-0.480744834$ & $-6.448440589$
		& $1.45724\phantom{0000}$ & $7.28681\phantom{0000}$ & $9\times 10^{-5}$ &
		$8\times 10^{-5}$
		\\
		& avg & $-0.481543289$ & $-6.447161427$
		& $1.45731\phantom{0000}$ & $7.28763\phantom{0000}$ & - & -
		\\ \hline
		$e05.12$ & $0$ & $-0.364726314$ & $-2.915401042$
		& $0.590229\phantom{000}$ & $3.85672\phantom{0000}$ & $3\times 10^{-5}$ &
		$2\times 10^{-5}$
		\\
		& $-\pi/2$ & $-0.361475840$ & $-2.919310921$
		& $0.589990\phantom{000}$ & $3.85408\phantom{0000}$ & $3\times 10^{-5}$ &
		$2\times 10^{-5}$
		\\
		& avg & $-0.363102237$ & $-2.917355986$
		& $0.590110\phantom{000}$ & $3.85540\phantom{0000}$ & - & -
		\\ \hline
		$e05.23$ & $0$ & $\phantom{-}0.092800200$ &
		$-0.713304108$
		& $0.12832694\phantom{0}$ & $1.3990094\phantom{00}$ & $3\times 10^{-7}$ &
		$2\times 10^{-7}$
		\\
		& $-\pi/2$ & $\phantom{-}0.090053088$ & $-0.710893569$
		& $0.12855813\phantom{0}$ & $1.4017751\phantom{00}$ & $3\times 10^{-7}$ &
		$2\times 10^{-7}$
		\\
		& avg & $\phantom{-}0.091426409$ & $-0.712098512$
		& $0.12844253\phantom{0}$ & $1.4003923\phantom{00}$ & - & -
		\\
		\hline
		\hline
	\end{tabular*}
\end{table*}
}

In a similar way the average torque $\mathcal{T}$ applied by the SSF balances the sum of the angular momentum flux
at the horizon $\langle \dot{L}_z \rangle^\mathcal{H}$ and at infinity $\langle \dot{L}_z \rangle^\infty$,
\begin{align}
\label{eqn:fluxLzBalance}
	-\mathcal{T} &= \langle \dot{L}_x \rangle^\mathrm{tot} \equiv
	\langle \dot{L}_z \rangle^\mathcal{H}
	+ \langle \dot{L}_z \rangle^\infty,
	\\
	\langle \dot{L}_z \rangle^\mathcal{H} & = \frac{1}{4\pi}
	\sum_{\lhat = 0}^\infty \sum_{m=-\lhat}^{\lhat} \sum_{N=-\infty}^\infty
	m \g_{mN} |\bar{C}^-_{\lhat mN}|^2,
	\\ \label{eqn:fluxLzBalanceLast}
	\langle \dot{L}_z \rangle^\infty & = \frac{1}{4\pi}
	\sum_{\lhat = 0}^\infty \sum_{m=-\lhat}^{\lhat} \sum_{N=-\infty}^\infty
	m\o_{mN} |\bar{C}^+_{\lhat mN}|^2 .
\end{align}
Recall from \eqref{eqn:ClmN2Clmkn} that the net amplitude $\bar{C}^\pm_{\lhat mN}= \bar{C}^\pm_{\lhat mN}(\bar{q}_0)$
is a function of $\bar{q}_0$, which captures the effect on the fluxes of the phase of the resonant orbit.

In a numerical calculation the fluxes tend to converge exponentially with increasing numbers of modes and only
require calculation of the matching coefficients $\bar{C}^\pm_{\lhat mN}$, not the full SSF.  The effect is that the
fluxes can usually be computed to high accuracy.  In Table \ref{tab:fluxes} we report numerical values for the
total fluxes (at infinity and the horizon) for two different phase parameters, $q_{\th 0} = \b_\th \bar{q}_0 = 0$
and $-\pi/2$, and for each of the models outlined in Table \ref{tab:orbits}.  Consistent with calculations of
gravitational fluxes \cite{Misn72}, most of the horizon fluxes are negative due to superradiant scattering (each
model has primary spin of $a/M = 0.9$).  As expected, orbits with smaller pericentric distances $r_\text{min}$ tend
to produce larger fluxes, while eccentricity has a smaller effect.

In Table \ref{tab:fluxes} we also list the computed average over $\bar{q}_0$ of the resonant-orbit fluxes
\begin{equation} \label{eqn:fluxAvg}
	\langle\langle \dot{X}\rangle\rangle_{\bar{q}_0}
	\equiv \frac{1}{2\pi}\int_0^{2\pi} \langle \dot{X} \rangle\;
	d\bar{q}_0,
\end{equation}
where $X = E, L_z$.  This average over $\bar{q}_0$ (i.e., a double averaging) gives the flux that would be seen in
a system with nearly the same orbital parameters but infinitesimally off resonance so that its motion ergodically
fills the torus.  It is equivalent to computing the fluxes with the normal incoherent sum of terms with
$\vert \hat{C}^\pm_{\lhat mkn} \vert^2$ over all $n$ and $k$.  With \eqref{eqn:fluxAvg} giving a background average,
we can define the residual variation (enhancement or diminishment) in the fluxes that arise on resonance,
\begin{equation} \label{eqn:resVar}
	\langle \d \dot{X} \rangle \equiv
	\langle \dot{X} \rangle - \langle\langle
	\dot{X}\rangle\rangle_{\bar{q}_0}.
\end{equation}
We plot $\langle \d\dot{E}\rangle^\mathrm{tot}$ and
$\langle \d\dot{L}_z\rangle^\mathrm{tot}$ for the $e02.12$ (solid black lines), $e02.23$ (dashed red lines), and
$e05.23$ (dot dashed lines) orbits in Fig.~\ref{fig:fluxRes}.  By comparing these figures to the values in Table
\ref{tab:fluxes}, we see that the residual variations are relatively small compared to the magnitudes of the total
fluxes, but they are still greater than the $\sim 10^{-8}$ fractional error in our numerical calculations.

\begin{figure*}[t]
	\centering
	\includegraphics[width=0.9\textwidth]{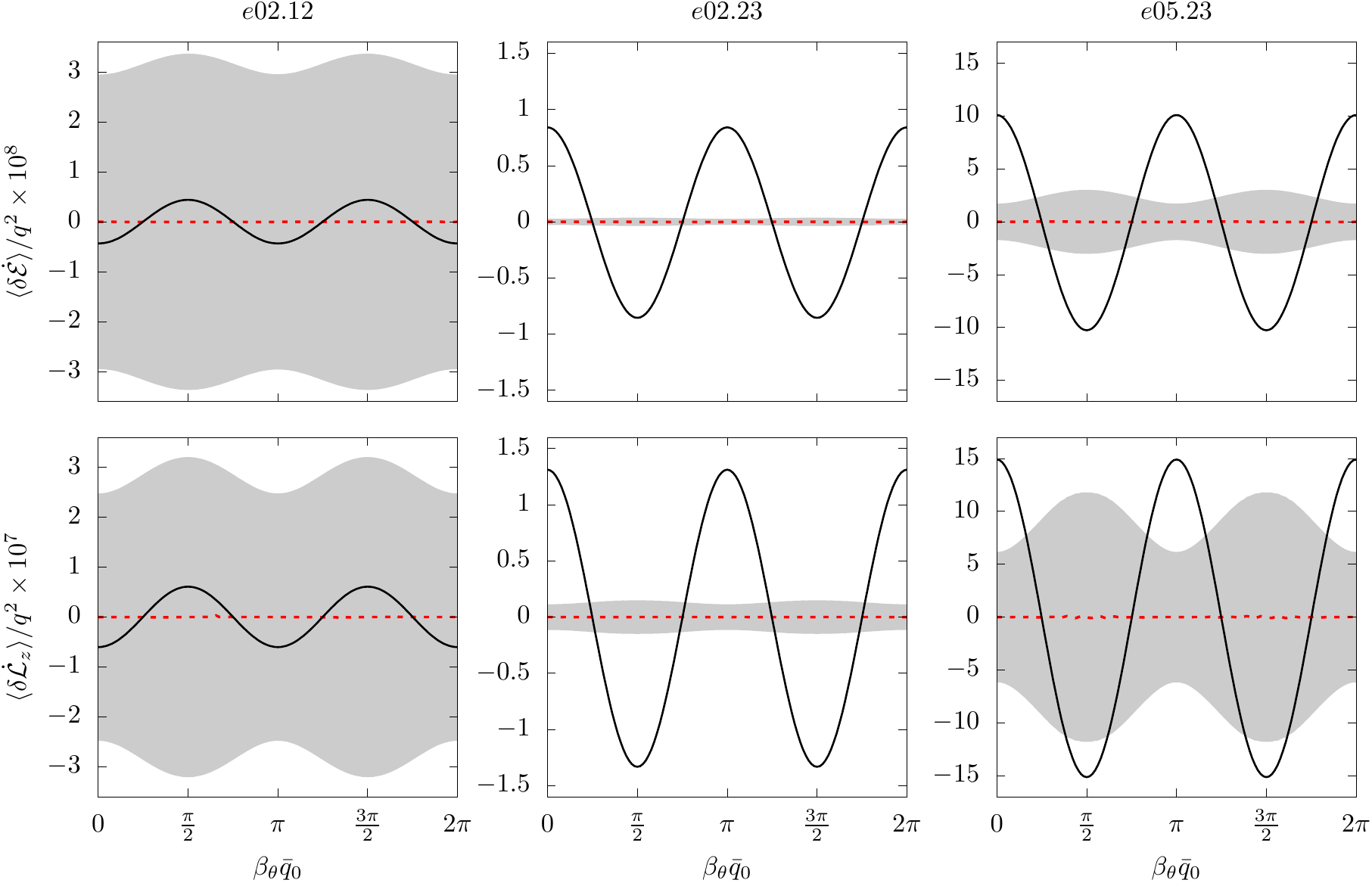}
	\caption{Separate contributions from the dissipative and conservative components of the self-force to the
	residual variations $\langle \d \dot{\mathcal{E}} \rangle$ in energy (top row) and
	$\langle \d \dot{\mathcal{L}}_z \rangle$ in angular momentum (bottom row) for the $e02.12$ (left),
	$e02.23$ (middle), and $e05.23$ (right) orbits plotted as functions of the resonant-orbit phase.
	Contributions from the dissipative self-force are given by the solid (black) curves, while contributions from
	the conservative self-force are depicted by the dashed (red) curves.  The (grey) shaded region represents the
	formal uncertainty in our calculation of the conservative SSF, $\langle \d \dot{\mathcal{Q}} \rangle^\mathrm{cons} \pm \s_\mathrm{cons}$, due to fitting for
	higher-order regularization parameters.  The calculation of $\s_\mathrm{cons}$ is discussed below
	(in Sec.~\ref{sec:EnAndLz}).  As expected, the residual variations in energy and angular momentum arising from
	the conservative SSF (dashed/red curves) are consistent with zero.
	}
	\label{fig:workTorque}
\end{figure*}

In the case of the 2:3 resonances plotted as dashed (red) and dot-dashed (blue) curves in Fig.~\ref{fig:fluxRes},
the energy and angular momentum fluxes are minimized when the motion possesses simultaneous turning points in $r$
and $\th$ (i.e., $\b_{\th}\bar{q}_{0}=0,\pi$).  While not plotted here, the other 2:3 resonance ($e05.23$) shares
this behavior.  On the other hand, for the 1:2 resonance in Fig.~\ref{fig:fluxRes}, the energy and angular momentum
fluxes are maximized when the motion possesses simultaneous turning points, a feature which is shared by the other
1:2- and 1:3-resonant orbits.  Evidence of this behavior can also be found in Table \ref{tab:fluxes}.  Following
the work of \cite{FlanHughRuan14}, we also report in Appendix \ref{app:fracVarFlux} the total fractional variations
$\D \dot{E}$ and $\D \dot{L}_z$ as defined by (4.5) in \cite{FlanHughRuan14}.

Additionally, we can make use of the flux-balance laws to test the accuracy of our SSF data.  The fractional errors
between the fluxes and the work $\mathcal{W}$ and torque $\mathcal{T}$, computed via \eqref{eqn:fluxEBalance}
and \eqref{eqn:fluxLzBalance}, are given in the last two columns of Table \ref{tab:fluxes}.  We find good agreement,
with fractional errors of $\sim 10^{-11}$-$10^{-5}$.  Recalling the numerical convergence of the SSF displayed in
Fig.~\ref{fig:convergenceFph}, these fractional errors are in line with the predicted numerical accuracy of our
SSF data.

The good agreement between our flux and SSF results is a further way of seeing that the conservative component of
the SSF does not contribute to $\langle \dot{\mathcal{E}} \rangle$ and $\langle \dot{\mathcal{L}}_z \rangle$.  The
fluxes are \emph{purely dissipative} quantities, and for the flux-balance laws to hold, only the dissipative component
of the SSF can contribute to the averages $\langle \dot{\mathcal{E}} \rangle$ and
$\langle \dot{\mathcal{L}}_z \rangle$,
even during resonances \cite{IsoyETC19}.

To verify this, we calculate separately the contributions of the
dissipative and conservative SSF to the residual variations $\langle \d \dot{\mathcal{E}} \rangle$ and
$\langle \d \dot{\mathcal{L}}_z \rangle$ by replacing $\bar{F}_\a^\mathrm{res}$ with $\bar{F}_\a^\mathrm{diss}$
and $\bar{F}_\a^\mathrm{cons}$ in \eqref{eqn:work} and \eqref{eqn:torque}.
In Fig.~\ref{fig:workTorque} we plot the residual variations for the $e02.12$ (left), $e02.23$ (middle), and
$e05.23$ (right) orbits.  The solid (black) curves correspond to the dissipative contributions, which share the same
varying behavior as seen in Fig.~\ref{fig:fluxRes} [note the opposite sign in \eqref{eqn:fluxEBalance} and
\eqref{eqn:fluxLzBalance}].  The dashed (red) curves correspond to the conservative contributions, and the filled
(grey) regions plot the estimated uncertainty of the conservative contributions due to truncation of the
regularization procedure, which affects the conservative SSF.

The uncertainty in the conservative contributions originates in our mode-sum regularization of the SSF.  As summarized
in Sec.~\ref{sec:ssfReg} [and discussed with additional detail in \cite{NasiOsbuEvan19}, Sec.~IV A, the paragraph
following Eq.~(4.11)], we regularize our SSF data via mode-sum regularization, but must numerically fit for
higher-order regularization parameters in order to improve the numerical convergence of our $l$-mode sum.  Without
extrapolating these higher-order terms, our regularized results would be dominated by truncation errors of
$O(l_\mathrm{max}^{-1})$.  Our extrapolation procedure typically enhances truncation error scaling to
$O(l_\mathrm{max}^{-7})$ but it also introduces systematic uncertainties, since our extrapolated results depend on
how many terms we include in our numerical fits and which multipole SSF modes we use to produce those fits.  These
systematic uncertainties tend to be much larger than what the improved truncation error scaling would naively suggest, and
thus become the dominant form of error in our numerical conservative SSF results.\footnote{Recall that only the
conservative component of the SSF needs to be regularized.}  Therefore, any quantity that depends on the conservative
self-force will also have an estimated uncertainty associated with the fitting.  We estimate the uncertainty of the
newly calculated quantity by propagating uncertainties in additive terms,
i.e.,
\begin{equation}
	\sigma_f^2 = \left\vert \frac{\partial f}{\partial x_{0}} \right\vert^2
	\sigma_{x_0}^2 + \left\vert \frac{\partial f}{\partial x_1} \right\vert^2
	\sigma_{x_1}^2 + \cdots
	+ \left\vert \frac{\partial f}{\partial x_n} \right\vert^2
	\sigma_{x_n}^2 ,
\end{equation}
where $\sigma_f$ is the propagated uncertainty of a quantity $f$ due to its dependence on $n$ (assumed independent)
parameters $(x_0, x_1, \dots, x_n)$ with (assumed uncorrelated) errors
$(\sigma_{x_0}, \sigma_{x_1}, \dots, \sigma_{x_n})$.  For example,
$\langle \d \dot{\mathcal{E}} \rangle^\mathrm{cons}$ and its uncertainty are explicitly computed via the sums
\begin{align}
	\langle \d \dot{\mathcal{E}} \rangle^\mathrm{cons}(\bar{q}_0) &= \frac{q^2}{N}
	\sum_{n = 0}^{N-1} \bar{\Sigma}_p\left(\frac{2\pi i n}{N};\bar{q}_0\right)
	\bar{F}^\mathrm{cons}_t\left(\frac{2\pi i n}{N}; \bar{q}_0\right), \notag
	\\
	\sigma_\mathrm{cons}^2(\bar{q}_0) &= \frac{q^4}{N^2} \sum_{n=0}^{N-1}
	\bar{\Sigma}_p^2\left(\frac{2\pi i n}{N};\bar{q}_0\right)
	\sigma_{t}^2\left(\frac{2\pi i n}{N};\bar{q}_0\right), \label{eqn:propErrorEn}
\end{align}
where $\sigma_{t}(\bar{q};\bar{q}_0)$ is the estimated uncertainty of $\bar{F}_t^\mathrm{cons}(\bar{q};\bar{q}_0)$
from our fitting procedure.  The first line of \eqref{eqn:propErrorEn} is obtained by replacing the integrand of
\eqref{eqn:work} with its discrete Fourier transform.\footnote{Recall that $\langle\langle \dot{\mathcal{E}} \rangle\rangle_{\bar{q}_0}$
vanishes exactly due to the symmetries of Kerr geodesics, and, thus,
$\langle\d \dot{\mathcal{E}}\rangle^\mathrm{cons} = \langle \dot{\mathcal{E}}\rangle^\mathrm{cons}$.}

No uncertainty estimates for the dissipative
contributions are included, as these are orders of magnitude smaller.  While the conservative
part leaves behind a nonzero numerical result, these variations fall well below our estimated uncertainty and are
thus consistent with zero, as expected.  The estimated uncertainty is much larger for the $e02.12$ model due to the
slower convergence of the SSF for orbits that lie deeper in the strong field (see Fig.~\ref{fig:convergenceFph}).
In that model, our formal uncertainty far exceeds not only the conservative contributions but even the dissipative
variations, which may point to the formal uncertainties being too conservative.

\subsection{Carter constant and the integrability conjecture}

\begin{figure*}[t]
	\centering
	\includegraphics[width=\textwidth]{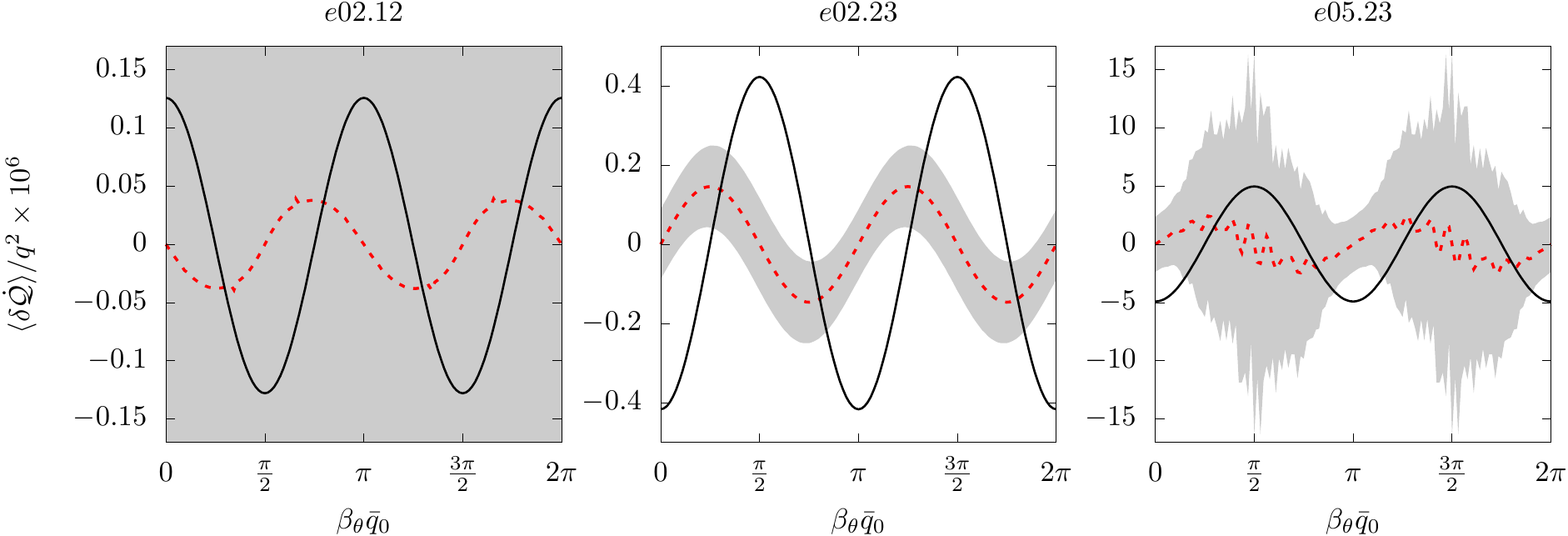}
	\caption{Contributions from the dissipative and conservative components of the self-force to
	$\langle \d \dot{\mathcal{Q}}\rangle$ for the $e02.12$ (left), $e02.23$ (middle), and $e05.23$ (right)
	orbits.  Contributions from the dissipative self-force are given by the solid (black) curves, while
	contributions from the conservative self-force are depicted by the dashed (red) curves.  The (grey) shaded
	region represents the formal uncertainty in our calculation of the conservative SSF, $\langle \d \dot{\mathcal{Q}} \rangle^\mathrm{cons} \pm \s_\mathrm{cons}$, due
	to fitting for higher-order regularization parameters.  The calculation of $\s_\mathrm{cons}$ is discussed in detail
	in Sec.~\ref{sec:EnAndLz}.
	}
	\label{fig:dQdot}
\end{figure*}

Unlike $\langle \dot{\mathcal{E}}\rangle$ and $\langle \dot{\mathcal{L}}_z\rangle$, $\langle \dot{\mathcal{Q}}\rangle$
is not associated with a radiation flux.  Instead, we must directly calculate the orbit-averaged rate of change of
the Carter constant from the self-force,
\begin{align} \label{eqn:QdotAvgMino}
	\langle \dot{\mathcal{Q}} \rangle
	= \frac{1}{\G} \left\langle \Sig \frac{d\mathcal{Q}}{d\tau}
	 \right\rangle_\la .
\end{align}
Here $\langle \mathcal{X}\rangle_\la$ refers to an average over $\mathcal{X}(\la)$ with respect to Mino time $\la$,
and the proper time derivative of $\mathcal{Q}$ is given by
\begin{align}
	\frac{\mu}{2}\frac{d\mathcal{Q}}{d\tau} &= \label{eqn:QTauDotTh}
	q^2 (\csc^2\th \mathcal{L}_z-a \mathcal{E})(F_\vp+a\sin^2\th F_t)
	\\ \notag
	& \qquad + q^2 u_\th F_\th - q^2(\mathcal{L}_z-a\mathcal{E})
	\left(F_\vp+aF_t\right)
	\\ \notag
	&\qquad \qquad \qquad \qquad
	- q^2(\mathcal{Q} -  a^2\cos^2\th) u^\a F_\a,
	\\
	&= \label{eqn:QTauDotR}
	q^2 \D^{-1}(a \mathcal{L}_z-\varpi^2 \mathcal{E})
	\left(a F_\vp+\varpi^2 F_t\right)
	\\ \notag
	&\qquad -  q^2 \D u_r F_r - q^2(\mathcal{L}_z-a\mathcal{E})
	\left(F_\vp+aF_t \right)
	\\ \notag
	&\qquad \qquad \qquad \qquad \qquad \qquad
	- q^2(\mathcal{Q}+  r^2 ) u^\a F_\a.
\end{align}
(See Appendix \ref{app:Qdot}.)

For nonresonant orbits, the conservative contributions to $\langle \dot{\mathcal{Q}} \rangle$ vanish due to
symmetries of the motion.  This can be seen by reexpressing \eqref{eqn:QdotAvgMino} as a two-dimensional integral
over $q_r$ and $q_\theta$ \cite{DrasHugh04,GrosLeviPere13},
\begin{equation} \label{eqn:QdotAverageNonRes}
	\langle \dot{\mathcal{Q}} \rangle = \frac{1}{\G} \int_0^{2\pi} \frac{dq_r}{2\pi}
	\int_0^{2\pi} \frac{dq_\th}{2\pi} \left(\Sig \frac{d\mathcal{Q}}{d\tau}
	\right).
\end{equation}
Recall from \eqref{eqn:consSSF} and \eqref{eqn:ssfRet2Adv} that the two components $F^\mathrm{cons}_{t,\vp}$ are
antisymmetric on the $q_r$-$q_{\th}$ torus, while the other two components $F^\mathrm{cons}_{r,\theta}$ are
symmetric on the torus, so that
\begin{align}
	F^\mathrm{cons}_{t,\vp}(2\pi-q_r, 2\pi-q_\theta)
&= - F^\mathrm{cons}_{t,\vp}(q_r, q_\theta),
\\
	F^\mathrm{cons}_{r,\theta}(2\pi-q_r, 2\pi-q_\theta)
&= +F^\mathrm{cons}_{r,\theta}(q_r, q_\theta).
\end{align}
Hence, if we only make use of the conservative components of the self-force in \eqref{eqn:QTauDotTh} or
\eqref{eqn:QTauDotR}, then $dQ^\mathrm{cons}/d\tau$ is also antisymmetric.  Because $\Sig$ is symmetric with respect
to the angle variables, the conservative contributions must vanish when integrated over the entire torus in
\eqref{eqn:QdotAverageNonRes}.  By disregarding these conservative perturbations and relating
$\langle \dot{\mathcal{Q}}\rangle$ to the purely radiative (dissipative) piece of the perturbing field
\cite{Mino03,SagoETC06}, \eqref{eqn:QdotAvgMino} reduces to a weighted mode sum over the field's asymptotic
amplitudes \cite{DrasFlanHugh05,SagoETC06,FlanHughRuan14}, akin to
\eqref{eqn:fluxEBalance}-\eqref{eqn:fluxLzBalanceLast}.

For $r\theta$ resonances, the Mino time average reduces to the single integral over $\bar{q}$ in \eqref{eqn:qdotAvg},
\begin{equation}
\label{eqn:QdotAverageRes}
	\langle \dot{\mathcal{Q}} \rangle = \frac{1}{\G} \int_0^{2\pi}
	\frac{d\bar{q}}{2\pi} \left(\bar{\Sig} \frac{d\mathcal{Q}}{d\tau}
	\right) .
\end{equation}
While the integrand in the above expression is still antisymmetric with respect to both $q_r$ and $q_\theta$, it is
not, for a general choice of the initial resonant phase $\bar{q}_0$, antisymmetric with respect to just $\bar{q}$
(though there may be special values of $\bar{q}_0$ where it is).  When integrated over a single closed track on the
torus, \eqref{eqn:QdotAverageRes} is not guaranteed to vanish, in contrast with the nonresonant case.

However, Flanagan and Hinderer \cite{FlanHind12} conjecture that dynamics driven by the conservative piece of the
self-force are always integrable in Kerr spacetime,\footnote{Conservative perturbations are integrable in
Schwarzschild spacetime, but their integrability has not been fully demonstrated for Kerr
\cite{VineFlan15,FujiETC17}.} and cannot therefore drive secular evolution of the perturbed system through a
resonance.  If this integrability conjecture for conservative perturbations holds true, then $r\th$-resonant dynamics
will be driven purely by the dissipative self-force (at adiabatic order), and $\langle \dot{\mathcal{E}}\rangle$,
$\langle \dot{\mathcal{L}}_z\rangle$, and $\langle \dot{\mathcal{Q}}\rangle$ can be computed via efficient mode-sum
expressions [e.g., \eqref{eqn:fluxEBalance}-\eqref{eqn:fluxLzBalanceLast}] during resonant (and nonresonant) motion,
as demonstrated in \cite{FlanHughRuan14}.\footnote{Note that the authors of \cite{FlanHughRuan14} made no claim about
the validity of the integrability conjecture, but instead adopted it as a matter of practicality, since calculations of
the conservative GSF were unavailable at that time (Hughes, private communications).}

On the other hand, Isoyama \textit{et al.}~\cite{IsoyETC13,IsoyETC19} also derived mode-sum expressions for
$\langle \dot{\mathcal{E}}\rangle$, $\langle \dot{\mathcal{L}}_z\rangle$, and $\langle \dot{\mathcal{Q}}\rangle$ during
resonances using a Hamiltonian formulation, but they found that their expression for $\langle \dot{\mathcal{Q}} \rangle$
depends on the conservative (or what they call the symmetric) component of the perturbed Hamiltonian (for the scalar
case, see Eqs.~(49)-(51) in \cite{IsoyETC13} and for the gravitational case, see Eqs.~(63) and (75) in
\cite{IsoyETC19}).  Unless this term vanishes upon averaging due to further symmetries, the integrability conjecture
must break down during resonances, and conservative perturbations will also drive the adiabatic evolution of $Q$.

This issue can in principle be tested using numerical modeling.  To date numerical calculations of
$\langle \dot{\mathcal{Q}} \rangle$ for $r\th$-resonant orbits \cite{FlanHughRuan14,RuanHugh14,BerrETC16} have not
incorporated the full first-order self-force.\footnote{Note that the results of Isoyama \textit{et al.}~\cite{IsoyETC13,IsoyETC19}
were purely analytical.  No one has made use of the formalisms outlined in \cite{IsoyETC13,IsoyETC19} to numerically
evaluate $\langle\dot{\mathcal{Q}}\rangle$, and our methods differ enough from the Hamiltonian formulation and Green's
function methods of \cite{IsoyETC13,IsoyETC19} that we cannot easily compare our numerical results to these works.}
Thus there is no numerical evidence to support or negate the integrability conjecture.  The scalar self-force model can
potentially provide some insight.  We test
the conjecture by measuring the relative contributions of the conservative and dissipative components of the SSF to
$\langle \dot{\mathcal{Q}} \rangle$.  As we did in Sec.~\ref{sec:EnAndLz}, we replace $\bar{F}^\mathrm{res}_\a$ in
\eqref{eqn:qdotAvg} with a breakdown in terms of $\bar{F}^\mathrm{cons}_\a$ and $\bar{F}^\mathrm{diss}_\a$ to
calculate separately the conservative and dissipative contributions to $\langle \dot{\mathcal{Q}}\rangle$.  To compare
these quantities, we then calculate their residual variations $\langle \d \dot{\mathcal{Q}}\rangle^\mathrm{cons}$
and $\langle \d \dot{\mathcal{Q}}\rangle^\mathrm{diss}$ as functions of the resonant-orbit phase using \eqref{eqn:resVar}.

In Fig.~\eqref{fig:dQdot}, we compare the variations $\langle \d \dot{\mathcal{Q}}\rangle^\mathrm{diss}$ (solid
black curves) and $\langle \d \dot{\mathcal{Q}}\rangle^\mathrm{cons}$ (dashed red curves) for the $e02.12$ (left),
$e02.23$ (middle), and $e05.23$ (right) orbits, along with our formal uncertainty estimate $\s_\mathrm{cons}$ due to
fitting for high-order regularization parameters in the conservative component of the self-force.  Across all three
orbits, the numerical calculation gives a smoothly varying conservative contribution to
$\langle \d \dot{\mathcal{Q}}\rangle$.  The amplitudes of the conservative contributions are slightly smaller but on
the same order as the dissipative variations.  \emph{The smooth variations of the conservative contributions and their
comparable magnitudes to the dissipative variations strongly suggest that $\langle \dot{\mathcal{Q}}\rangle$ does depend
on conservative scalar perturbations during a resonance.}  However, in each case the conservative contributions from our
SSF data fall below or nearly within the formal uncertainty estimates.  Furthermore, for all three orbits, the estimated
uncertainty in our calculations of $\langle \d \dot{\mathcal{Q}}\rangle^\mathrm{cons}$ are significantly larger than
the estimated uncertainties in $\langle \d \dot{\mathcal{E}}\rangle^\mathrm{cons}$ and
$\langle \d \dot{\mathcal{L}}_z\rangle^\mathrm{cons}$, as seen from comparing Figs.~\ref{fig:workTorque} and
\ref{fig:dQdot}.  In fact, for the $e02.12$ and $e05.23$ orbits, our estimated uncertainty is typically large enough
that even if the conservative contribution dominated the variations in the dissipative contributions, they could
still be consistent with zero.

In the case of the $e02.12$ model, the truncation in the regularization of the conservative part of the SSF suggests
a formal uncertainty that is an order of magnitude greater than the numerically determined values of both the
conservative and dissipative contributions to $\langle \d \dot{\mathcal{Q}}\rangle$, and the grey region engulfs
the entire left panel of Fig.~\ref{fig:dQdot}.  The size of the uncertainty in this model is similar to what was seen
in $\langle \d \dot{\mathcal{E}}\rangle$ and $\langle \d \dot{\mathcal{L}}_z \rangle$ in Fig.~\ref{fig:workTorque}.
In the $e05.23$ orbit, the calculated conservative contribution not only falls consistently below the formal
uncertainty, but contains high-frequency oscillations that are indicative of numerical noise.  This noise appears
also in the formal uncertainty estimate itself.  In the $e02.23$ model,
$\langle \d \dot{\mathcal{Q}}\rangle^\mathrm{cons}$ still primarily falls within the formal uncertainty, but is
closest, in this case, to being a significant nonzero result.  The numerical calculations, in this model at least,
are close to providing evidence of the integrability conjecture's failure, but a reduction in the
regularization errors by an order of magnitude or two would be required to be sure.

When compared to calculating $\langle \d \dot{\mathcal{E}}\rangle^\mathrm{cons}$ and
$\langle \d \dot{\mathcal{L}}_z\rangle^\mathrm{cons}$, the estimated uncertainty proves to be much higher when
analyzing the evolution of the Carter constant.  In part, this is because
$\langle \d \dot{\mathcal{Q}}\rangle^\mathrm{cons}$ depends on $F_\th$, which tends to be the most difficult
component of the SSF to accurately extrapolate with our numerical regularization procedure.  The $\th$ component
couples to even higher spheroidal modes than the other SSF components as a result of the window function described
in Sec.~\ref{sec:ssf}.  If we truncate all mode calculations at a particular $l_\mathrm{max}$, then we can only
calculate the multipoles of $F_\th$ up to $l_\mathrm{max}-3$, as seen from \eqref{eqn:ssfRetMino} and
\eqref{eqn:delTheta}.  Since the higher $l$ modes are beneficial for extracting the higher-order regularization
parameters, missing this higher-mode information for $F_\th$ hampers our ability to fit for its (regularized)
conservative component.  We still are able to calculate $F_\th$ to 3 or 4 significant digits, which in the
absence of resonances is accurate enough to provide its contribution to EMRI evolutions that are phase accurate to
less than a radian.  As this analysis indicates, additional accuracy will be needed to quantify with certainty
possible contributions from the conservative sector to the secular evolution of the Carter constant.  In principle,
we could calculate additional modes to improve the accuracy of $F_\th$.  In practice, this is difficult because
higher $l$ modes are harder to calculate accurately due to rapid oscillatory behavior of the integrands in the
source integrations.  This behavior is mirrored in the gravitational case \cite{Vand16}.

The plotted uncertainty regions in Fig.~\eqref{fig:dQdot} provide an estimate of how well we have fit for the
higher-order regularization parameters and regularized our self-force data.  Any nonvanishing, and potentially
smooth, variations within these uncertainty bounds could be a result of residual contributions of the singular field
that were not removed during regularization.  This issue is unique to averages over $r\th$-resonant orbit.  In the
nonresonant case, the singular field---much like the conservative contribution---vanishes when averaged over the
entire two-torus.  Even if we do not regularize the conservative component of the SSF,
$\langle \d \dot{\mathcal{Q}}\rangle^\mathrm{cons}$ will still exactly vanish for nonresonant orbits.  On the other
hand, averages over the singular contributions, or any antisymmetric function on the two-torus, are not guaranteed
to vanish for an arbitrary $r\th$ resonance.

We conclude that our conservative SSF results are not yet accurate enough to show a definitive conflict with the
integrability conjecture.  That being said, our uncertainty regions may overestimate residual contributions of the
singular field that were not properly regularized.  If this is the case, then the nondissipative variations that we
see in Fig.~\ref{fig:dQdot} may very well be physical.  Further tests of the integrability conjecture are necessary
but will require improved regularization of the self-force.  This might come from either more extensive numerical
calculations and regularization parameter fitting or from the added input of analytically determined higher-order
regularization parameters.

\section{Summary}
\label{sec:conclusion}

We considered point scalar charges following $r\th$-resonant geodesics in Kerr spacetime and calculated, for the
first time, the resulting strong-field SSF experienced by these charges. This work serves as a
first step in understanding the still unquantified behavior of the gravitational self-force during transient orbital
$r\th$ resonances.  To calculate the SSF we used a \textsc{Mathematica} code previously developed in
\cite{NasiOsbuEvan19}.  In constructing our SSF data, we derived a simple shifting relation, \eqref{eqn:ssfqq0}, that
allows us to calculate the self-force during $r\th$ resonances using self-force data that assumes the geodesic
sources are nonresonant.  This mapping provides an efficient method for analyzing the self-force as a function of
the initial phase at which a system enters resonance.

When calculating the SSF, we focused on six different $r\th$-resonant orbits: each scalar charge followed a 1:3, 1:2,
or 2:3 $r\th$-resonant geodesic and each orbit either had an eccentricity of $0.2$ or $0.5$.  The full set of source
parameters can be found in Table \ref{tab:orbits}.  In Figs.~\ref{fig:ssfRVar}-\ref{fig:ssfPhVar}, we demonstrated
how varying the initial phase of $r\th$-resonant orbits impacts the evolution of the self-force.  We then projected
our SSF data onto invariant two-tori in Figs.~\ref{fig:ssfTTorus}-\ref{fig:ssfPhTorus} to display the dependence of
the self-force on the radial and polar motion of the source, regardless of initial conditions and phases.  We
validated our SSF data by analyzing the convergence properties of our regularized self-force multipoles, as shown
in Fig.~\ref{fig:convergenceFph}, and by comparing the radiation fluxes to the rate of work and torque done on each
scalar-charge source by the SSF via flux-balance laws, which are reported in Table \ref{tab:fluxes}.

With these novel self-force calculations we also analyzed the impact of the conservative scalar self-force on the
orbit-averaged evolution of the orbital energy $\langle \dot{\mathcal{E}} \rangle$, $z$ component of the angular
momentum $\langle \dot{\mathcal{L}}_z \rangle$, and Carter constant $\langle \dot{\mathcal{Q}} \rangle$.  As expected
from flux-balance arguments, the contributions to $\langle \dot{\mathcal{E}} \rangle$ and
$\langle \dot{\mathcal{L}}_z \rangle$ from the conservative SSF are negligible and consistent with zero, as shown
in Fig.~\ref{fig:workTorque}.  On the other hand, our conservative SSF data substantially contribute to
$\langle \dot{\mathcal{Q}} \rangle$, as displayed in Fig.~\ref{fig:dQdot}, though these contributions are on the
order of, or much less than, our estimated uncertainty.  This uncertainty is a result of the numerical regularization
of the conservative SSF.  Because these contributions fall within these uncertainty bounds, we cannot distinguish
whether these nondissipative contributions are due to the residual singular field or the regularized conservative
self-force.  If these contributions are in fact from the conservative SSF, then this would indicate that the
integrability conjecture proposed by Flanagan and Hinderer \cite{FlanHind12} breaks down during $r\th$ resonances,
and conservative contributions will need to be considered as predicted by Isoyama \textit{et al.} \cite{IsoyETC19}.  The
presence of these conservative contributions would then introduce a new numerical challenge for adiabatic
calculations: to account for the evolution through resonances, adiabatic codes would need to incorporate
regularization procedures to accurately quantify $\langle \dot{\mathcal{Q}} \rangle$.  Furthermore, these regularized
contributions would need to be known to high levels of precision, at least to $O(\eps^{-1/2})$, which we have not
been able to achieve with our current numerical implementations.

In future work, we will determine whether or not these nondissipative contributions to
$\langle \dot{\mathcal{Q}} \rangle$ in the scalar case are physical, or merely systematic errors, by implementing
alternative, improved regularization schemes and by deriving analytic expressions for the higher-order regularization
parameters.  If the conservative SSF contributions that we have observed are indeed physical, then we hypothesize that
during $r\theta$ resonances the conservative GSF will also contribute to $\langle \dot{\mathcal{Q}} \rangle$, as first suggested in \cite{IsoyETC19}. Therefore, we are also constructing a code to calculate the GSF experienced by EMRI systems as they encounter $r\th$ resonances to test this hypothesis.

\acknowledgements

This work was supported by NSF Grant Nos.~PHY-1806447 and PHY-2110335 to the University of North Carolina--Chapel Hill,
and by the North Carolina Space Grant Graduate Research Fellowship.  Z.N.~acknowledges additional support from NSF Grant
No.~DMS-1439786 while in residence as a postdoctoral fellow for the Institute for Computational and Experimental
Research in Mathematics in Providence, Rhode Island, during the Advances in Computational Relativity semester program.
Z.N.~also acknowledges support by appointment to the NASA Postdoctoral Program at the Goddard Space Flight Center,
administered by Universities Space Research Association under contract with NASA.

\appendix
\section{Normalization coefficients}
\label{app:normC}

We review the definitions of the normalization coefficients $C^\pm_{\lhat mkn}$ and $\bar{C}^\pm_{\lhat mN}$ and
their dependence on the initial conditions of a geodesic source.

\subsection{Nonresonant sources}
We first look at nonresonant geodesic orbits and make use of the angle variables introduced
in Sec.~\ref{sec:angleVar}.  The $C^\pm_{\lhat mkn}(q_{(\a)0})$ are defined by the integrals
\begin{equation}
\label{eqn:normC}
	C^\pm_{\lhat mkn}(q_{(\a)0}) = \int_{r_\text{min}}^{r_\text{max}}
	\frac{\varpi^2\ti{X}^\mp_{\lhat mkn}(r)
	Z_{\lhat mkn}(r;q_{(\a)0})}{W_{\lhat mkn}\D(r)} dr,
\end{equation}
where $Z_{\lhat mkn}$ is the radial decomposition of the source in the frequency domain
\begin{equation}
		\rho = -\frac{\varpi^3}{4\pi\Sig\D}\sum_{\lhat mkn}
		Z_{\lhat mkn}(r;q_{(\a)0})
		S_{\lhat mkn}(\th)e^{im\vp}e^{-i\o_{mkn}t},
\end{equation}
and $W_{\lhat mkn}$ is the Wronskian
\begin{equation} \label{eqn:Wlmkn}
	W_{\lhat mkn} = \frac{\D}{\varpi^2}
	\left( \ti{X}^-_{\lhat mkn}\frac{d\ti{X}^+_{\lhat mkn}}{dr}
	-\ti{X}^+_{\lhat mkn}\frac{d\ti{X}^-_{\lhat mkn}}{dr} \right).
\end{equation}
We have now explicitly included the dependence on initial conditions, which we
represent with the four-tuple $q_{(\a)0} \equiv (t_0,q_{r0},q_{\th0},\vp_0)$.

For the scalar point-particle source described by \eqref{eqn:pointSource}, $Z_{\lhat mkn}$ takes the form
\begin{multline}
Z_{\lhat mkn}(r;q_{(\a)0}) = \frac{1}{4\pi^2}\int_0^{2\pi}dq_r
	\int_0^{2\pi} dq_\th\, e^{i(kq_\th+nq_r)}
	\\
	\times B_{mkn}(q_r,q_\th;q_{(\a)0})S_{\lhat mkn}(\th_p)\d(r-r_p),
\end{multline}
where
\begin{multline}
	B_{mkn}(q_r,q_\th;q_{(\a)0})\equiv -\frac{4\pi q}{\G}\frac{\Sig_p \D_p}{\varpi_p^3}
	\\
	\times e^{i\o_{mkn}(\D t + t_0)
	-im(\D \vp + \vp_0)} ,
\end{multline}
$\Sig_p = r_p^2+a^2\cos^2\th_p$, $\varpi_p^2=r_p^2+a^2$,
$\D_p = \varpi_p^2-2M r_p$,
and the geodesic functions $\D t$, $r_p$, $\th_p$, and $\D \vp$ are understood to be functions of
$q_r$, $q_\th$, $q_{r0}$, and $q_{\th 0}$,
\begin{align}
	\D t &= \D \hat{t}^{(r)}(q_r+q_{r0})- \D \hat{t}^{(r)}(q_{r0})
	\\ \notag
	&\qquad + \D \hat{t}^{(\th)}(q_\th+q_{\th 0})
	- \D \hat{t}^{(\th)}(q_{\th 0}),
	\\
	r_p &= \hat{r}_p(q_r+q_{r0}),
	\\
	\th_p &= \hat{\th}_p(q_\th+q_{\th0}).
	\\
	\D \vp &= \D \hat{\vp}^{(r)}(q_r+q_{r0})- \D \hat{\vp}^{(r)}(q_{r0})
	\\ \notag
	&\qquad + \D \hat{\vp}^{(\th)}(q_\th+q_{\th 0})
	- \D \hat{\vp}^{(\th)}(q_{\th 0}).
\end{align}
\eqref{eqn:normC} then simplifies to
\begin{multline} \label{eqn:Clmkn2D}
	C^\pm_{\lhat mkn}(q_{(\a)0}) = \int_0^{2\pi}\frac{dq_r}{2\pi}
	\int_0^{2\pi} \frac{dq_\th}{2\pi}\, e^{i(kq_\th+nq_r)}
	\\
	\times
	D^\pm_{\lhat mkn}(q_r,q_\th;q_{(\a)0}),
\end{multline}
where
\begin{multline} \label{eqn:Dlmkn}
	D^\pm_{\lhat mkn}(q_r,q_\th;q_{(\a)0})\equiv -\frac{4\pi q}{\G}
	\ti{X}^\mp_{\lhat mkn}(r_p)S_{\lhat mkn}(\th_p)
	\\
	\times \frac{\Sig_p}{W_{\lhat mkn}\varpi_p}
	 e^{i\o_{mkn}(\D t + t_0)-im(\D \vp + \vp_0)}.
\end{multline}
\eqref{eqn:Clmkn2D}
separates into four one-dimensional integrals $I^{(i)}_{\lhat mkn}$
\begin{multline}
	C^\pm_{\lhat mkn}(q_{(\a)0}) = I^{(1)\pm}_{\lhat mkn}(t_0,q_{r0})
	I^{(2)}_{\lhat mkn}(q_{\th 0},\vp_0)
	\\
	+ I^{(3)\pm}_{\lhat mkn}(t_0,q_{r0})
	I^{(4)}_{\lhat mkn}(q_{\th 0},\vp_0),
\end{multline}
where the integrals $I^{(i)}_{\lhat mkn}$ are given in Sec.~III C of \cite{NasiOsbuEvan19}, but with the appropriate
initial parameters retained.  These integrals are amenable to spectral integration techniques that provide
exponentially convergent, discrete representations of these integrals.  These techniques are outlined in
\cite{NasiOsbuEvan19}, and we make use of these spectral integration techniques in this work as well.

Using \eqref{eqn:Clmkn2D}, we can relate the normalization coefficients for an arbitrary orbit
$C^\pm_{\lhat mkn}(q_{(\a)0})$ to the coefficients for a fiducial geodesic
$\hat{C}^\pm_{\lhat mkn}
\equiv C^\pm_{\lhat mkn}(t_0=\vp_0=q_{r0}=q_{\th 0}=0)$.
Transforming to the shifted angle variables $w_r\equiv q_r+q_{r0}$ and
$w_\th\equiv q_\th+q_{\th 0}$, \eqref{eqn:Clmkn2D} becomes
\begin{multline} \notag
	C^\pm_{\lhat mkn}(q_{(\a)0}) =
	\int_{q_{r 0}}^{2\pi+q_{r 0}}\frac{dw_r}{2\pi}
	\int_{q_{\th 0}}^{2\pi+q_{\th 0}} \frac{dw_\th}{2\pi}\,
	e^{i(kw_\th+nw_r)}
	\\
	\times
	D^\pm_{\lhat mkn}(w_r-q_{r 0},w_\th-q_{\th 0};q_{(\a)0})
	e^{-i(kq_{\th 0}+nq_{r 0})}.
\end{multline}
Because the integrand is periodic on the intervals $q_r,\,
q_\th \in [0,2\pi)$, we can shift the limits of integration,
e.g.,
\begin{align} \notag
	\int_{q_{r 0}}^{2\pi+q_{r 0}} \rightarrow
	\int_{0}^{2\pi} + \int_{2\pi}^{2\pi+q_{r 0}}
	- \int_{0}^{q_{r 0}} \rightarrow \int_{0}^{2\pi}.
\end{align}
Recalling \eqref{eqn:Dlmkn}, we see then that $D^\pm_{\lhat mkn}$ can be
rewritten as
\begin{multline}
	D^\pm_{\lhat mkn}(w_r-q_{r 0},w_\th-q_{\th 0};q_{(\a)0}) =
	\\
	e^{i\xi_{mkn}(q_{(\a)0})}
	D^\pm_{\lhat mkn}(w_r,w_\th; 0, 0, 0, 0),
\end{multline}
where we have defined the phase factor $\xi_{mkn}$,
\begin{multline} \label{eqn:ximknAppendix}
	\xi_{mkn}(q_{(\a)0})\equiv - k q_{\th 0} - nq_{r 0}
	\\
	-\o_{mkn}(\D \hat{t}^{(r)}(q_{r0})+
	\D \hat{t}^{(\th)}(q_{\th 0})-t_0)
	\\
	+ m(\D \hat{\vp}^{(r)}(q_{r0})+\D \hat{\vp}^{(\th)}(q_{\th 0})-\vp_0).
\end{multline}
Combining these results, we see that different initial conditions will only
alter the normalization coefficients by an overall phase,
\begin{equation} \label{eqn:ClmknGenAppendix}
	C^\pm_{\lhat mkn}(q_{(\a)0}) = e^{i\xi_{mkn}(q_{(\a)0})}
	\hat{C}^\pm_{\lhat mkn}.
\end{equation}

\subsection{Resonant sources}

We now look at $r\th$-resonant geodesics, and make use of the resonant angle variable $\bar{q}$ and initial resonant
phase $\bar{q}_0$ introduced in Sec.~\ref{sec:angleVar}.  The resonant normalization constants
$\bar{C}^\pm_{\lhat mN}$ are defined by the integrals
\begin{equation}
\label{eqn:normCRes}
	\bar{C}^\pm_{\lhat mN}(\bar{q}_0) = \int_{r_\text{min}}^{r_\text{max}}
	\frac{\varpi^2\ti{X}^\mp_{\lhat mN}(r)
	\bar{Z}_{\lhat mN}(r;\bar{q}_0)}{W_{\lhat mN}\D}
	dr,
\end{equation}
where $\bar{Z}_{\lhat mN}$ is the radial decomposition of the resonant source in the frequency domain
\begin{equation} \notag
		\rho = -\frac{\varpi^3}{4\pi\Sig\D}\sum_{\lhat mN}
		\bar{Z}_{\lhat mN}(r;q_0)
		S_{\lhat mN}(\th)e^{im\vp}e^{-i\o_{mN}t},
\end{equation}
and $W_{\lhat mN}$ is the Wronskian, defined similarly to \eqref{eqn:Wlmkn} with $\lhat mkn \rightarrow \lhat mN$.
The radial dependence $\bar{Z}_{\lhat mN}$ takes the form
\begin{multline}
\bar{Z}_{\lhat mN}(r;\bar{q}_0) = \frac{1}{2\pi}\int_0^{2\pi}d\bar{q}\,
	\bar{B}_{mN}(\bar{q};\bar{q}_0) e^{iN\bar{q}}
	\\
	\times S_{\lhat mN}(\bar{\th}_p)\d(r-\bar{r}_p),
\end{multline}
where
\begin{equation}
	\bar{B}_{mN}(\bar{q};\bar{q}_0)\equiv-\frac{4\pi q}{\G}
	\frac{\bar{\Sig}_p\bar{\D}_p}{\bar{\varpi}^3_p}
	e^{i\o_{mN}\D \bar{t}-im\D \bar{\vp}},
\end{equation}
$\bar{\Sig}_p =
\bar{r}_p^2+a^2\cos^2\bar{\th}_p$, $\bar{\varpi}^2=\bar{r}_p^2+a^2$, $\bar{\D}_p = \bar{\varpi}^2-2M\bar{r}_p$,
and the geodesic functions $\D \bar{t}$, $\bar{r}_p$, $\bar{\th}_p$, and $\D \bar{\vp}$ are understood to be
functions of $\bar{q}$ and $\bar{q}_{0}$,
\begin{align}
	\D \bar{t} &\equiv \D \hat{t}^{(r)}(\bar{q})
	+ \D \hat{t}^{(\th)}(\bar{q}+\bar{q}_{0})
	- \D \hat{t}^{(\th)}(\bar{q}_{0}),
	\\
	\bar{r}_p &\equiv \hat{r}_p(\bar{q}),
	\\
	\bar{\th}_p &\equiv \hat{\th}_p(\bar{q}+\bar{q}_{0}).
	\\
	\D \bar{\vp} &\equiv \D \hat{\vp}^{(r)}(\bar{q})
	+ \D \hat{\vp}^{(\th)}(\bar{q}+\bar{q}_{0})
	- \D \hat{\vp}^{(\th)}(\bar{q}_{0}).
\end{align}
Equation \eqref{eqn:normCRes} simplifies to
\begin{equation} \label{eqn:ClmN}
	\bar{C}^\pm_{\lhat mN}(\bar{q}_0) = \frac{1}{2\pi}\int_0^{2\pi}d\bar{q}\,
	\bar{D}^\pm_{\lhat mN}(\bar{q};\bar{q}_0) e^{iN\bar{q}},
\end{equation}
where
\begin{multline} \label{eqn:DlmN}
	\bar{D}^\pm_{\lhat mN}(\bar{q},\bar{q}_0)\equiv -\frac{4\pi q}{\G}
	\ti{X}^\mp_{\lhat mN}(\bar{r}_p)S_{\lhat mN}(\bar{\th}_p)
	\\
	\times \frac{\bar{\Sig}_p}{\bar{\varpi}_p W_{\lhat mN}}
	 e^{i\o_{mN}\D \bar{t}-im\D \bar{\vp}}.
\end{multline}
Equation \eqref{eqn:ClmN}, like \eqref{eqn:Clmkn2D} is amenable to spectral integration, though the dependence on
$\bar{q}$ and $\bar{q}_0$ does not separate.  Each value of $\bar{q}_0$ (modulus $2\pi$) leads to a unique
source integral.

\section{Dependence on initial conditions for the self-force}
\label{app:ssfInvariant}

We derive the shifting relation \eqref{eqn:ssfFid2Gen} for the SSF and the GSF in outgoing radiation gauge (though
we expect this derivation can be extended to other gauges as well).

\begin{widetext}
\subsection{Scalar self-force}

We begin by generalizing our angle variable expression for the multipole moments of the retarded self-force given
in \eqref{eqn:angleVarSSF} by retaining all dependence on initial conditions
$q_{(\a)0}$,
\begin{equation}
	{F}^{\text{ret},l}_{\a\pm}(q_r,q_\th; q_{(\a)0}) =
	\sum_{m=-l}^l (D^{lm}_\a {\phi}_{lm}^{\pm})(q_r,q_\th;q_{(\a)0})
	\times {Y}_{lm}(q_r,q_\th; q_{(\a)0}).
\end{equation}
where the functions $Y_{lm}$ and $\phi^\pm_{lm}$ can be expressed in terms of their fiducial forms $\hat{Y}_{lm}$
and $\hat{\phi}^\pm_{lm}$ defined in \eqref{eqn:angleVarYlm} and \eqref{eqn:angleVarEHS}, respectively,
\begin{equation}
\label{eqn:angleVarYlm3}
	Y_{lm}(q_r,q_\th;q_{(\a)0})\equiv \hat{Y}_{lm}(q_r+q_{r0},q_\th+q_{\th 0})
	\times e^{-im(\D\hat{\vp}(q_{r0},q_{\th 0})-\vp_0)},
\end{equation}
and
\begin{align} \label{eqn:angleVarEHS3}
	{\phi}^\pm_{lm}(q_r,q_\th; q_{(\a)0})&\equiv \sum_{\lhat kn}
	\hat{\phi}^\pm_{l\lhat mkn}(q_r+q_{r0})e^{-i(kq_{\th}+nq_{r})}
	e^{-i\o_{mkn}\D\hat{t}(q_r+q_{r0},q_\th+ q_{\th 0})}
	e^{i\o_{mkn}(\D\hat{t}(q_{r0},q_{\th 0})-t_0)}
	e^{i\xi_{mkn}(q_{(\a)0})},
	\\
	&= \sum_{\lhat kn}
	\hat{\phi}^\pm_{l\lhat mkn}(q_r+q_{r0})e^{-i(kq_{\th}+nq_{r})}
	e^{-i(kq_{\th 0}+nq_{r 0})} e^{im(\D\hat{\vp}(q_{r0},q_{\th 0})-\vp_0)},
	\\
	&= \hat{\phi}^\pm_{lm}(q_{r}+q_{r0},q_{\th}+q_{\th 0})
	e^{im(\D\hat{\vp}(q_{r0},q_{\th 0})-\vp_0)},
\end{align}
where the exponential factor of $\xi_{mkn}(q_{(\a)0})$ comes from the dependence of the normalization coefficient
on initial conditions in \eqref{eqn:ClmknGenAppendix}.
\end{widetext}

We see that, upon combining our results for $Y_{lm}$ and $\phi^\pm_{lm}$, the exponential dependence on the
initial conditions will cancel, leaving us with the shifting relation
\begin{equation}
\label{eqn:ssfRetFid2Gen}
	F^{\text{ret},l}_{\a\pm}(q_r,q_\th;q^\mu_0) =
	\hat{F}^{\text{ret},l}_{\a\pm}(q_r+q_{r0},q_\th+q_{\th 0}).
\end{equation}
The same result also holds true for $F^{\text{S},l}_\a$.  The singular contributions along the particle worldline
are only functions of $r_p$, $\th_p$, $u^r$, and $u^\th$, and all of these functions are related to their fiducial
counterparts via
\begin{align}
	r_p(q_r;q_{r0})&=\hat{r}_p(q_r+q_{r0}),
	\\
	\th_p(q_\th;q_{\th0})&=\hat{\th}_p(q_\th+q_{\th0}),
	\\
	u^r(q_r,q_\th;q_{r0},q_{\th 0})&=\hat{u}^r(q_r+q_{r0},q_\th+q_{\th 0}),
	\\
	u^\th(q_r,q_\th;q_{r0},q_{\th 0})&=\hat{u}^\th(q_r+q_{r0},q_\th+q_{\th 0}).
\end{align}
Consequently, \eqref{eqn:ssfRetFid2Gen} also holds true for the regularized SSF $F_\a$.

\begin{widetext}
\subsection{Gravitational self-force}

Using the results presented by van de Meent in his calculation of the GSF \cite{Vand18}, we find that
\eqref{eqn:ssfRetFid2Gen} also extends to the gravitational case, at least for the form of the GSF in the outgoing
radiation gauge (ORG) presented in \cite{Vand18}.  The unregularized $l$-mode contributions to the GSF in the ORG can
be written in the form (see (44) in \cite{Vand18})\footnote{Our expression slightly differs from (44) of \cite{Vand18},
which is missing a factor of $Y_{lm}$.  In our expression, we also reformatted indices to more closely reflect the
notation used in our SSF calculations.}
\begin{multline}
\label{eqn:GSFFullMode}
	F^{\mu,l\pm}_{\text{Rad}}(q_r,q_\th; q_{(\a)0}) = \sum_{\substack{m k n s i j \\ l_1 l_2 \lhat}}
	C^\mu_{m k n s i j}(\hat{r}_p(q_r+q_{r0}),\hat{\th}_p(q_\th+q_{\th 0}))
	\Psi^\pm_{\lhat m k n}(q_{r0}, q_{\th 0})
	\\
	\times
	\prescript{}{2}{R}^{\pm,(i)}_{\lhat m kn}(\hat{r}_p(q_r+q_{r0}))
	{}_{2} b^{l_1}_{\lhat mkn}
	\prescript{m}{s}{\mathcal{A}}_{l_1}^{l_2}
	\prescript{j}{m}{\mathcal{B}}^{l}_{l_2} Y_{lm}(\hat{\th}_p(q_\th+q_{\th 0}),0)
	e^{im\D\hat{\vp}^{(r)}(q_r+q_{r0})}
	e^{im\D\hat{\vp}^{(\th)}(q_\th+q_{\th 0})}
	\\
	\times
	e^{-i\o_{mkn}\D\hat{t}^{(r)}(q_r+q_{r0})}
	e^{-i\o_{mkn}\D\hat{t}^{(\th)}(q_\th+q_{\th 0})}
	e^{-ik(q_\th+q_{\th 0})} e^{-in(q_r+q_{r0})} e^{-i\xi_{mkn}(q_{(\a)0})}
	+ \text{c.c.},
\end{multline}
where the functions and coefficients are defined in \cite{Vand18} and
c.c.~denotes complex conjugation of the previous terms.

As van de Meent and Shah demonstrated in \cite{VandShah15}, the asymptotic amplitudes of the ORG Hertz potential
$\Psi^\pm_{\lhat mkn}(q_{r0},q_{\th 0})$ are proportional to the normalization coefficients for a gravitational
source, also known as Teukolsky amplitudes, $Z_{\lhat mkn}^\pm(q_{r0},q_{\th 0})$.  The Teukolsky amplitudes are also
related to their fiducial forms via \eqref{eqn:ClmknGenAppendix} \cite{FlanHughRuan14}.  Consequently, the
$\xi_{mkn}$ phase in the third line of \eqref{eqn:GSFFullMode} will cancel with the phase dependence that arises
from $\Psi^\pm_{\lhat mkn}(q_{r0},q_{\th 0})=e^{i\xi_{mkn}}\hat{\Psi}^\pm_{\lhat mkn}$.  The GSF in the ORG can then be
expressed in terms of the fiducial GSF $\hat{F}^{\mu,l\pm}_{\text{Rad}}(q_r,q_\th)\equiv{F}^{\mu,l\pm}_{\text{Rad}}
(q_r,q_\th;t_0=q_{r0}=q_{\th 0}=\vp_0=0)$ through the shifting relation
\begin{equation}
\label{eqn:GSFinvariant}
	F^{\mu,l\pm}_{\text{Rad}}(q_r,q_\th; q_{r 0}, q_{\th 0}) =
	\hat{F}^{\mu,l\pm}_{\text{Rad}}(q_r+ q_{r 0},q_\th+q_{\th 0}).
\end{equation}
\end{widetext}

\section{Evolution of the orbital parameters}

For completeness, we provide derivations of the orbit-averaged time derivatives of the orbital parameters $\mu$,
$\mathcal{E}$, $\mathcal{L}_z$, and $\mathcal{Q}$ due to perturbations from the SSF.  Similar derivations can be
found in \cite{DrasFlanHugh05,FlanHughRuan14}.

\subsection{Rest mass}
\label{app:RestMass}

Recall that the SSF is not purely orthogonal to the four-velocity and thus contributes to the evolution of the
particle's rest mass during its evolution,
\begin{equation}
\label{eqn:dmudtauApp}
	\frac{d\mu}{d\tau} = -q^2 u^\a F_\a.
\end{equation}
Because $q^2 F_\a = q\nabla_\a \Phi^\text{R}$, we can directly integrate \eqref{eqn:dmudtauApp},
\begin{equation}
	\mu(\tau) = \mu_0 - q \Phi^{\text{R}}(x_p^\mu(\tau)),
\end{equation}
where the integration constant $\mu_0$ is commonly refereed to as the \textit{bare mass}.  Note that the second term
is of $O(q^2/M)$ so that $\mu/M$ is constant at leading order in $q/M$.  After a full orbital period, the value of
the regular field will return to itself.  Thus, via the second fundamental theorem of calculus, the orbit average of
the time derivative of the rest mass vanishes,
\begin{equation} \label{eqn:muDotAverage}
	\langle \dot{\mu} \rangle = 0,
\end{equation}
where $\dot{x} \equiv dx/dt$ and the angle-bracket averages are defined in \eqref{eqn:orbitAvg}.

\subsection{Orbital energy and angular momentum}
\label{app:EnAndLz}

In Kerr spacetime, orbital quantities are typically calculated with respect to Mino time $\la$, rather than
coordinate time $t$.  Therefore, we reexpress the orbit-averaged time derivative of the orbital energy
$\langle \dot{\mathcal{E}}\rangle$ as an average over Mino time $\la$ \cite{DrasHugh04,DrasFlanHugh05}
\begin{equation}
\label{eqn:drasHugh}
	\langle \dot{\mathcal{E}} \rangle
	= \frac{1}{\G} \left\langle \frac{d\mathcal{E}}{d\la} \right\rangle_\la
	= \frac{1}{\G} \left\langle \Sig \frac{d\mathcal{E}}{d\tau} \right\rangle_\la,
\end{equation}
where $\langle \mathcal{X} \rangle_\la$ refers to an orbit average of the quantity $\mathcal{X}(\la)$ with respect
to $\la$.  The right-hand side of \eqref{eqn:drasHugh} can be further expanded by taking the proper-time derivative
of \eqref{eqn:EAndLzConstants},
\begin{align}
\label{eqn:dEdtauLong}
	\frac{d\mathcal{E}}{d\tau} &= u^\a\nabla_\a\left(g_{\mu\nu}
	\xi^\mu_{(t)} u^\nu \right), \\
	& = g_{\mu\nu}
	\left(u^\nu u^\a\nabla_\a\xi^{\mu}_{(t)}
	+ \xi^{\mu}_{(t)}u^\a \nabla_\a u^\nu\right), \\
	& = g_{\mu\nu} \xi^{\mu}_{(t)}u^\a \nabla_\a u^\nu,
\end{align}
where $u^\nu u^\a\nabla_\a \xi^\mu_{(t)}$ vanishes due to $\xi^\mu_{(t)}$ satisfying Killing's equation.  Defining
the four-acceleration $a^\mu$ and perpendicular self-force $f^\mu$,
\begin{equation}
	f^\mu = \mu a^\mu \equiv (g^{\mu\nu}+u^\mu u^\nu)F_{\nu},
\end{equation}
\eqref{eqn:dEdtauLong} then takes the compact form
\begin{equation}
\label{eqn:dEdtau}
	\mu \frac{d\mathcal{E}}{d\tau} = - q^2 f_t.
\end{equation}
Combining \eqref{eqn:dEdtau} and \eqref{eqn:drasHugh}, we find that
\begin{align}
	\mu\langle \dot{\mathcal{E}} \rangle
	&= -\frac{q^2}{\G} \left\langle \Sig f_t \right\rangle_\la,
	\\
	&= -\frac{q^2}{\G} \left\langle \Sig F_t \right\rangle_\la,
\end{align}
where in the second line, we have taken into account that
\begin{align}
	\left\langle \Sig f_t \right\rangle_\la
	&= \left\langle \Sig (F_t + u_t u^\a F_\a) \right\rangle_\la,
	\\
	&= \left\langle \Sig F_t \right\rangle_\la
	+ q^{-2}\mathcal{E} \left\langle \frac{d\mu}{d\la} \right\rangle_\la,
	\\
	& = \left\langle \Sig F_t \right\rangle_\la.
\end{align}
The orbit-averaged rate of change of the $z$ component of the orbital angular momentum,
$\langle \dot{\mathcal{L}_z} \rangle$, can be derived in a similar manner,
\begin{equation}
	\mu \langle \dot{\mathcal{L}_z} \rangle
	= \frac{q^2}{\G} \left\langle \Sig F_\vp \right\rangle_\la.
\end{equation}

\subsection{Carter constant}
\label{app:Qdot}

{\renewcommand{\arraystretch}{1.45}
\begin{table*}[tb!]
	\caption{Fractional variation in the resonant fluxes for the scalar models listed in Table
	\ref{tab:orbits}. Fractional variations are reported to four decimal places for brevity.}
	\label{tab:fluxesVar}
	\centering
	\begin{tabular*}{\textwidth}{c @{\extracolsep{\fill}} c c c c c c c}
		\hline
		\hline
		\multicolumn{2}{c }{\quad source} & $\D\dot{E}^\mathcal{H}$
		& $\D\dot{L}_z^\mathcal{H}$ & $\D\dot{E}^\infty$
		& $\D\dot{L}_z^\infty$ & $\D\dot{E}^\text{tot}$
		& $\D\dot{L}_z^\text{tot}$
		\\
		\hline
		\multicolumn{2}{c}{\quad $e02.13$} & $0.0011\%$ & $0.0004\%$
		& $0.0001\%$ & $0.0002\%$ & $0.0001\%$ & $0.0003\%$
		\\
		\multicolumn{2}{c}{\quad $e02.12$} & $0.0204\%$ & $0.0047\%$
		& $0.0017\%$ & $0.0023\%$ & $0.0016\%$ & $0.0029\%$
		\\
		\multicolumn{2}{c}{\quad $e02.23$} & $0.2008\%$ & $0.0252\%$
		& $0.0156\%$ & $0.0138\%$ & $0.0129\%$ & $0.0158\%$
		\\
		\multicolumn{2}{c}{\quad $e05.13$} & $1.1214\%$ & $0.0400\%$
		& $0.0109\%$ & $0.0227\%$ & $0.0092\%$ & $0.0287\%$
		\\
		\multicolumn{2}{c}{\quad $e05.12$} & $2.3850\%$ & $0.1352\%$
		& $0.0427\%$ & $0.0684\%$ & $0.0352\%$ & $0.0849\%$
		\\
		\multicolumn{2}{c}{\quad $e05.23$} & $5.8510\%$ & $0.3400\%$
		& $0.1835\%$ & $0.1975\%$ & $0.1575\%$ & $0.2262\%$
		\\
		\hline
		\hline
	\end{tabular*}
\end{table*}
}

The orbit-averaged rate of change of the Carter constant $\langle \dot{\mathcal{Q}} \rangle$ can be derived using
the process outlined in the previous section, \ref{app:EnAndLz}.  Like $\langle \dot{\mathcal{E}}\rangle$ and
$\langle \dot{\mathcal{L}}_z\rangle$, $\langle \dot{\mathcal{Q}} \rangle$ can be reexpressed in terms of an
average over $\la$,
\begin{align}
	\mu\langle \dot{\mathcal{Q}} \rangle
	= \frac{\mu}{\G} \left\langle \Sig \frac{d\mathcal{Q}}{d\tau}
	 \right\rangle_\la .
\end{align}
Once again we can expand the right-hand side by taking the proper time derivative of the Carter constant as defined
by \eqref{eqn:Q},
\begin{multline}
\label{eqn:dQdtauApp}
	\frac{d\mathcal{Q}}{d\tau}= \mu u^\mu u^\nu u^\a \nabla_\a K_{\mu\nu} +
	2 u^\mu K_{\mu\nu} u^\a \nabla_\a u^\nu
	\\
	- 2(\mathcal{L}_z-a\mathcal{E})
	\left(\frac{d\mathcal{L}_z}{d\tau}-a\frac{d\mathcal{E}}{d\tau}\right),
\end{multline}
which, after utilizing the properties of the Killing tensor (i.e., $\nabla_{(\a}K_{\mu\nu)}=0$) and recalling the
equations of motion, further reduces to
\begin{align}
\label{eqn:dKdt1}
	\frac{d\mathcal{Q}}{d\tau} &= 2  K_{\mu\nu} u^\mu a^\nu - 2(\mathcal{L}_z-a\mathcal{E})
	\left(\frac{d\mathcal{L}_z}{d\tau}-a\frac{d\mathcal{E}}{d\tau}\right).
\end{align}
The Killing tensor $K^{\mu\nu}$ can be expressed in terms of the Kinnersley basis vectors
\begin{align}
	l^{\mu} &= \frac{1}{\D}(\varpi^2,\D,0,a),
	\\
	n^\mu &= \frac{1}{2\Sig}(\varpi^2,-\D,0,a),
	\\
	m^\mu &= \frac{1}{\sqrt{2}(r+ia\cos\th)}(ia\sin\th,0,1,i\csc\th),
\end{align}
giving
\begin{align}
	K^{\mu\nu} &= \Sig (m^\mu ({m}^\nu)^* + ({m}^\mu)^* m^\nu) - a^2\cos^2\th g^{\mu\nu},
	\label{eqn:KmunuMM}
	\\
	&= \Sig(l^\mu n^\nu + n^\mu l^\nu) + r^2 g^{\mu\nu},\label{eqn:KmunuLN}
\end{align}
where the star denotes complex conjugation.  The equivalency of \eqref{eqn:KmunuMM} and \eqref{eqn:KmunuLN} is
clearly demonstrated by expressing the metric in terms of this new basis
\begin{equation}
	g^{\mu\nu} = -(l^\mu n^\nu + n^\mu l^\nu) + (m^\mu ({m}^\nu)^* + ({m}^\mu)^* m^\nu).
\end{equation}
Plugging \eqref{eqn:KmunuMM} into \eqref{eqn:dKdt1} and multiplying by $\mu/2$, we find that
\begin{align}
\label{eqn:QdotTh}
	\frac{\mu}{2}\frac{d\mathcal{Q}}{d\tau} &= \notag
	q^2(\mathcal{L}_z\csc^2\th  -a \mathcal{E})(f_\vp+a\sin^2\th f_t)
	\\
	& \qquad + q^2 u_\th f_\th - q^2(\mathcal{L}_z-a\mathcal{E})
	\left(f_\vp+af_t\right),
	\\
	&= \notag
	q^2(\csc^2\th \mathcal{L}_z-a \mathcal{E})(F_\vp+a\sin^2\th F_t)
	\\ \label{eqn:QdotTh2}
	& \qquad + q^2 u_\th F_\th - q^2(\mathcal{L}_z-a\mathcal{E})
	\left(F_\vp+aF_t\right)
	\\ \notag
	&\qquad \qquad \qquad \qquad
	- q^2(\mathcal{Q}- a^2\cos^2\th)\, u^\a F_\a.
\end{align}
Alternatively, plugging \eqref{eqn:KmunuLN} into
\eqref{eqn:dKdt1}, we find that
\begin{align} \label{eqn:QdotR}
	\frac{\mu}{2}\frac{d\mathcal{Q}}{d\tau} &= \notag
	q^2\D^{-1}(a \mathcal{L}_z-\varpi^2 \mathcal{E})
	\left(a f_\vp+\varpi^2 f_t\right)
	\\
	&\qquad - q^2 \D u_r f_r - q^2(\mathcal{L}_z-a\mathcal{E})
	\left(f_\vp+af_t \right),
	\\
	&= \notag
	q^2\D^{-1}(a \mathcal{L}_z-\varpi^2 \mathcal{E})
	\left(a F_\vp+\varpi^2 F_t\right)
	\\ \label{eqn:QdotR2}
	&\qquad - q^2 \D u_r F_r - q^2(\mathcal{L}_z-a\mathcal{E})
	\left(F_\vp+aF_t \right)
	\\ \notag
	&\qquad \qquad \qquad \qquad \qquad \quad
	- q^2(\mathcal{Q} + r^2) \, u^\a F_\a .
\end{align}
Regardless of which expression we use, due to the time variation of the rest mass \eqref{eqn:dmudtauApp}, all four
components of the SSF must be known to evaluate $\langle \dot{\mathcal{Q}}\rangle$, though the terms proportional
to $u^\a F_\a$ will vanish after taking the orbit average of \eqref{eqn:QdotTh2} and \eqref{eqn:QdotR2}.  In this
work, we evaluate $\langle \dot{\mathcal{Q}}\rangle$ using \eqref{eqn:QdotTh}-\eqref{eqn:QdotR2} and check for their
consistency as one way of validating our numerical results.

\section{Fractional variations of the fluxes}
\label{app:fracVarFlux}

In Table \ref{tab:fluxesVar} we report the total fractional variations $\D \dot{E}$ and $\D \dot{L}_z$, where
\begin{equation}
\label{eqn:fracVar}
	\D X \equiv 2\,\left\vert\frac{\text{max}\left[\langle X\rangle\right]
	-\text{min}\left[\langle X\rangle\right]}
	{\text{max}\left[\langle X\rangle\right]
	+\text{min}\left[\langle X\rangle\right]}\right\vert.
\end{equation}
The total fractional variations primarily increase as $p$ and $e$ increase.  In fact, the fractional flux variations
for the $e02.13$ orbit are negligible compared to the other sources.  This suggests that the dependence of the
fluxes on the initial phase is dampened as a source moves closer to the more massive primary.  While $e05.13$ has a
smaller pericenter value than $e02.13$, the larger variation in the radial motion could then be responsible for the
larger variation in the fluxes with respect to $\bar{q}_0$.  The behavior of these flux fluctuations and the
magnitude their variations is consistent with the gravitational fluxes of \cite{FlanHughRuan14}.

\bibliography{resPaper}

\end{document}